\documentclass{agujournal}
\journalname{JGR-Planets}

\usepackage[finalnew]{trackchanges}
\usepackage{nameref}
\begin{document}

\title{Impact Dynamics of Moons Within a Planetary Potential}
\authors{R. Rufu\affil{1,2 }  and O. Aharonson\affil{1,3}}
\affiliation{1}{Weizmann Institute of Science, Department of Earth and Planetary Sciences, Rehovot, Israel}
\affiliation{2}{Southwest Research Institute, 1050 Walnut Street, Suite 300, Boulder, CO 80302, USA}
\affiliation{3}{Planetary Science Institute, Tucson, AZ, USA}
\correspondingauthor{R. Rufu}{raluca@boulder.swri.edu}


\begin{keypoints}
\item Impacts within the planetary potential are more erosive.
\item Dynamical studies of the evolution of multiple satellite, must consider fragmentation.
\item Significant portions of the lunar mantle are melted during a companion impact.
\end{keypoints}

%

\begin{abstract}
\change{Earth's Moon shows indications that it may have resulted from a merger between two (or more) smaller moonlets}{Current lunar origin scenarios suggest that Earth's Moon may have resulted from the merger of two (or more) smaller moonlets}. Dynamical studies of multiple moons find that these satellite systems are not stable, resulting in moonlet collision or loss of one or more of the moonlets. We perform Smoothed Particle Hydrodynamic (SPH) impact simulations of two orbiting moonlets inside the planetary gravitational potential and find that the classical outcome of two bodies impacting in free space is altered as erosive mass loss is more significant with decreasing distance to the planet. Depending on the conditions of accretion, each moonlet could have a distinct isotopic signature, therefore, \remove{in this work} we assess the initial mixing during their merger\add{, in order to estimate whether future measurements of surface variations could distinguish between lunar origin scenarios (single vs. multiple moonlets)}. We find that for comparable-size impacting bodies in the accretionary regime, surface mixing is efficient, but in the hit-and-run regime, only small amount of material is transferred between the bodies. However, sequences of hit-and-run impacts are expected, which will enhance the surface mixing. \remove{For smaller impactors, disruption of the impactor and the reaccretion of its material on the surface of the larger moonlet induces surface mixing.}Overall, our results show that large scale heterogeneities can arise only from the merger of drastically different component masses. Surfaces of moons resulting from merger of comparable-sized components have little material heterogeneities, and such impacts are preferred, as the relatively massive impactor generates more melt, extending the lunar magma ocean phase. 

\textbf{Plain Language Summary} 

\add{Lunar origin scenarios suggest that} Earth's single Moon may be a result of merger between smaller moonlets. In this work we test different impact scenarios of two orbiting moonlets, and include the effect of Earth's gravitational potential. We find that impacts within Earth's gravitational potential are different from impacts of two bodies in free space. For example, the amount of debris generated during such impacts is enhanced, and thus the mass retained in the final Moon is smaller than previously estimated. As each moonlet could inherit distinct isotopic signature, we test the initial surface mixing between the two components\add{, in order to estimate whether future measurements of surface variation could differentiate between lunar origin theories.} \change{and}{We} find that surfaces of moonlets resulting from the merger of two comparable-size components are efficiently mixed. For the origin of the Moon, comparable-sized components are preferred, as they are able to generate enough melt and extend the lunar magma ocean phase to be compatible with the observed lunar crust. 
\end{abstract}
%

\section{Introduction}

\label{sec:intro}
Impacts between two orbiting satellites may be an integral part of satellite formation and specifically lunar formation. Mergers between moonlets are especially interesting for the newly proposed multiple-impact hypothesis as these moonlets form from different debris disks and merge together to form the final Moon \citep{Rufu:2017aa}. However, this process is also relevant for the single giant impact, as previous work shows that multiple moonlets can form from the same debris disk  \citep{ida1997lunar,salmon2012lunar,Salmon20130256}. 

Satellite pairs were found to be mostly unstable \citep{Canup1999,citron2018role}, leading to moonlet-moonlet collisions or the loss of one (or both) of the moonlets. In the context of the canonical giant impact, where two moonlets are accreted in the same debris disk, merging occurs \add{rapidly}\remove{after significant tidal evolution} when the inner moonlet is larger \add{than the outer moonlet} \citep{Canup1999}. In the context of the multiple impact hypothesis,
\cite{citron2018role} demonstrated that a preexisting moonlet can remain stable during subsequent impacts onto the protoplanet and later merge with the newly accreted moonlet. While these studies were able to evaluate the dynamical evolution of the two moonlets up to their impact, they do not estimate the impact outcome, and rather assume perfect merger of the two components. However, the collisional outcomes vary substantially from perfect merger of the two colliding bodies at low velocities, to hit-and-run where the two bodies graze each other but have sufficient relative velocity to escape the mutual gravitational well \citep{Leinhardt}. In the hit-and-run regime little mass is transferred between the two colliding bodies. 

The dynamics of impacts between two orbiting bodies is substantially different from previously heavily studied planetary-sized impacts \citep{hartmann1975satellite, cameron1976origin, BENZ1989113,canup2001origin,Canup:2004aa,canup2008lunar, cuk2012making,Canup23112012, Rufu:2017aa,lock2018origin}. Firstly,\add{ for orbiting sub-lunar mass bodies around Earth,} the impact velocities are smaller and limited to $\sim1\,{\rm km/s}$, thus heating is limited. Secondly, multiple satellite systems that typically lead to close encounters are moonlets with comparable sizes \citep{Canup1999}. Therefore, moonlet-moonlet collisions offer an interesting, and only marginally explored, impact size-distribution \citep{Canup:2005aa,asphaug2010similar} where both moonlets would contribute similarly and substantially to the final satellite, as opposed to the planetary-scale impacts, that were thoroughly studied in the context of the Moon formation \citep{BENZ1989113,Canup:2004aa,cuk2012making,Rufu:2017aa}. Thirdly, in the context of planetary scale impacts the  solar tidal environment is neglected because the planetary Hill sphere is substantially large, compared to the radius of the components, such that the impact dynamics are not influenced by solar tides. However, this approximation is not necessarily true for moonlet impacts, especially if the impact occurs close to the planet. Therefore, moonlet impacts can be more erosive than planetary impacts as the velocity of ejected material required to reach the mutual Hill sphere is smaller than the gravitational escape velocity, altering the merger efficiency \citep{canup1995accretion}. 

Moonlet-moonlet collisions may be relevant for other satellite systems, for example, in the Neptunian system, where typical impact velocities between Triton and a primordial satellite system are expected to be high \citep{RufuCanup}, however, the impacts occur farther away from the planet, where tidal effects are minimal. In the Saturnian system, collisions among small and close-in satellites, could have created their peculiar shapes \citep{leleu2018peculiar}. Moreover, collision among satellites or within a potential well may be important for KBO objects as well. For example, in the formation of Haumea's system (consisting of two satellites, collisional family members, \citealt{Brown:2007aa}, and a newly observed ring,
\citealt{Ortiz:2017aa}) by a collision from an unbound KBO onto a pre-existing satellite \citep{SchlichtingSari}. For this system it has been proposed that the same impact is able to disrupt the satellite\change{and}{,} form the current two companions and probably the ring \citep{SchlichtingSari}. 

In this work we estimate the merger efficiency of impacts within the planetary gravitational potential, where tidal forces alter the amount of mass that comprises the final moon. In order to directly compare our results with the standard impacts of two bodies in free space, we perform reference simulations with no central potential. 

Previous simulations show that protolunar debris disks and their accreted moonlets have different isotopic signatures, depending on the parameters of the collision with the planet and the impactor's isotopic signature \citep{Rufu:2017aa}. Moreover, in the single giant impact \add{scenario}, although moonlets are formed from the same debris disk, the moonlets are not sourced from the same regions in the debris disk. Material from the inner debris disk (inside the Roche limit), comprising most of the secondary moonlet, may experience some equilibration with the proto-Earth \citep{Pahlevan:2007aa,salmon2012lunar,Salmon20130256}, therefore isotopic signatures between the two moonlets may still arise. After the merger of the two moonlets, the surface solidified in the the first $1000\,\rm{yr}$, forming a conductive lid and prolonging the complete lunar solidification to $\sim10\,\rm{Myr}$ \citep{Elkins2011}. Therefore, the surface records the oldest stages of the lunar thermal evolution. \add{Efficient post-impact convection}, subsequent impacts \add{and/}or subsequent melting due to tidal heating \citep{meyer2010coupled},  could disrupt the initial crust, but some initial heterogeneities \change{could}{may} still be preserved on older terrains, such as, the  lunar farside. In this study, we estimate the resulting initial surface mixing between moonlets after their merger, in order to test whether initial lunar crustal heterogeneities could emerge from the last global accretionary event \citep{Robinson2016244}\add{ and estimate whether future observed heterogeneities in lunar samples could distinguish between lunar origin scenarios (single vs. multiple moonlets)}.\par
Anorthositic material is widely observed on old lunar terrains and \change{estimated}{considered} to represent the primordial lunar crust, formed from the flotation of plagioclase minerals from the lunar magma ocean  (\citealt{Elkins2011}\add{; current  crystallization studies require a depth of $\sim500\,\rm km$ to explain the observed lunar crustal thickness;} \citealt{zuber2013gravity,charlier2018crystallization}). An impact that occurs after the formation of the anorthositic crust (less than $1000\ {\rm yr}$; \citealt{Elkins2011}) can cause massive resurfacing, disrupting the previous anorthositic crust. Therefore, the amount of melt that is generated from moonlet impacts is important in order to access whether an anorthositic global crust could reform from the secondary magma ocean phase.\par
In this work we perform hydrodynamic simulations of impacting moonlets within a planetary potential (described in section 2) to asses: (1) the merger efficiencies and to compare them to impacts of bodies in free space (section 3.1); (2) the initial surface mixing between the two moonlets (section 3.2); (3) the amount of melt that occurs during these impacts (section 3.3).

%
\section{Methods}\label{sec:Methods}

We perform Smoothed Particle Hydrodynamic (SPH) impact simulations of two orbiting moonlets, using GADGET2 \citep{Springel2005} with a tabulated M-ANEOS equation of state \add{for forsterite} \citep{melosh2007hydrocode} \add{and ANEOS for iron} \citep{thompson1990aneos} \add{(code modifications were performed by }\citealt{Marcus2009,marcus2011role} \add{and available in the supplemental material of }\remove{as described in previous work}\citealt{cuk2012making}). In this implementation material strength is not introduced as previous studies of impacts with \change{similar}{lunar}-sized bodies did not find significant differences when adding material forces \citep{Jutzi:2011aa}. Note that recent work suggests that material strength effects can increase the required energy to disrupt the body, particularly for small masses \citep{jutzi2015sph}. However, the expected impacting velocities in this scenario \citep{citron2018role} are lower than the catastrophic disruption threshold. For planetary scale impacts, material strength was shown to more effectively disperse the heating in the mantle induced by the impact shock \citep{emsenhuber2018sph}, therefore purely hydrodynamic simulation \change{will}{may} underestimate the \add{initial} amount of  melting in the mantle.\add{ Further studies are required in order to estimate the effect of material strength on the initial melt distribution and later stage deformation. }While the exact role of material strength is somewhat unknown, we expect it to have a minimal effect on the impact results in this study, especially in regards to the final moonlet mass.

The impacting material consists of $\sim2\times10^{5}$ particles while the planet is simulated using a small number of particles ($\sim10^{4}$). The small planetary resolution efficiently represents the gravitational potential, however discontinuities in the hydrodynamic calculation could emerge if particles with distinctly different masses, as is the case for planetary and impacting material, are in close vicinity. In this current setup, these two types of particles rarely interact directly. 

As the cores of the colliding bodies \add{may} erode during impact\remove{(because portions of their mass are incorporated into the mantle)}, we assume that the colliding bodies have 10\% iron material, similar to the current upper estimate for the lunar iron fraction \citep{Wood1986,canup2004simulations}. To cover the range of possible Moon formation scenarios, we tested two extreme cases of initial thermal states, "cold" and "hot" (from previous studies of lunar evolution, \citealt{Laneuville2013}, see Text S1 in Supplementary material). Small differences were observed between the outcomes of the two states (see Figure S3 and S5 in Supplementary material), therefore the following results assume a "cold" initial thermal state. The bodies are simulated in isolation for 10 hours to allow their initial relaxation and to establish gravitational equilibrium. The temperature is then corrected to the initial thermal profile and simulated in isolation again for 10 hours for further relaxation. We verify this configuration is stable by ensuring that the RMS velocity of the particles is $<1\%$ of the typical impact velocity, $\sim1\ {\rm km/s}$. 

\begin{figure*}
 \begin{centering}
 \includegraphics[scale=0.25]{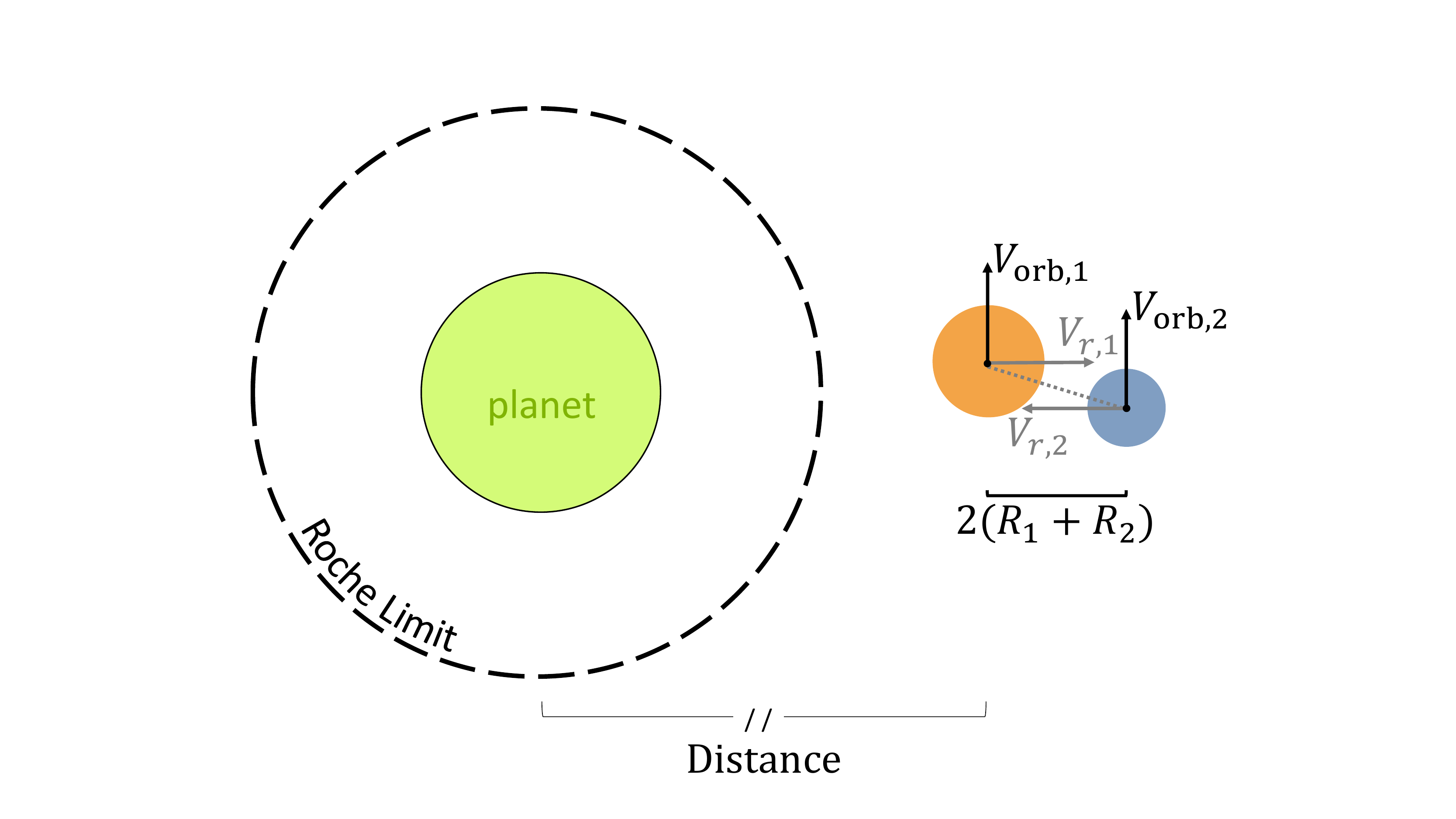}
 \par\end{centering}
 \caption[Figure 1]{\label{Figure 1} Schematic of the initial setup of the simulations. Each moonlet is given a circular orbital velocity $(V_{{\rm orb,{i}}})$ with an additional radial velocity component $(V_{r,{i}})$. The black dashed line represents the Roche limit of the planet. Sizes and distances are not to scale..}
 \end{figure*}

The moonlets are placed on a orbit around an Earth-mass planet ($M_\oplus$) at three given distances ($1.5,\,3,\,5\,R_{{\rm Roche}}$, where $R_{\rm Roche}= 18,381\ \rm{km}$ is Earth's fluid Roche limit). The smaller distance represents a collision of two moonlets that are formed from the same debris disk and will impact early in their evolution \citep{Canup1999,Jutzi:2011aa}. The larger distances represent the collision of two moonlets that are formed in different debris disks, as predicted by the multiple impact hypothesis. These impacts will occur farther away from the planet as the moonlets experienced some tidal evolution before impact \citep{citron2018role}.
The moonlets are placed at $2(R_1+R_2)$  (where $R_i$ is the radius of moonlet $i$) distance from each other at the start of the simulation in order to allow for tidal forces to change the shape of the moonlets prior to impact (see \nameref{Figure 1}). We vary the mass ratio of the impacting moonlets ($\gamma=M_{2}/M_{{\rm tot}}$, where $M_{2}$ is the smaller body and $M_{\rm tot}$ is the total impacting mass). In this work we choose to focus on the last impact that formed the Moon, hence the total impacting mass is constant and equal to one lunar mass ($M_{{\rm tot}}=1M_{{\rm moon}}$).  Initially, each moonlet is given a circular orbital velocity $(V_{{\rm orb,{i}}})$ with an additional radial velocity component $(V_{r,{i}})$, such that the radial velocity of the center of mass of the moonlets is zero. The maximum allowed impact velocity is the maximum radial velocity that maintains each moonlet on closed orbit around the planet with a perigee larger than the Roche limit, $R_{\rm Roche}$, therefore, for smaller mass ratios or for larger planetary distance the maximum allowed velocity decreases. Because the moonlets move slightly around the planet before the impact (and the exact moment of impact is not recorded by the SPH output), we numerically estimate the impact angle and velocity by evolving the two moonlets until contact is established (by computing the 3-body gravitational interactions), assuming the bodies retain their spherical shape. 

The simulations cover 48 hours of simulated time after impact. In some cases where the resulting moons did not reach a stable state, the simulations are continued for an additional 24 hours. It should be noted that for eccentric moonlets with perigees close to Roche limit or moonlets that continue to pass through a debris disk, the shape and surface of the moonlet will constantly change and equilibrium cannot be reached. The realistic simulated time is limited to a few days due to artificial viscosity that is introduced in the SPH code in order to mimic shock dissipation \citep{canup2004simulations}.

At the end of the simulation, we use K-means clustering machine learning algorithm \citep{spath1985cluster} to divide the mass into two clusters and obtain an initial guess on the mass and radius of the largest cluster. Subsequently, we use an iterative algorithm to find the bounded mass to that specific cluster. The algorithm uses the initial guess and classifies particles as bounded to the moonlet if they are within the Hill radius of the moonlet ($r_i<R_{{\rm Hill,}i}$, where $r_i$ is the distance of the particle to moonlet $i$, $R_{{\rm Hill,}i}=a_i(M_i/3(M_\oplus+M_i))^{1/3}$ is the Hill radius of moonlet $i$ and $a_i$ its semimajor axis) and have a velocity smaller than that required to reach the Hill sphere of the moonlet \citep{Bierhaus_2012,Alvarellos_2002}:
\begin{equation}
V_{\rm{Hill}}^2=\frac{2GM_i}{r_i}\left(\frac{R_{{\rm Hill,}i}^2-R_{{\rm Hill,}i}\cdot r_i}{R_{{\rm Hill,}i}^2-r_i^2\sin^2 \zeta_i}\right) \label{eq:V_hill}
\end{equation}
where $G$ is the gravitational constant, $M_i$ is the mass of the moonlet $i$, and $\zeta$ is the angle of the particle relative to the moonlet (following \citealp{Bierhaus_2012}, we assume that $\sin^2 \zeta\approx 0.5$). The moonlet's mass, Hill radius, and center of mass are adjusted and bounded particles are recalculated until convergence is achieved (typically a few iterations). Clustering algorithm is repeated on the material that is not bounded to the previously classified cluster, until the mass with high-density ($\rho>1\,{\rm g/cm^{3}}$) is smaller than $0.01\,M_{{\rm moon}}$ or the two most massive clusters have been classified. 

In order to directly compare our results with the standard impacts of two bodies in free space we perform additional simulations with the same impact parameters but without a central potential. To insure that the angle and velocity of the impact at contact are equal to the previous set of simulations, we perform a backward integration (2-body interaction) from the moment of contact until a distance of 2 radii. We set these new positions and velocities as the new initial setup. We verified that the impact parameters in both cases (with/without central potential) are similar by comparing the positions and velocities of the moonlets at contact, as simulated by the hydrodynamic code. The resulting impact angle and velocity difference is $<1^{o}$ and $<0.001\,V_{\rm esc}$, respectively (where the mutual escape velocity is defined as $V_{\rm esc}=\sqrt{2G(M_{\rm 1}+M_{\rm 2})/(R_{\rm 1}+R_{\rm 2})}$.

%
\section{Results}


\subsection{Merger Efficiency}
\label{subsec:MergerEfficiency}
\begin{figure}
\hspace{-1cm}
\begin{tabular}{lll}
\ \ \ \ a) $1.5\,R_{{\rm Roche}}$ & b) $3\,R_{{\rm Roche}}$ & c) Free space \\
\includegraphics[height=4.05cm]{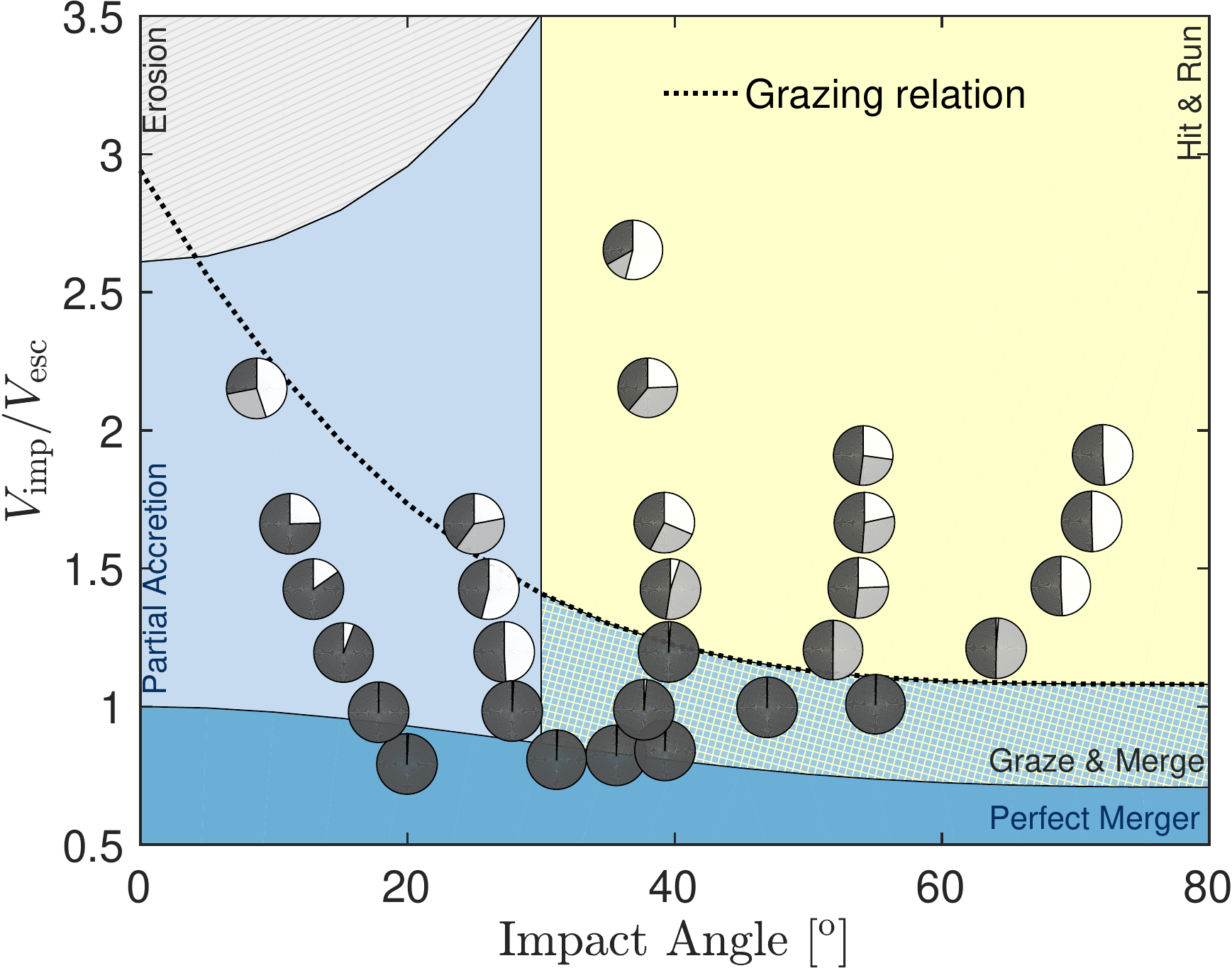} & \includegraphics[height=4cm]{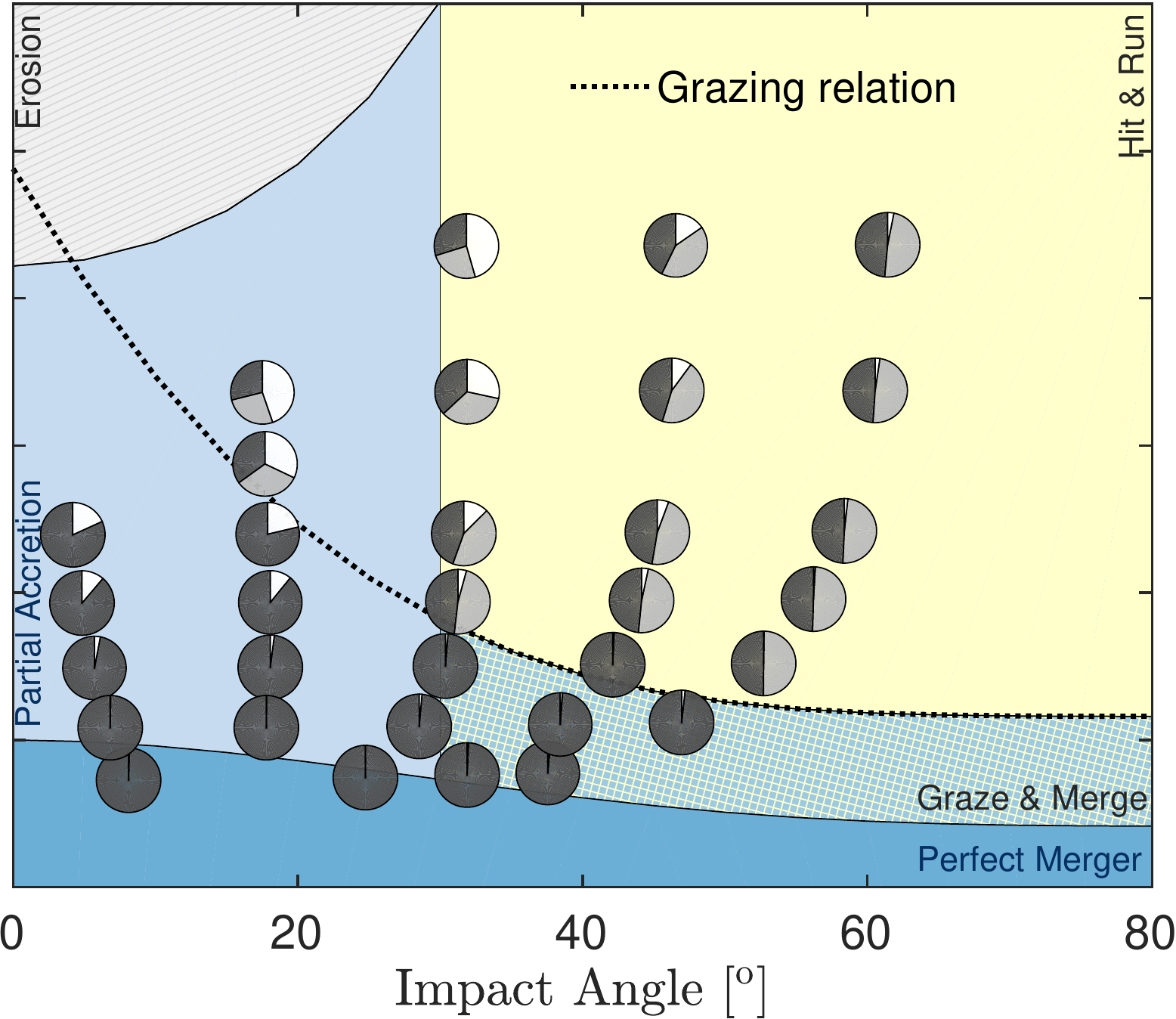} & \includegraphics[height=4cm]{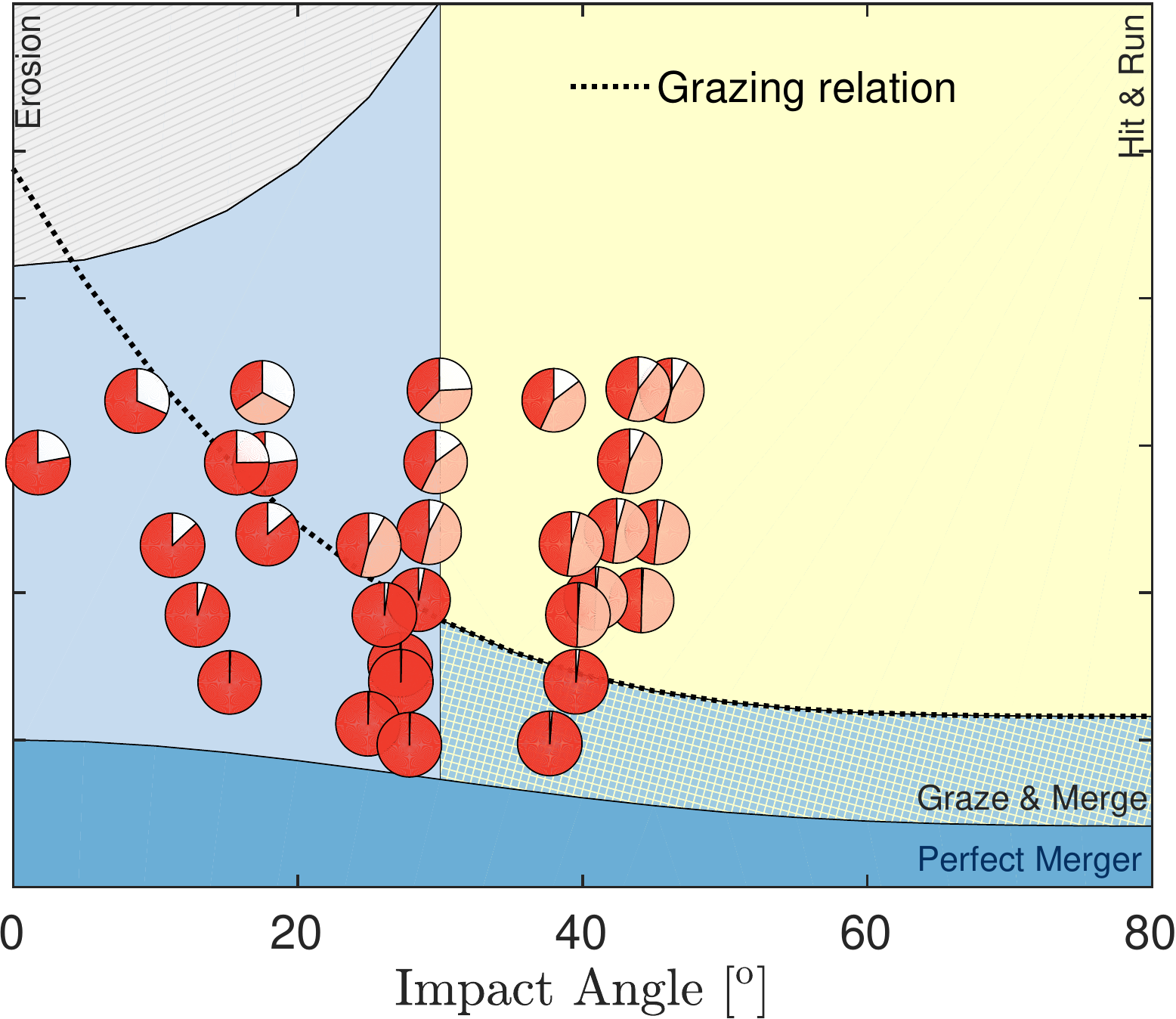}
\end{tabular}
 \caption[Figure 2] {\label{Figure 2} Merger efficiencies for impacts between two half-moon-sized bodies in angle-velocity phase space. Impacts occurred at a distance of a) $1.5~R_{{\rm Roche}}$; b) $3~R_{{\rm Roche}}$ from the planet; for comparison, c) impacts of two bodies in free space (no central potential). Each pie plot represents the distribution of the final impacting mass (the largest remnant - dark gray/red; if present the second largest remnant - light gray/red; debris material - white). The different regions of the plot represent different impacting regimes previously defined by \citet{Leinhardt} for two bodies in free space (with the best fitted parameter of energy dissipation within the target, $c^{*}$=2.8). The dashed line represents the critical velocity for which grazing occurs, previously defined by \citet{Genda}.}
\end{figure}

Previous works of ring material accretion near the Roche zone emphasized the difference between the accretion under the influence of tidal forces and accretion of material far\ from a central potential \citep{canup1995accretion}. However, as these studies estimate the accretion rate of debris composed of many small components, they do not consider the effect of the angle for each impact, but rather average the accretion efficiency for all angles, or assume radial alignment of the components, which promotes accretion \citep{canup1995accretion}. Our results show \add{the} transitions between accretion and hit-and-run regime at different angles and velocities. Estimating when grazing impacts occur is important because the merger efficiencies in this region is low and interactions between grazing moonlets can destabilize them towards the Roche limit, where they will be disintegrated by tidal forces (\nameref{Figure 2}-a; \textit{e.g.} simulations in the hit-and-run regime with high angles and velocities above $1.4\ V_{{\rm esc}}$). 

\begin{figure}
\begin{raggedright}
a) 
\par\end{raggedright}
\begin{centering}
\includegraphics[width=1\linewidth]{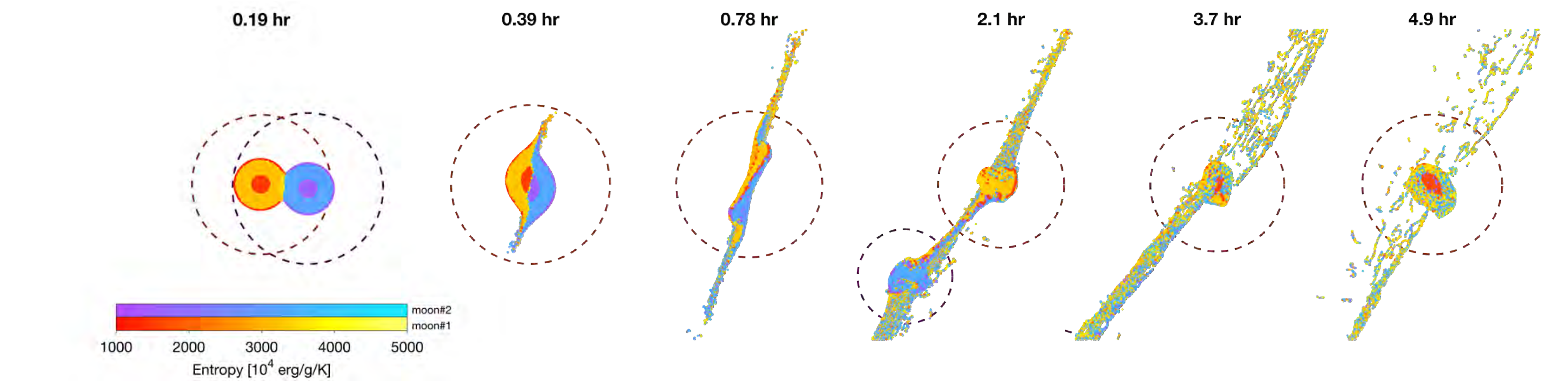}
\par\end{centering}
\begin{raggedright}
b) 
\par\end{raggedright}
\begin{centering}
\includegraphics[width=1\linewidth]{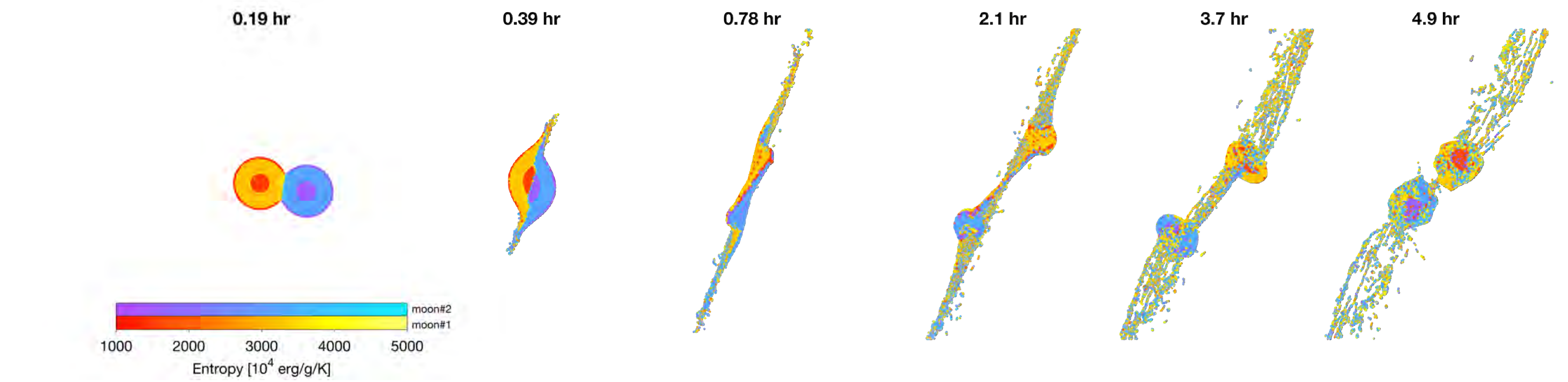}
\par\end{centering}
\caption [Figure 3] {\label{Figure 3}Differences between impact with and without the central potential. Several snapshots with similar initial conditions ($\beta=8.75^{o}$; $V_{{\rm imp}}=2.15\,V_{{\rm esc}}$) of a) two orbiting moonlets inside the planetary gravitational potential; b) two bodies in free space. The different color bars represent the entropy of the material originating from different moonlets. All projections are on the equatorial plane with one hemisphere removed. The Hill sphere of each orbiting moonlet is represented by the dashed lines. For simplicity, only material with density of $\rho>3\,[{\rm g/cm^{3}}]$ is shown.}
 \end{figure}

\begin{figure}
\hspace{-1cm}
\begin{tabular}{ll}
\ \ \ \ a) \ & \ b) \\ 
\includegraphics[width=0.5\linewidth]{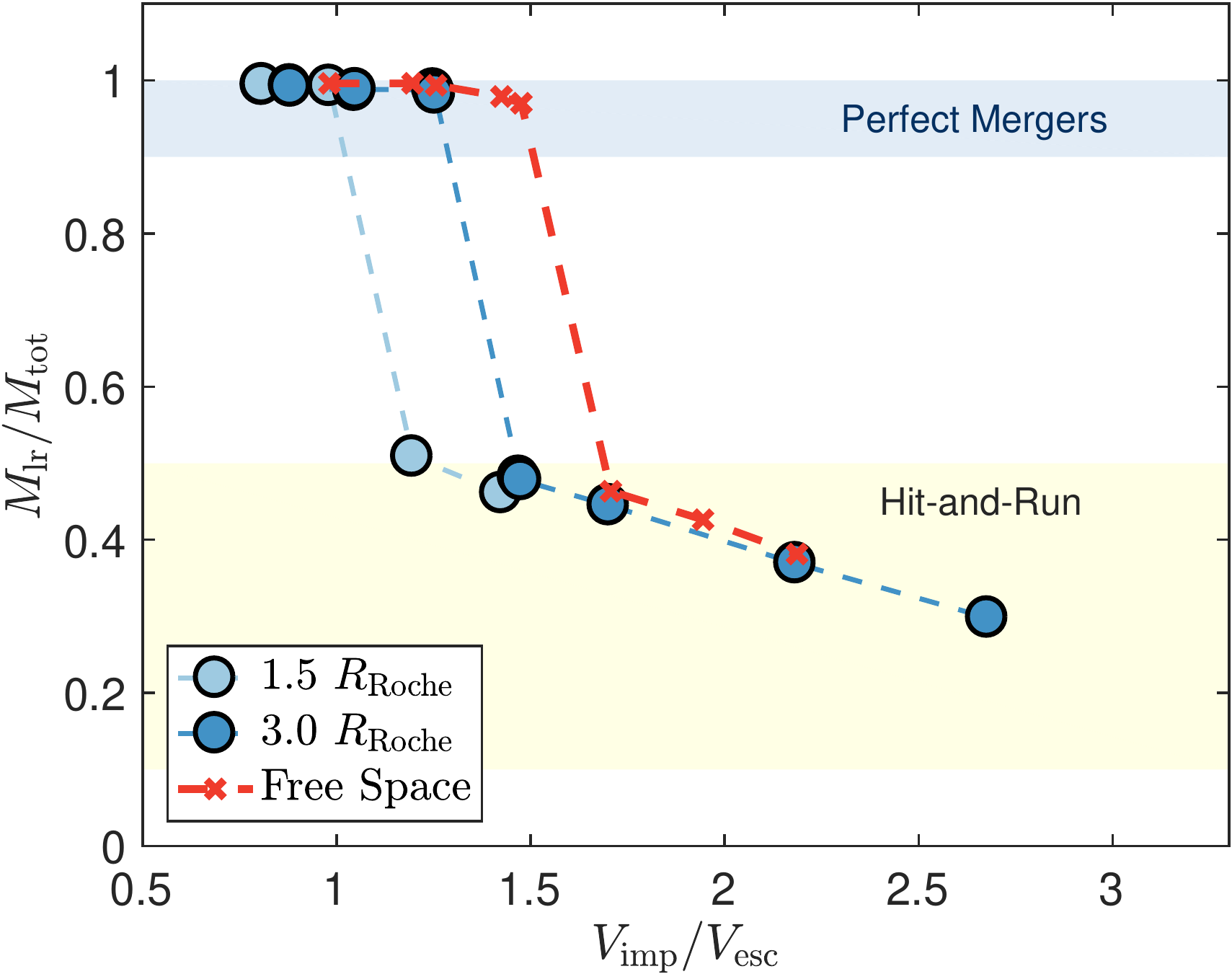}
&
\includegraphics[width=0.5\linewidth]{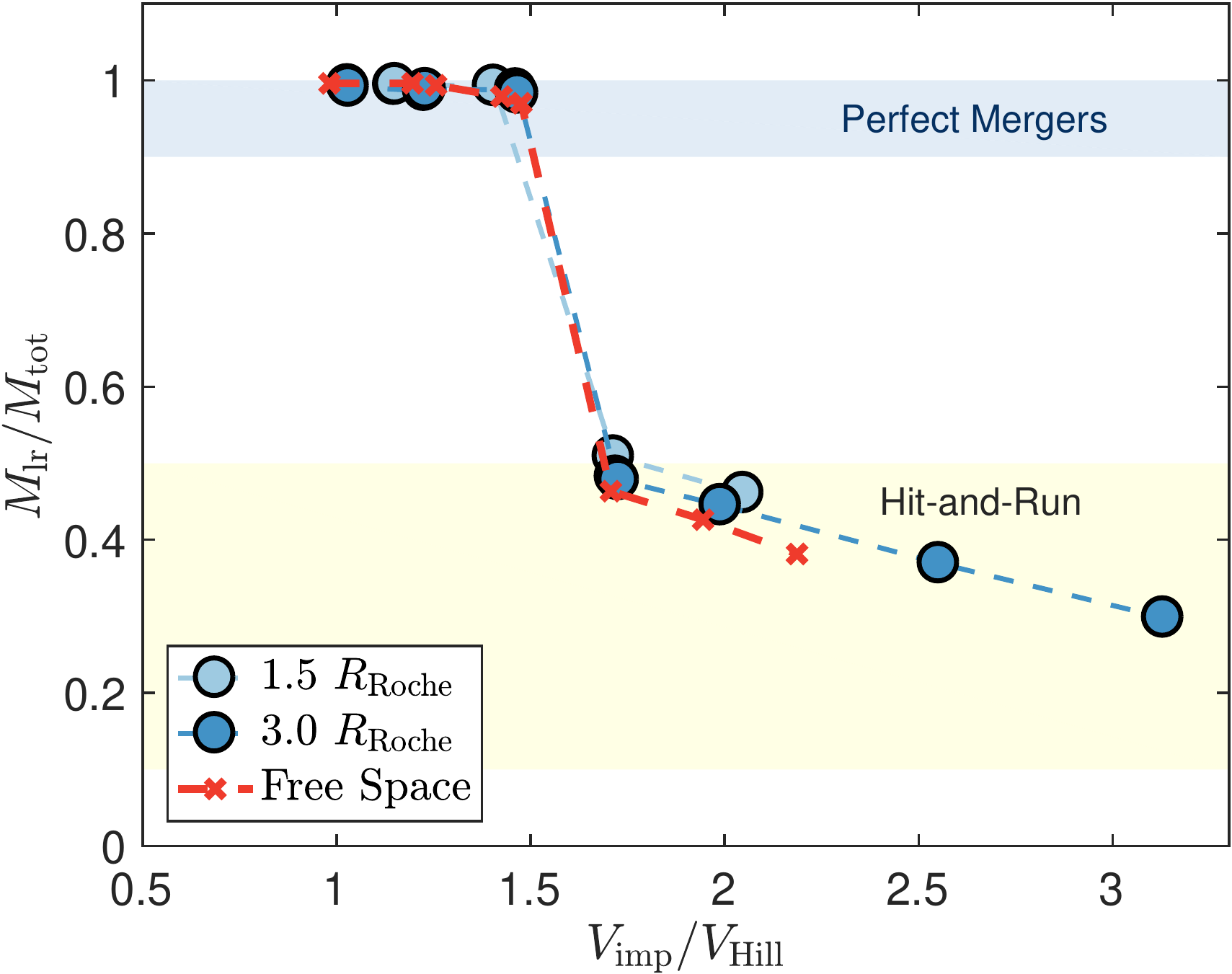}
\end{tabular}
 \caption [Figure 4] {\label{Figure 4}Transition to hit-and-run regime. Normalized mass of largest remnant vs. impact velocity normalized by the a) mutual escape velocity; b) Hill escape velocity (for free space cases we set $V_{\rm Hill}=V_{\rm esc}$ ) for impact angle of $\beta\sim30^{o}$. Circled markers represent moonlet impacts at different distances from the planet whereas the crosses represent impacts of two bodies in free space (no central potential).}
\end{figure}

Importantly, in a tidal environment, the boundary of the hit-and-run regime is expanded and such impacts are more abundant. For example, the low angle and high velocity impact in \nameref{Figure 2}-a resulted in two large remnants (largest remnant mass, $M_{{\rm lr}}=0.28\,M_{{\rm tot}};$ second largest remnant mass, $M_{{\rm sr}}=0.27\,M{\rm _{tot}}$). This impact is below the grazing curve \citep{Genda}, therefore a single final body was expected. We confirmed this disparity by simulating the same impact in free space, which resulted in a single final body of mass $0.71M_{{\rm tot}}$ (\nameref{Figure 2}-c). In both cases the bodies graze after the first impact. In the case with the central potential, the bodies expand beyond the mutual Hill radius (dashed lines in \nameref{Figure 3}-a) and gravitationally separate, as opposed to the free space case where the bodies do not have enough energy to escape the the mutual gravitational pull and re-impact again after 5 hours (\nameref{Figure 3}-b). Overall, the transition to the hit-and-run regime occurs at a lower velocity when the impact occurs closer to the planet (\nameref{Figure 4}-a). The Hill escape velocity (Eq. \ref{eq:V_hill}, here we defined $R_{\rm Hill}$ as the mutual Hill radius of both impacting moons and set $r=R_1+R_2$) at $1.5\ R_{\rm Roche}$ is $\sim 0.7\, V_{\rm esc}$, whereas at $5.0\ R_{\rm Roche}$ it is $\sim 0.9 V_{\rm esc}$. By normalizing the impact velocity to the local Hill escape velocity, $V_{\rm Hill}$, rather than the escape velocity \citep{Genda}, we find that the transition to the hit-and-run regime occurs at the same normalized velocity (\nameref{Figure 4}-b).

\citet{citron2018role} found that most impacts are at a velocity $<1.75V_{{\rm esc}}$, and these will have a high merger efficiency if the impact angle is low enough for accretion, but these also include hit-and-run impacts for higher angles. Because grazing angles are the most probable ($\sim 45^o$), most similar size impacts will usually not accrete at velocities $>V_{{\rm esc}}$. Hit-and-run impacts transfer little angular momentum between the orbiting components. Therefore, moonlets will remain on approximately similar orbits, and will likely impact several times \citep{asphaug2010similar} until the final accretion or the loss of one moonlet\add{ (\textit{e.g.}, Supplemental Movie S1)}. Perfect merger is often assumed in dynamical studies  \citep{Kokubo2006,Raymond2009644,RufuCanup,citron2018role}, however resolving hit-and-run impacts is important, as each impact can raise the melting fraction of the final moon (see section \ref{subsec:Melting}) and could potentially dynamically destabilize moonlets (as is the case for the impacts near the Roche limit).

\begin{figure}
\hspace{-1cm}
\begin{tabular}{ll}
\ \ \ \ a) \ & b) \\ 
\includegraphics[width=0.5\linewidth]{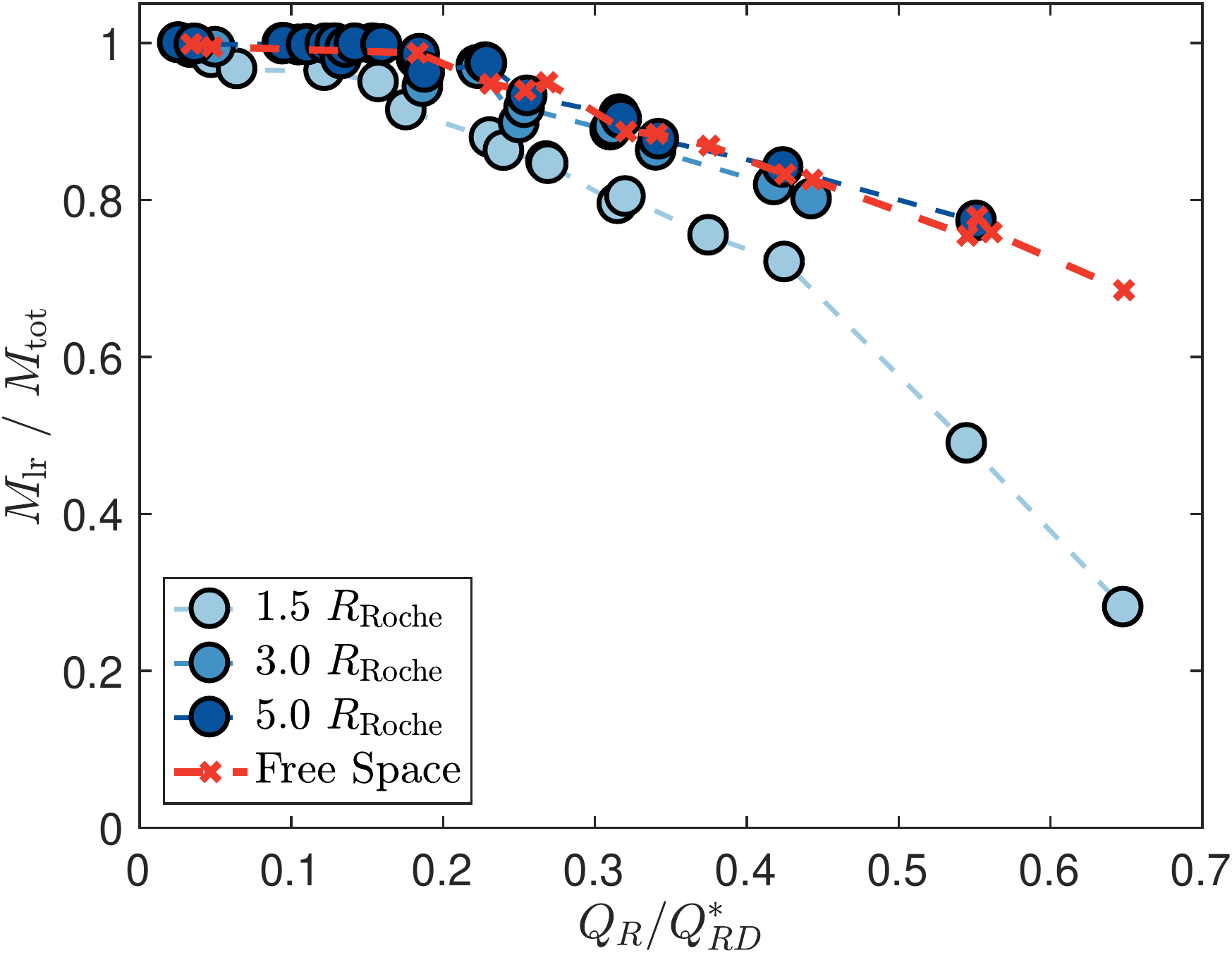}
&
\includegraphics[width=0.5\linewidth]{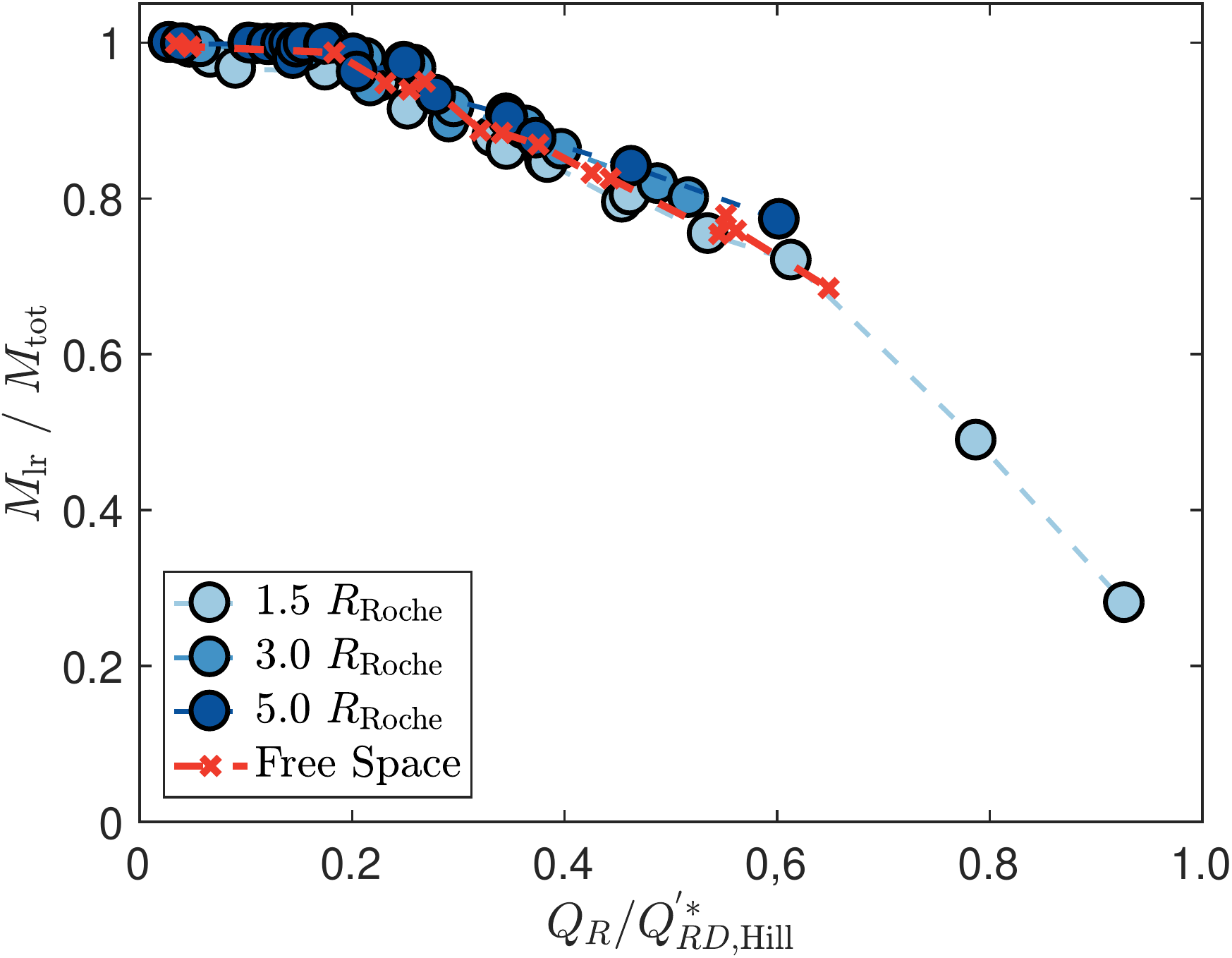}
\end{tabular}
\caption[Figure 5]{\label{Figure 5} Mass of largest remnant post impact for low impact angles. Normalized mass of largest remnant vs. normalized impact energy for low impact angles ($\beta<15^{o}$). The impact energy is normalized by a) the catastrophic disruption criteria   $Q_{RD}^{'*}$, defined by \citeauthor{Leinhardt} (2012; with the best fitted parameter of energy dissipation within the target, $c^{*}=2.8$); b) the corrected Hill catastrophic disruption criteria, $Q_{RD,\rm{Hill}}^{'*}=Q_{RD}^{'*}(V_{\rm Hill}/V_{\rm esc})^{2-3\bar{\mu}}$ (where $\bar{\mu}=0.36$ is a dimensionless material constant \citep  {housen1990fragmentation,Leinhardt}; for free space cases we set $V_{\rm Hill}=V_{\rm esc}$). Circle markers represent impacts at different distances from the planet whereas the crosses represent impacts of two bodies in free space (no central potential)}.
\end{figure}

For impacts in free space and small impacting angles, partial accretion is expected (\nameref{Figure 2} - light blue region; most of the impacting mass should be retained within a single body). Previous studies found that the mass of the largest remnant in accretionary impacts, $M_{\rm lr}$, is linearly proportional to the kinetic energy of the impact, $Q_{R}$,
\begin{equation}
    M_{\rm lr}/M_{\rm tot}\propto{-Q_{R}/Q_{RD}^{'*}}\label{eq:Mlr}
\end{equation}

where, $Q_{RD}^{'*}$, is defined as the disruptive (catastrophic) energy required for $M_{\rm lr}=0.5M_{\rm tot}$ \citep{stewart2009velocity}. However, for impacts within a central potential, we observe that with decreasing planetary distance, debris generation is enhanced because the Hill radius of the orbiting moonlet is smaller. Hence, ejected material requires less energy to escape the gravitational pull of the moonlet (Eq.  \ref{eq:V_hill}) and typical merger efficiencies are lower (light blue line in \nameref{Figure 5}-a) than predicted from previously defined scaling laws in free space (red line in \nameref{Figure 5}-a; \citealp{Leinhardt}). The disruptive energy scales as,  $Q_{RD}^{'*}\propto V^{*2-3\bar{\mu}}$, where $\bar{\mu}$ is a dimensionless material constant \citep{housen1990fragmentation,Leinhardt}, and $V^*$ is the catastrophic velocity and  proportional to the escape velocity, $V^*\propto V_{\rm esc}$. Hence, the scaling between the catastrophic energy in free space scales to the escape velocity as, $Q_{RD}^{'*}\propto V_{\rm esc}^{2-3\bar{\mu}}$. Similarly, for impacts within a central potential we scale the disruptive energy to the velocity required to escape from the Hill sphere (rather the mutual escape velocity) hence:
\begin{equation}
Q_{RD,\rm Hill}^{'*}=Q_{RD}^{'*}\left(V_{\rm Hill}/V_{\rm esc}\right)^{2-3\bar{\mu}}
\end{equation}
We find that with the corrected disrupting energy, one slope fits well with the results from different planetary distances (\nameref{Figure 5}-b).

Overall, in both grazing and non-grazing cases, the merger efficiency is lower near the Roche limit and erosion of the moonlet is enhanced. The generated debris typically remain in orbit and can reaccrete on the surface of the surviving moonlet or fall onto the planet. 

\subsection{Surface Properties}

Different moonlets can inherit different isotopic compositions, either because each moonlet is sourced from a distinct region in the heterogeneous debris disk \citep{salmon2012lunar}, or because each moonlet is accreted from a distinct debris disk generated by different impacts \citep{Rufu:2017aa}. In order to estimate the initial surface heterogeneities in the moon after the last global impact event, we checked whether large areas containing contributions from a single moonlet exist on the surface. Our motivation being that, if surface heterogeneities are present, they could be preserved only in the upper layer, as it quickly solidifies.  

For each particle at the surface (material in the upper $5\%$ layer \add{in the  volume enclosed by a longitude-latitude rectangle of $15^o\times15^o$,} and typically represented by $\sim1000$ particles\add{; see Text S2 for other definitions}) of the surviving moonlet we calculate $f_{\rm moon1}-f_{\rm moon2}$ (where $f_{\rm moon,i}$ is the fraction of surface particles in the neighborhood sourced from moon $i$). The width defining the surface layer is chosen as it \change{represents}{samples} the upper $80\,\rm{km}$ \add{layer}, therefore only large subsequent impacts (of the order of the South-Pole Aitken-forming impact, \citealt{melosh2017south}) could drastically disrupt this layer and excavate enough material to bury these signatures. SPH lacks the ability to resolve the small scales required for chemical equilibration, therefore, the length defining each neighbourhood is $400\,\rm km$, ensuring that only large scale heterogeneities are resolved (\add{typically resulting in $\sim 100$ neighbours for each particle}; see Text S2 for sensitivity analysis on this value).\add{ Moreover, we emphasizes that this is an estimation of the immediate post-impact surface variation, and additional post-impact process (\textit{e.g.}, resurfacing due to impacts, vigorous convection or small scale instabilities) could remix the surface and remove any prior variations. However, the motivation of this study is to estimate an upper limit on the heterogeneities emerging from the last global event.}\remove{ We choose to represent the overall surface variation as
where, $\sigma$ is the standard variation of the surface particles and $\sigma_{\rm rand}$ is the standard variation of a random  distribution of the surface particles. Under this definition, small values of $\Delta_{\rm surf}$ are expected for surfaces that are are well mixed between two bodies (Figure S5-a) or surfaces that include material from one moonlet (Figure S5-b). Whereas, large values of $\Delta_{\rm surf}$ are expected for surfaces that include material from both bodies but their material is not mixed, creating regions where the material is unproportionally sourced (Figure S5-c, d). The statistic $\Delta_{\rm surf}$ should be regarded as an estimate of heterogeneity, designed to provide a convenient summary of the results, other definitions are certainly possible.}\par

\begin{figure}
\begin{centering}
\includegraphics[width=0.8\textwidth]{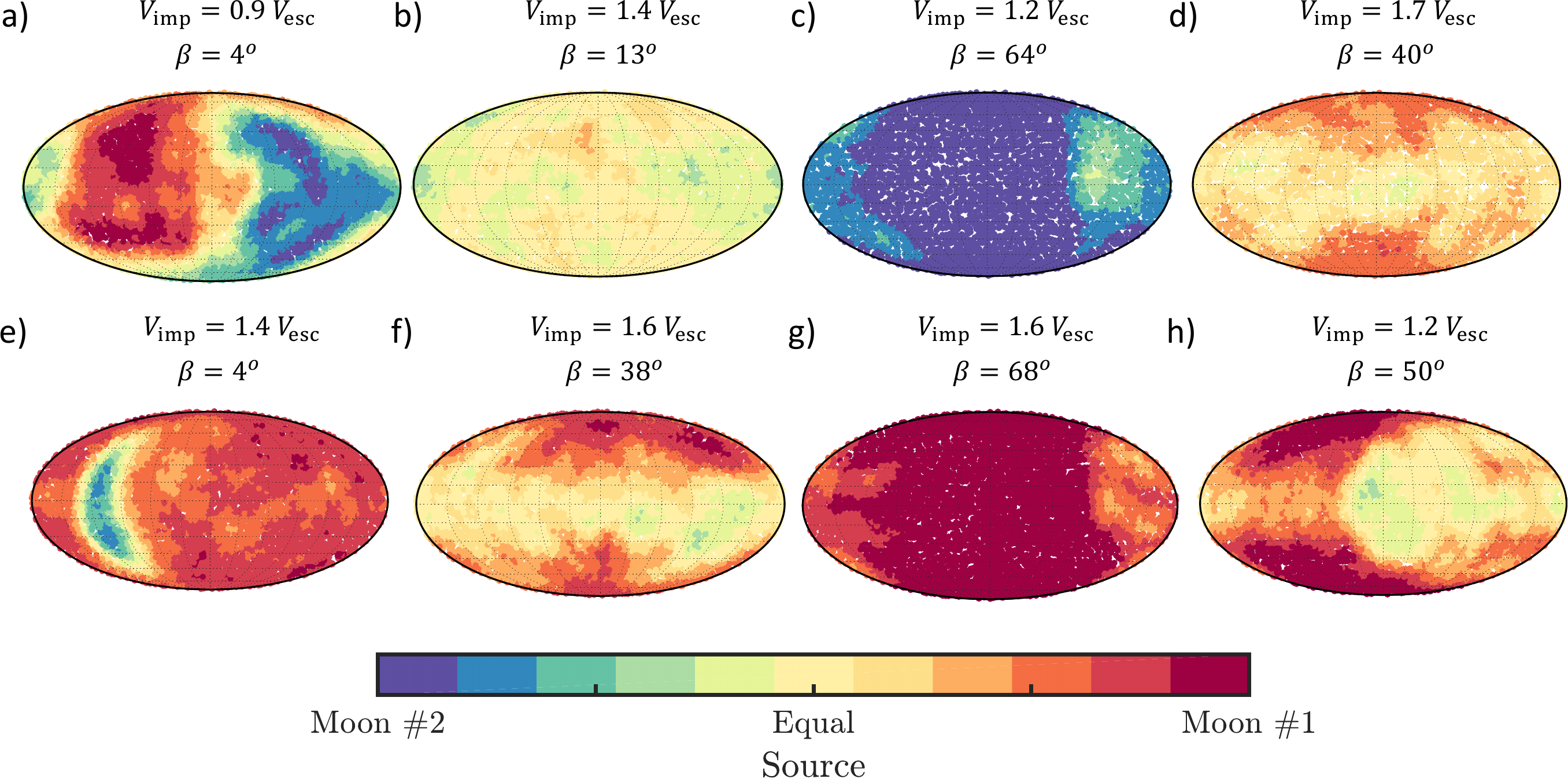}
\par\end{centering}

 \caption[Figure 6] {\label{Figure 6}Examples of source material distribution on surfaces resulting from impacts between components with different mass ratios,  $\gamma$=0.5 (first row) and  $\gamma$=0.1 (second row). The velocity ($V_{\rm imp}$) and impact angle ($\beta$) are denoted above each panel. The examples span the different regions of impact outcome: a)/e) low impact angle resulting in a perfect merger; b)/f) partial accretion; c)/g) single hit-and-run impact; d) sequence of two hit-and-run impacts; h) disruption and reacretion of the smaller component on the larger component.}
  \end{figure}

The \add{resulted surface patterns}\remove{results} show that low velocity and head-on impacts do not efficiently mix the surface of the resulting product because differential rotation post impact is limited. In these rare cases, large surface heterogeneities occur (\remove{$\Delta_{\rm surf}>0.3$; }\nameref{Figure 6}-\remove{c}a, e). More typical accretionary impacts of comparable-sized components have a larger amount of mixing (\remove{$\Delta_{\rm surf}<0.1$; }\nameref{Figure 6}-\remove{c}b, f). Specifically, surface variations for smaller mass ratios is decreased by the disruption of the small body and the reaccretion of material on the surface of the surviving moonlet (\nameref{Figure 6}-\remove{c}f, h; with similar initial condition and findings of \citealt{Jutzi:2011aa}). In hit-and-runs impacts, which are more probable for larger mass ratios, little mass is transferred by the impact, therefore, only small-scale heterogeneities are created (\nameref{Figure 6}-\remove{c }c, g). However, because the large components usually survive intact after impact, subsequent impacts could transfer additional material to the surface, creating a surface that is sourced almost equally from both moonlets  (\nameref{Figure 6}-\remove{c }d). 

It should be noted that for the highly erosive impacts or cases where one moonlet is disrupted due to passage through the planetary Roche limit, large amount of debris is generated and reaccreted on the surface of the surviving moonlet. The reaccretion timescale is larger than computationally attainable with SPH methods. Therefore, in these cases, we expect the overall surface variation to decrease, supporting our conclusions that typically surfaces are mixed.

\subsection{\label{subsec:Melting}Melting}

The highland terrain comprises about $77\%$ of the lunar surface area and is composed mostly of anorthositic material \citep{taylor2014lunar}. Anorthosites are rarely present in the Procellarum KREEP \change{terrance}{terrane} and remote sensing identified anorthosites only in the outer edges of the South Pole Aitken Basin \citep{taylor2014lunar}. Anorthositic-poor regions, comprising $23\%$ of the lunar area, are correlated with younger surfaces, therefore the anorthositic materials are considered to represent the initial crust. The leading theory of the formation of the anorthositic crust is the flotation of plagioclase minerals from a lunar magma ocean \citep{Elkins2011}. Gravity data from GRAIL \citep{zuber2013gravity} reveal igneous intrusions that provide evidence for a lunar radial expansion and consistent with a solidification of a 200-300 km-deep magma ocean \citep{andrews2013ancient}. However, recent lunar crystallization studies \citep{charlier2018crystallization} reveal that magma ocean of $\sim500\,\rm km$ is required to explain the lunar crustal thickness observed by GRAIL. Moreover, deeper magma oceans are possible if plagioclase flotation is imperfect or trapped liquids in cumulates are considered. Due to the uncertainties on the magma ocean depth required to reconcile the observations (crust thickness and lunar expansion limits), and the uncertainties on the initial thermal state of the mooonlets, we estimate that a magma ocean depth $>200\,\rm km$ resulting from a "cold" initial thermal state is compatible with recrystallizing a global anorthositic crust.

We consider that material melted if the final entropy of the particle is larger than the value defined by the liquidus curve  \add{(calculated using the M-ANEOS code,} \citealp{melosh2007hydrocode}\add{; see Figure S4 in Supplementary material)} at that density. We assume a "cold" thermal initial state\add{ and ignore partially melted material}, therefore this is a lower boundary for the total amount of melt after impact (see Figure S5 in Supplementary material for differences between the two thermal states).

\begin{figure}
\begin{raggedright}
\hspace{1.1cm}a) $\gamma=0.5$\hspace{2.3cm}b) $\gamma=0.3$\hspace{2.3cm}c) $\gamma=0.1$
\par\end{raggedright}
\begin{centering}
\includegraphics[height=3.23cm]{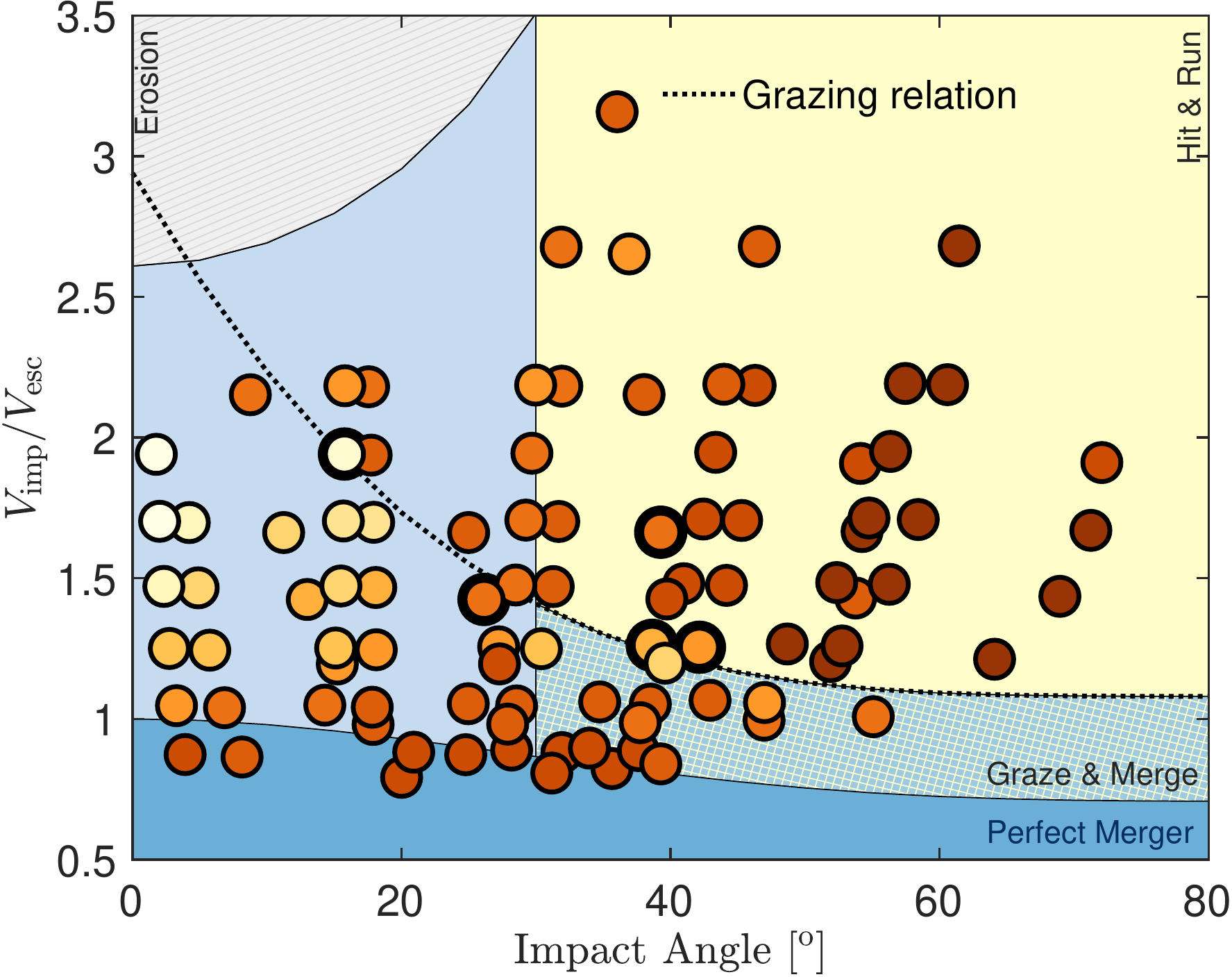}\hspace{0.1cm}\includegraphics[height=3.2cm]{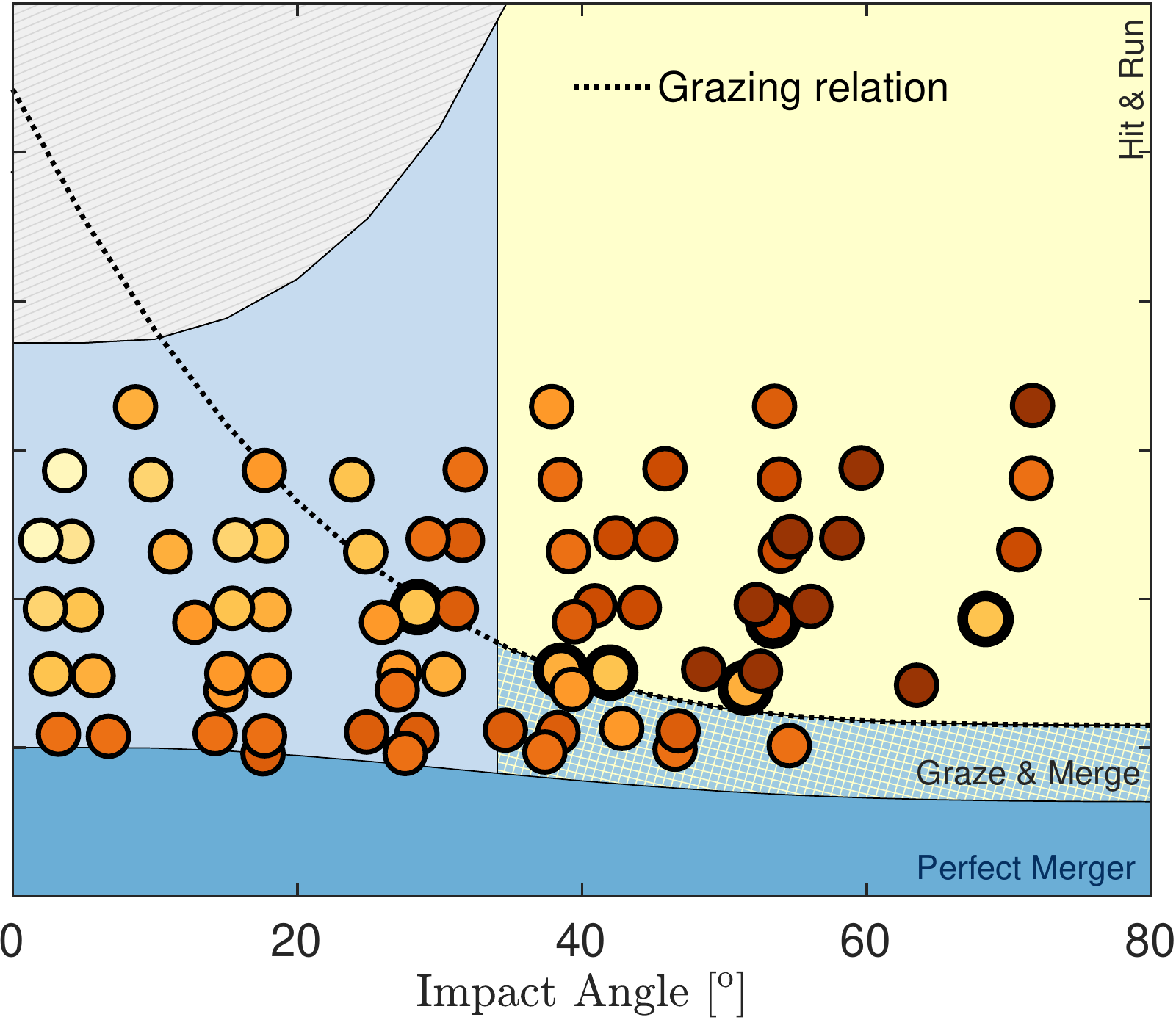}\hspace{0.1cm}\includegraphics[height=3.25cm]{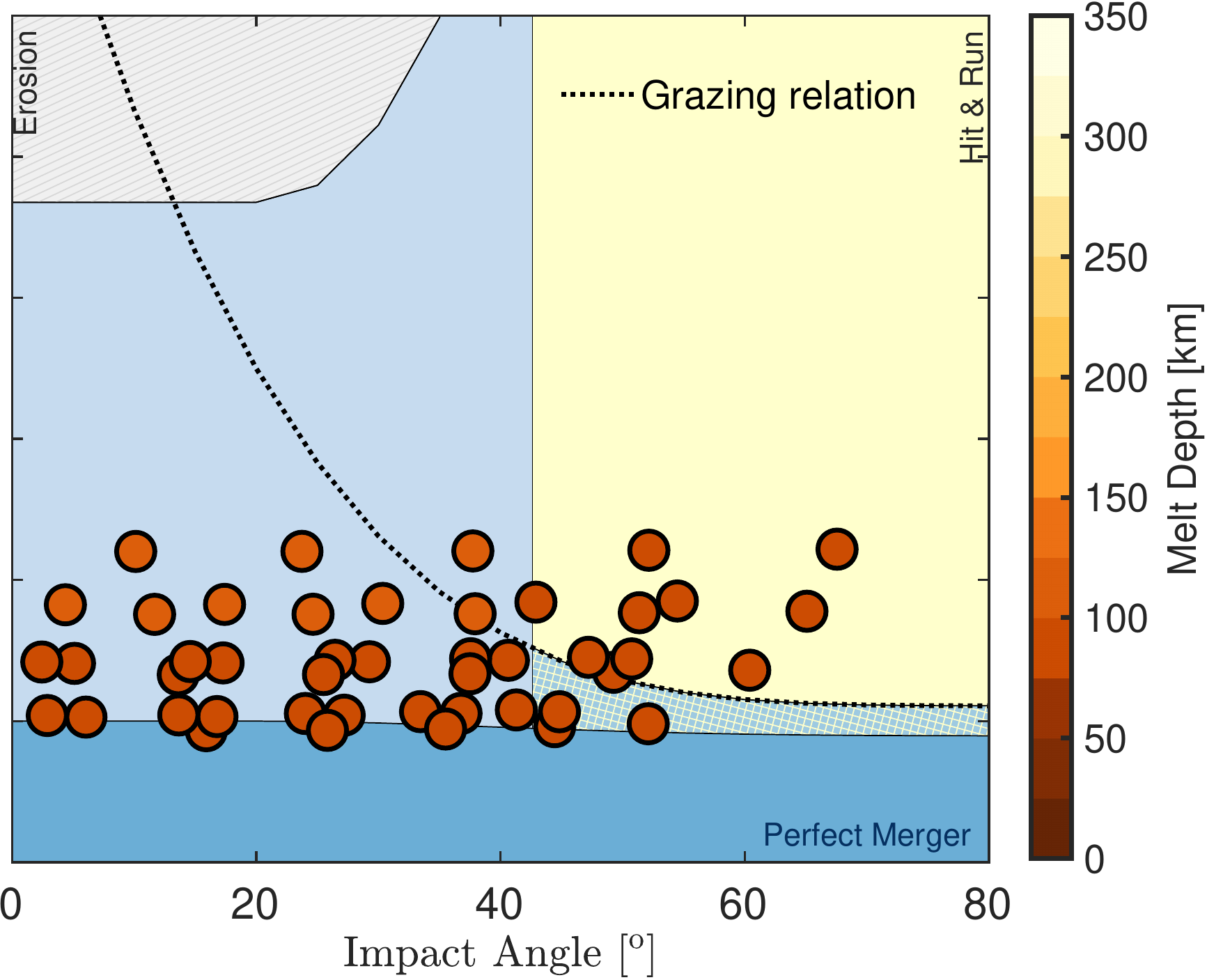}
\par\end{centering}
\caption[Figure 7]{\label{Figure 7} Depth of magma ocean in the angle-velocity phase space. Impacts between components with different mass ratios, a) $\gamma=0.5$; b) $\gamma=0.3$; c) $\gamma=0.1$. Colors represent the overall magma-ocean-depth, assuming that the melt is homogeneously distributed on the top layers of the moonlet. The different regions of the plot represent different impacting regimes previously defined by \citet{Leinhardt} for two bodies in free space (with the best fitted parameter of energy dissipation within the target, $c^{*}$=2.8). The dashed line represents the critical velocity for which grazing occurs, previously defined by \citet{Genda}. The bullets with the black thick line represent simulations where the bodies impact twice, therefore, in these cases the magma ocean depth is higher.}
\end{figure}

\begin{figure}
\begin{centering}
\includegraphics[height=0.2\linewidth]{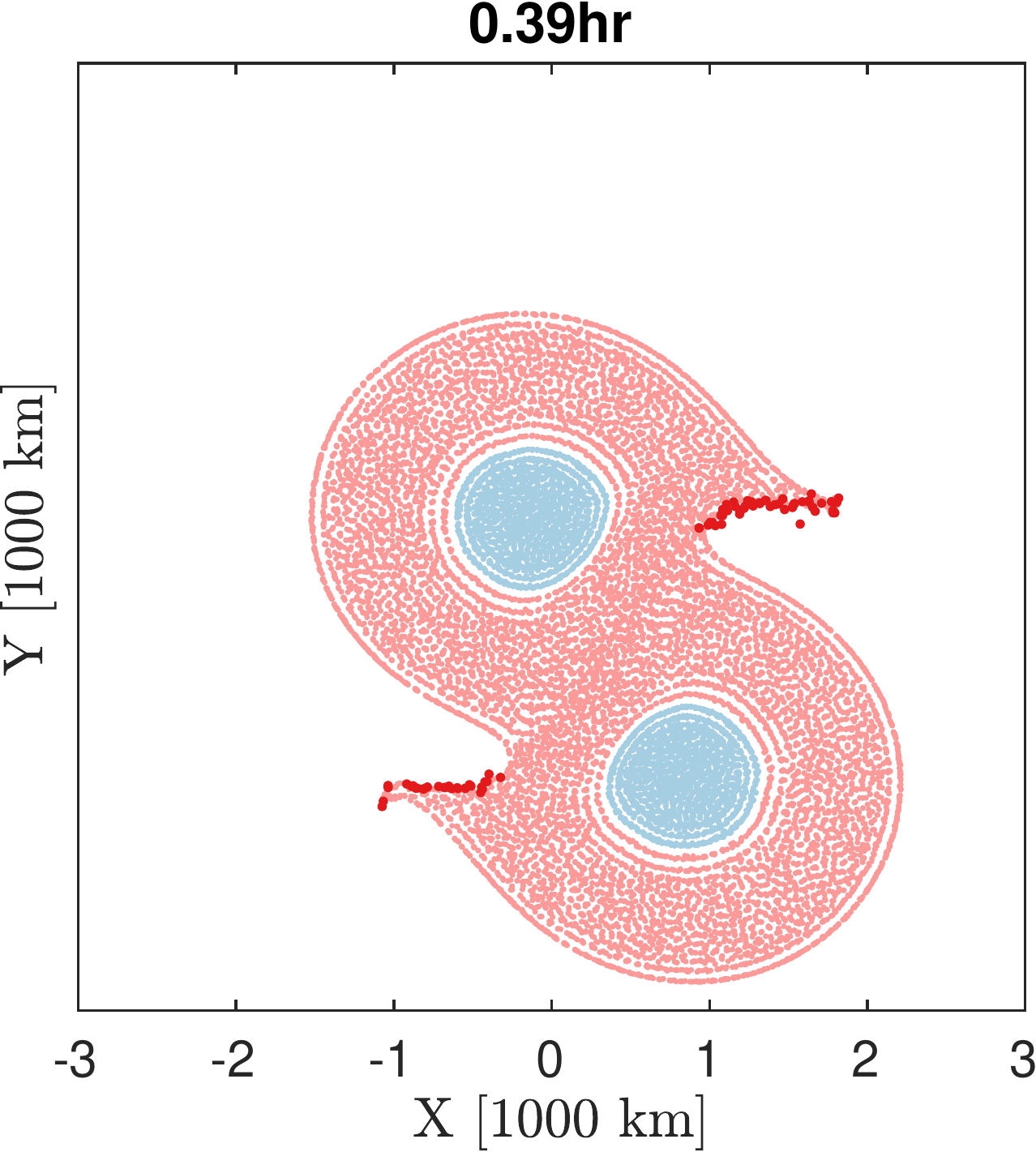} \includegraphics[height=0.2\linewidth]{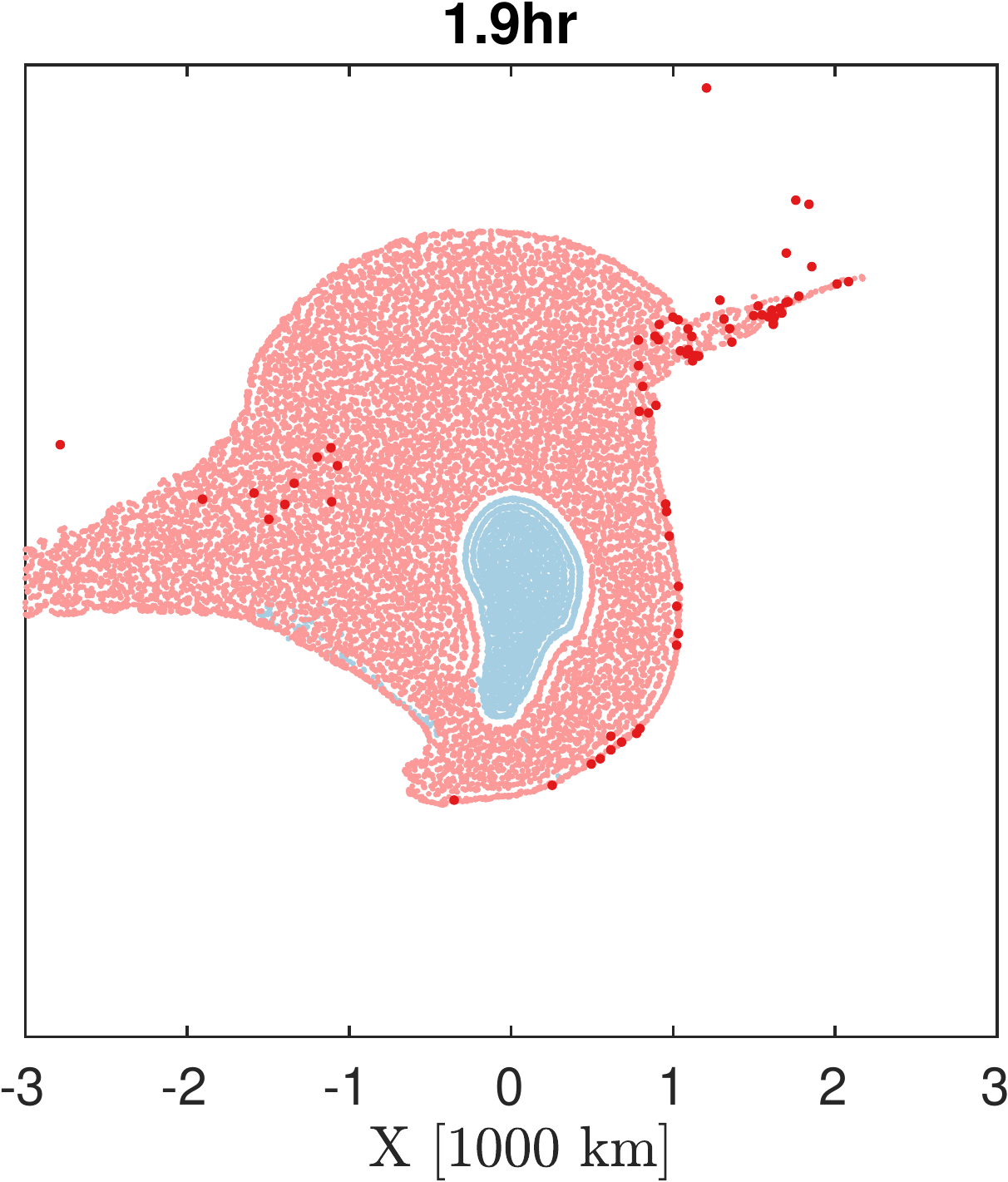} \includegraphics[height=0.2\linewidth]{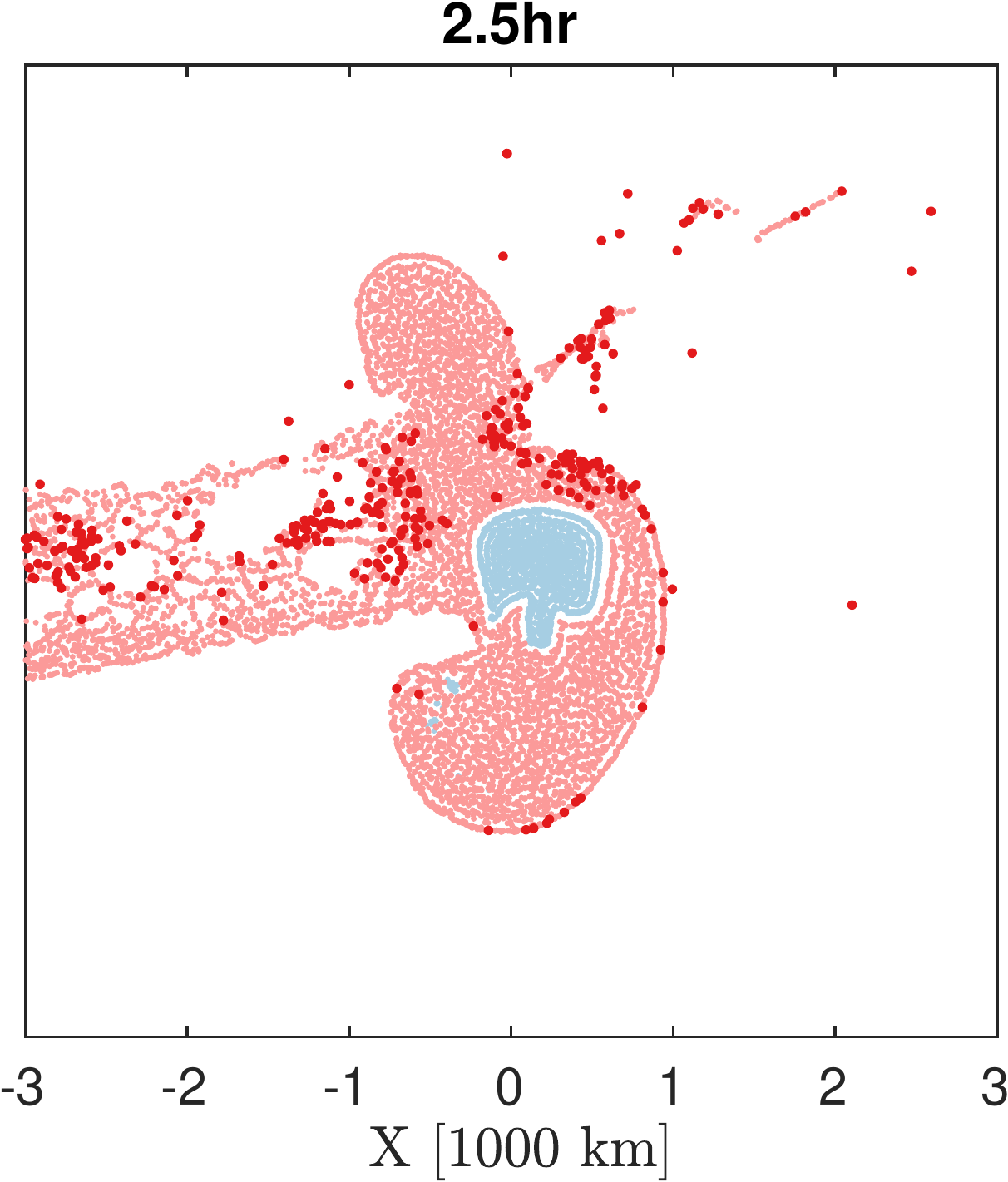} \includegraphics[height=0.2\linewidth]{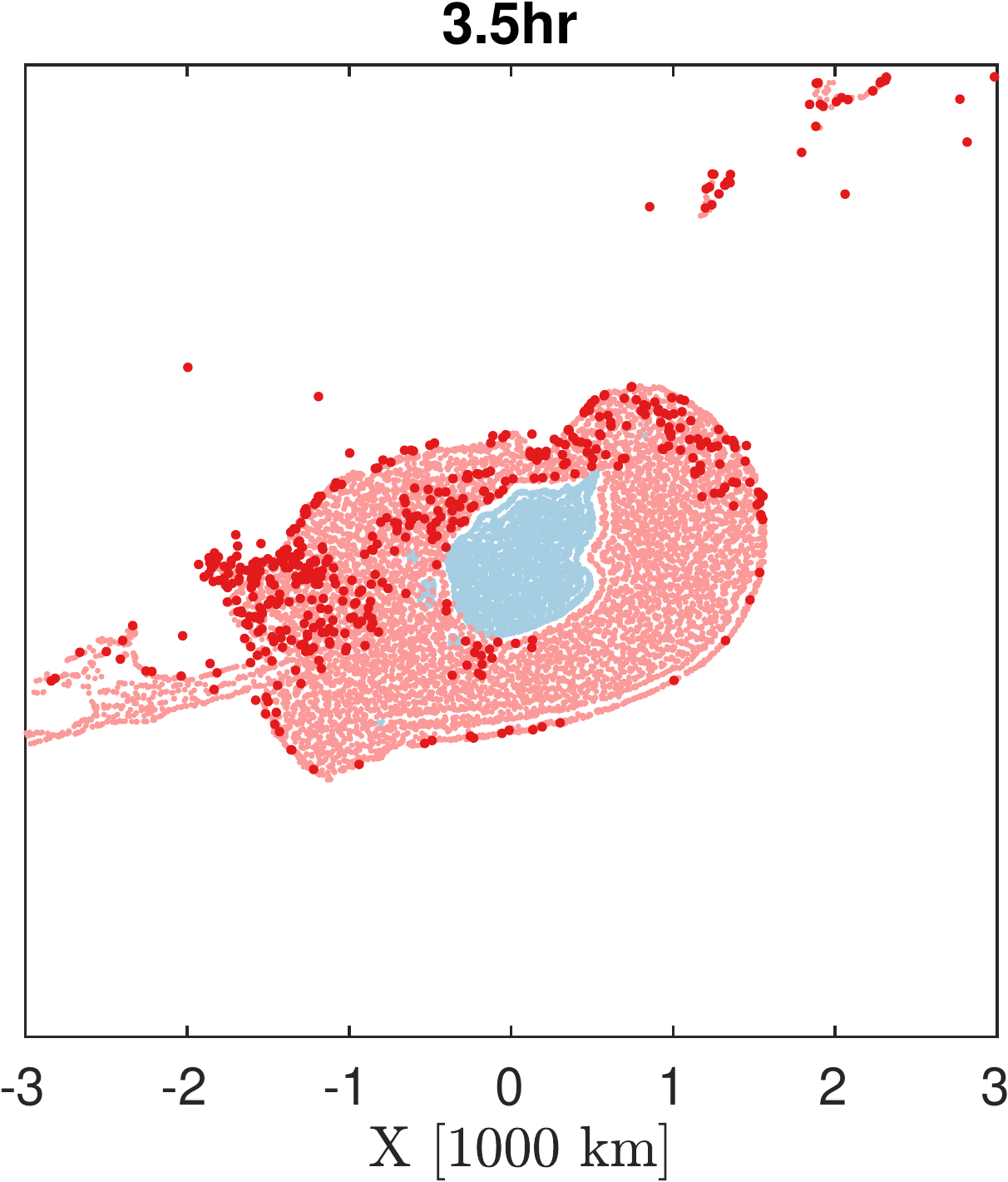} \includegraphics[height=0.2\linewidth]{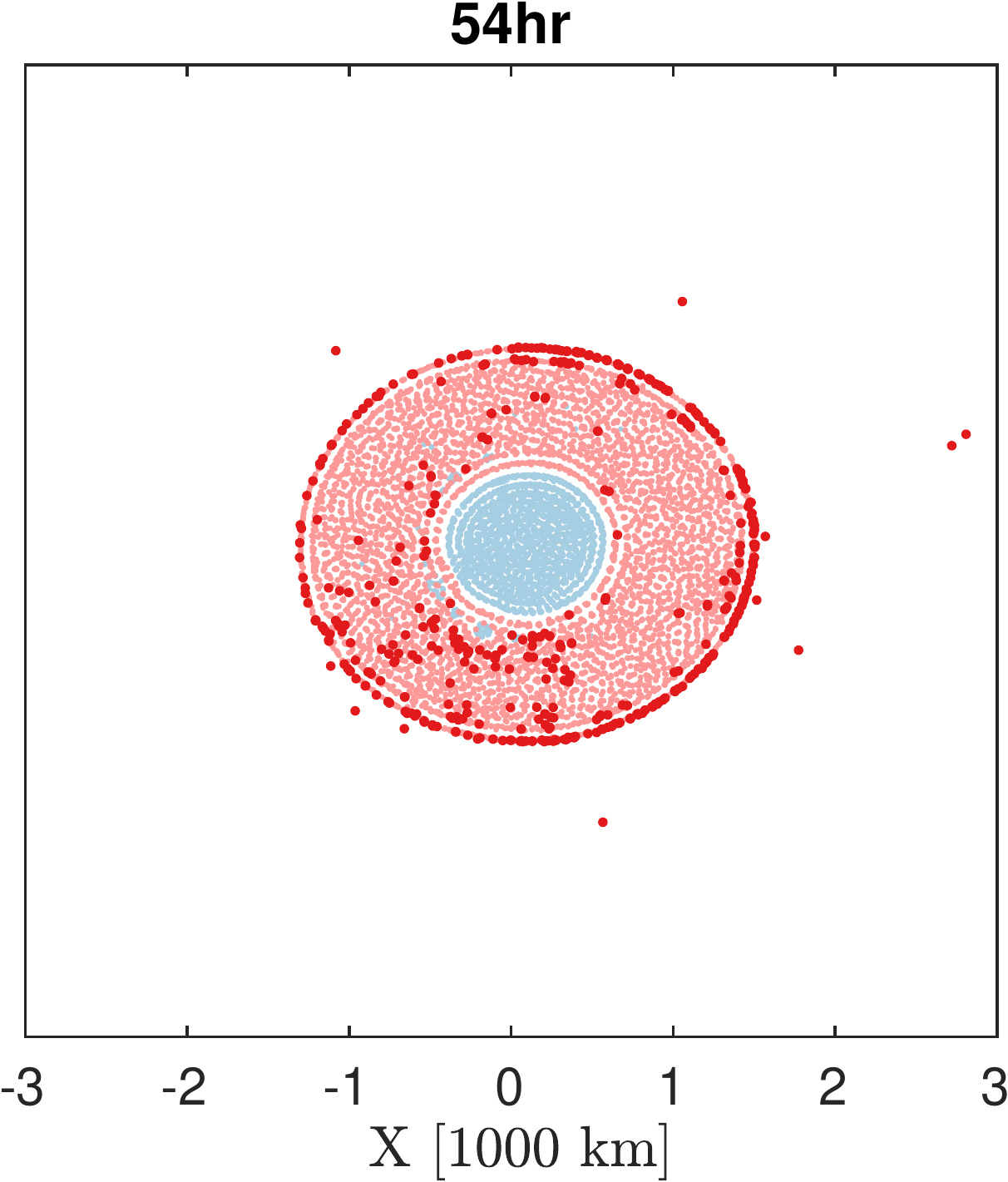}
\par\end{centering}
\begin{centering}
\includegraphics[height=0.2\linewidth]{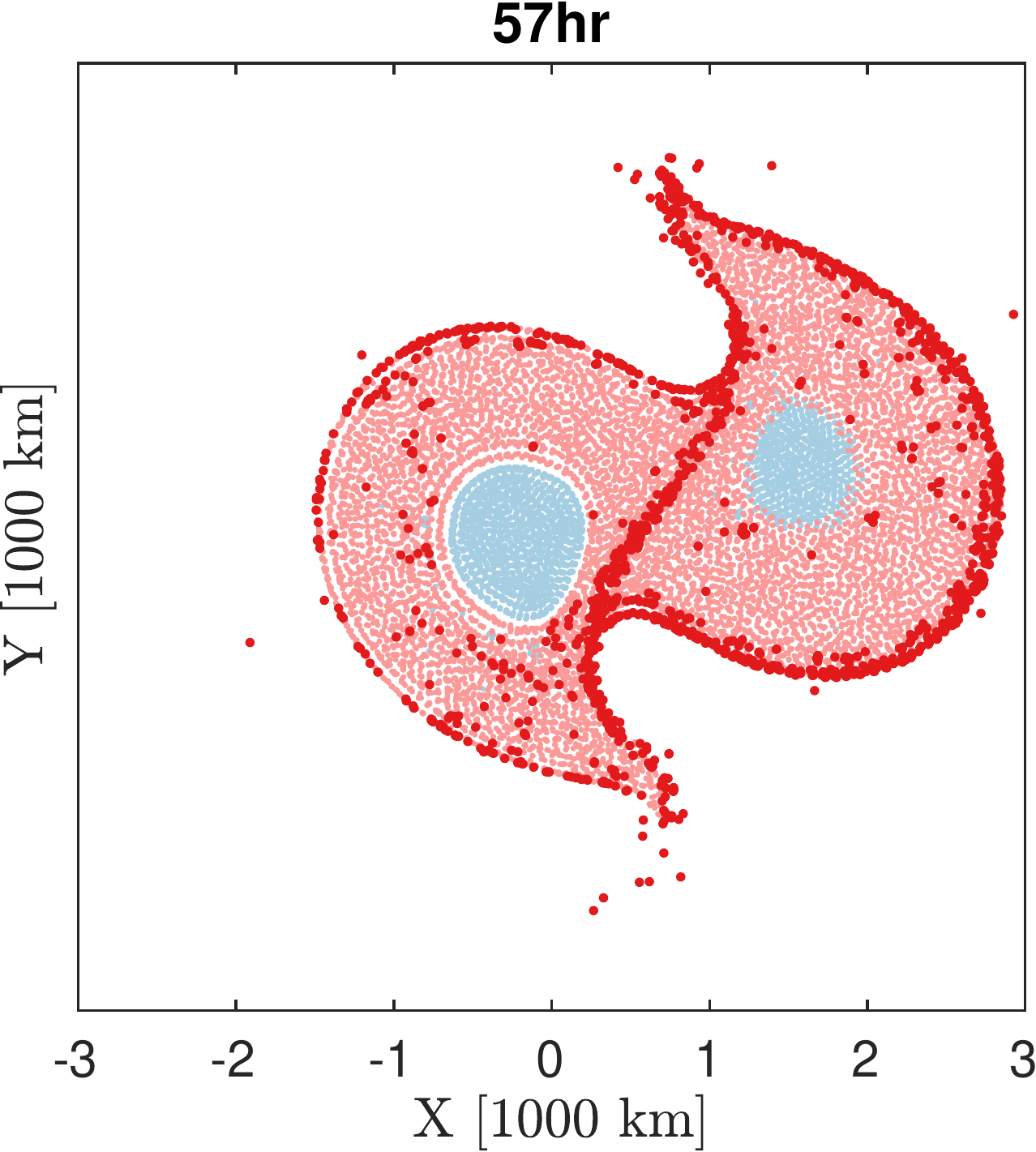} \includegraphics[height=0.2\linewidth]{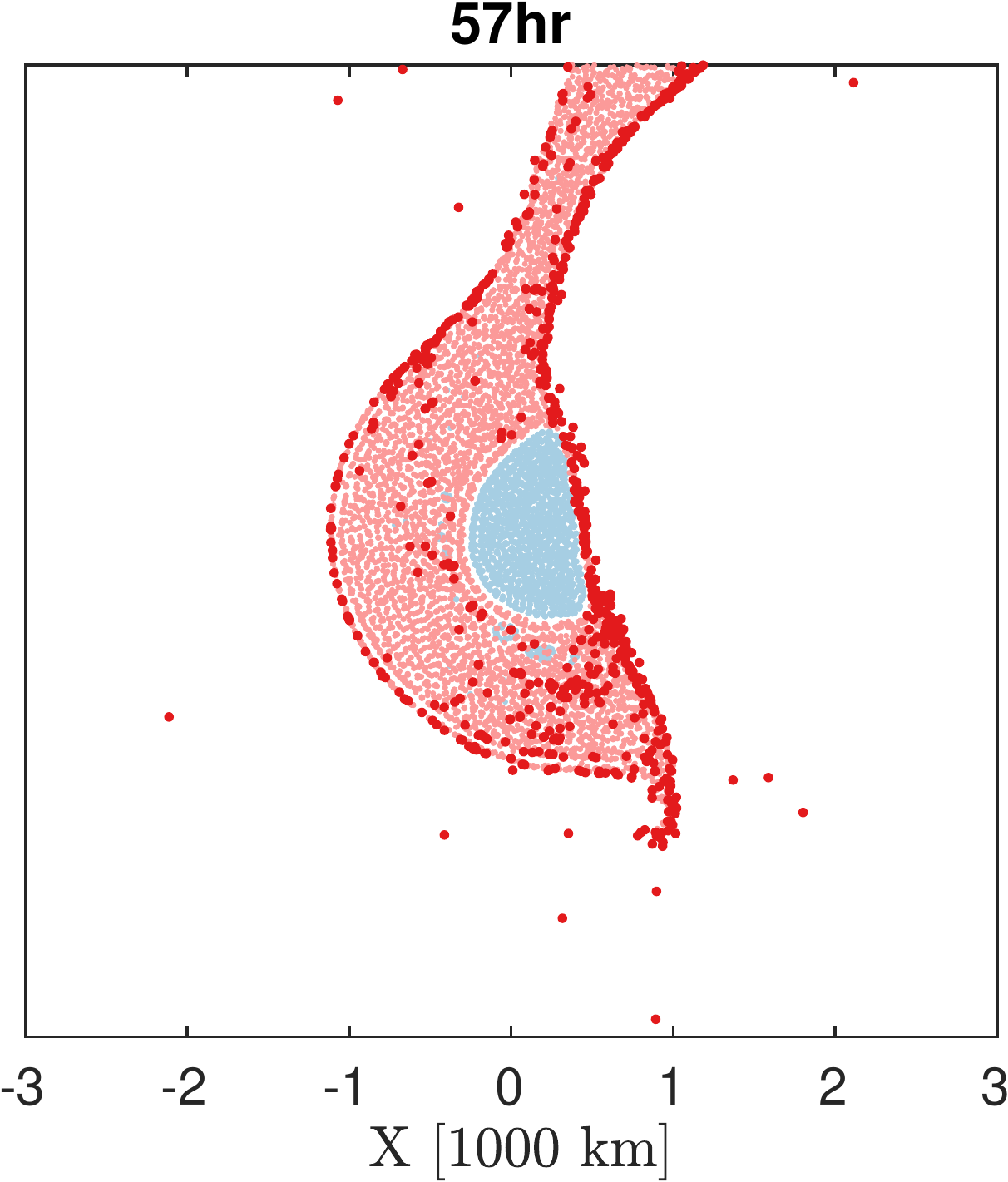} \includegraphics[height=0.2\linewidth]{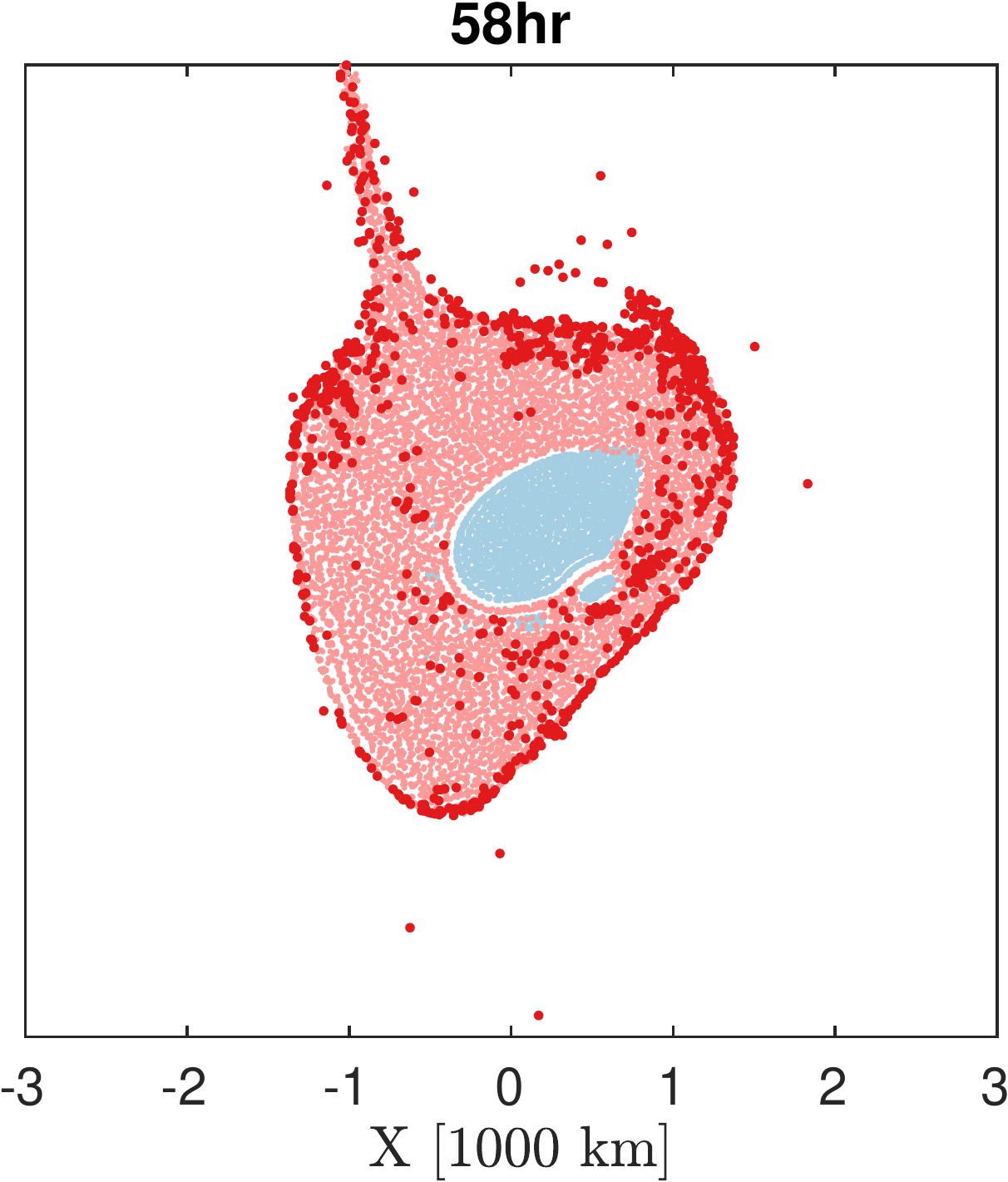} \includegraphics[height=0.2\linewidth]{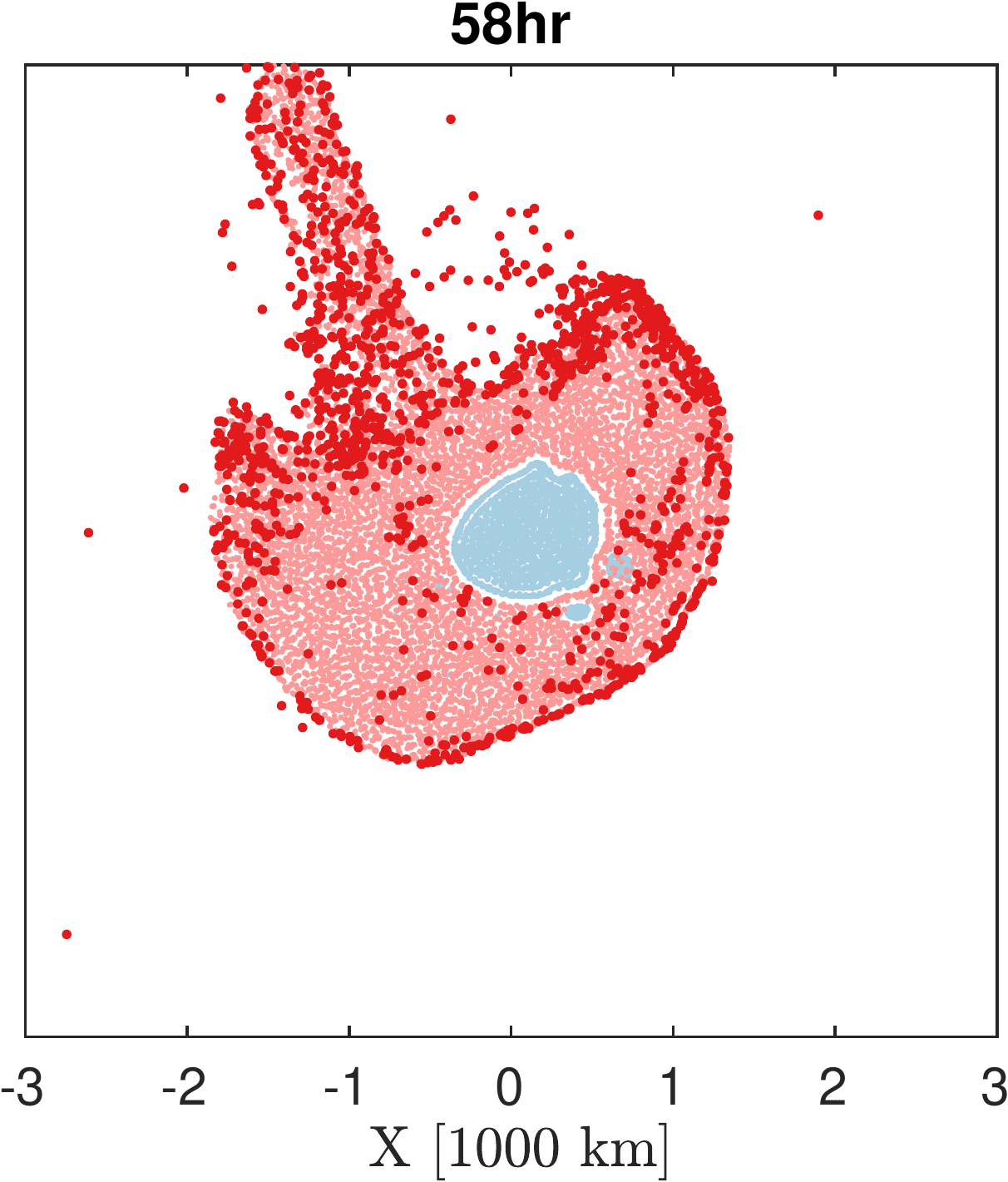} \includegraphics[height=0.2\linewidth]{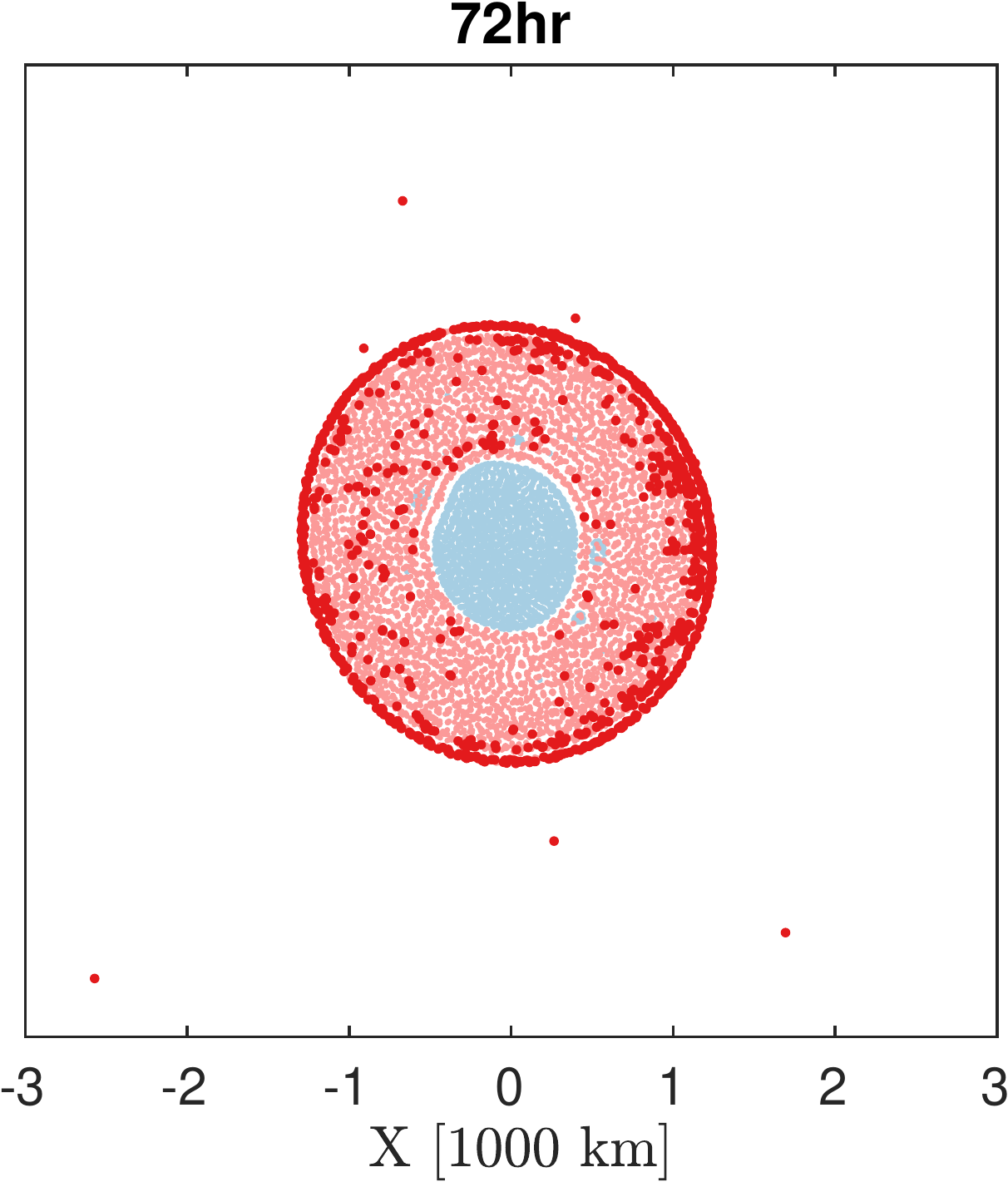}
\par\end{centering}
\caption[Figure 8]{\label{Figure 8}Melt evolution of a sequence of two hit-and-run impacts. Snapshots of two colliding half moonlets at a distance of $1.5\,R_{{\rm Roche}}$ from the planet (first impact: $\beta=40^{o}$; $V_{{\rm imp}}=1.66\,V_{{\rm esc}}$; second impact: $\beta=44^{o}$; $V_{{\rm imp}}=1.56\,V_{{\rm esc}}$). The snapshots show a slice of width 100 km centered in the equatorial plane of the surviving moon. The blue particles represent iron material and light/dark red particles represent the unmelted/melted magma material. Note that for emphasizing purposes, the melted particles are larger. In the first (first row) and second impact (second row), melt is created mainly in the secondary stage of the impact, where material falls back to the surface. Full movie can be found in the Supplementary material (Movie S2).}
 \end{figure}

\begin{figure}
\begin{centering}
\includegraphics[height=0.2\linewidth]{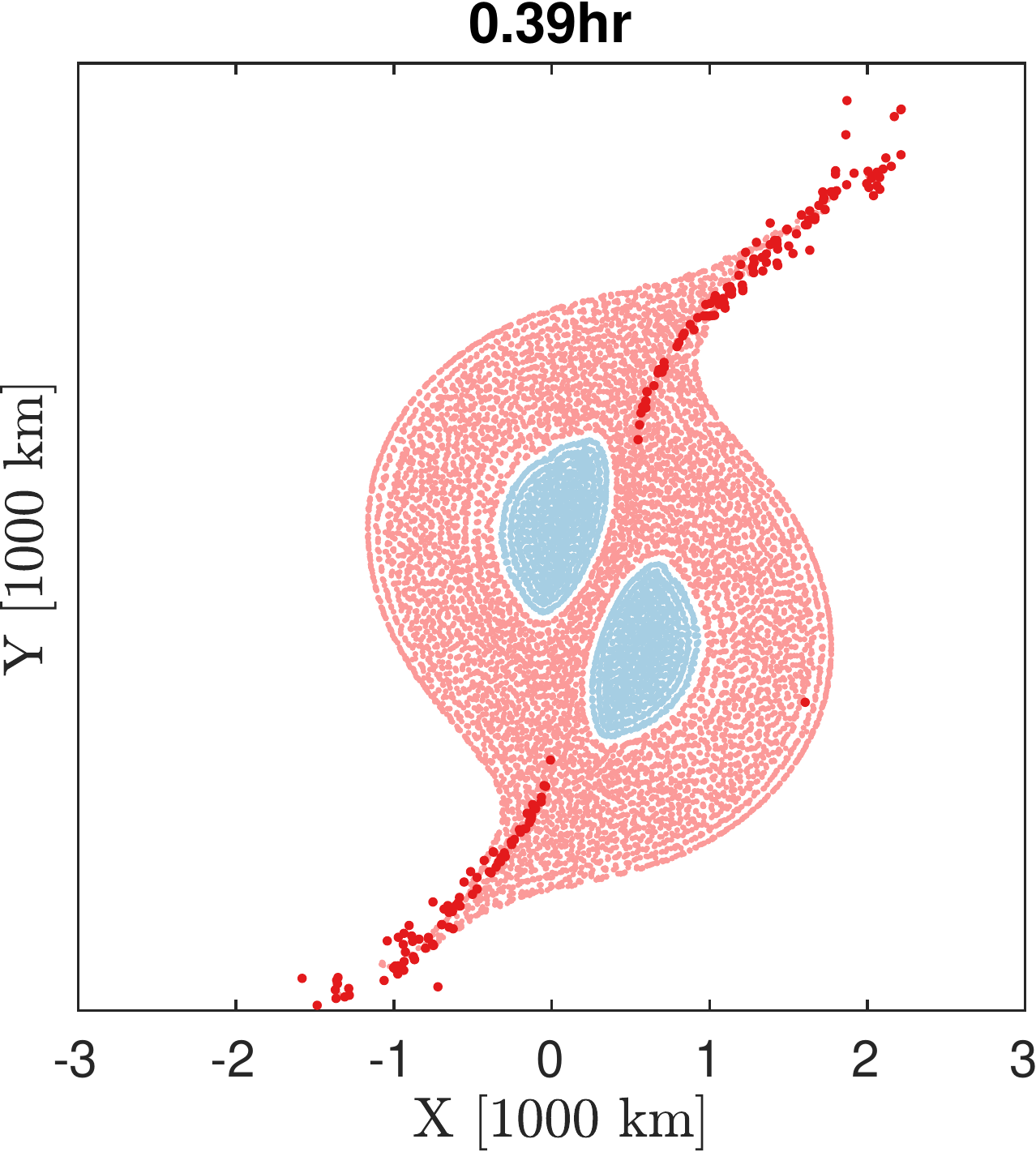} \includegraphics[height=0.2\linewidth]{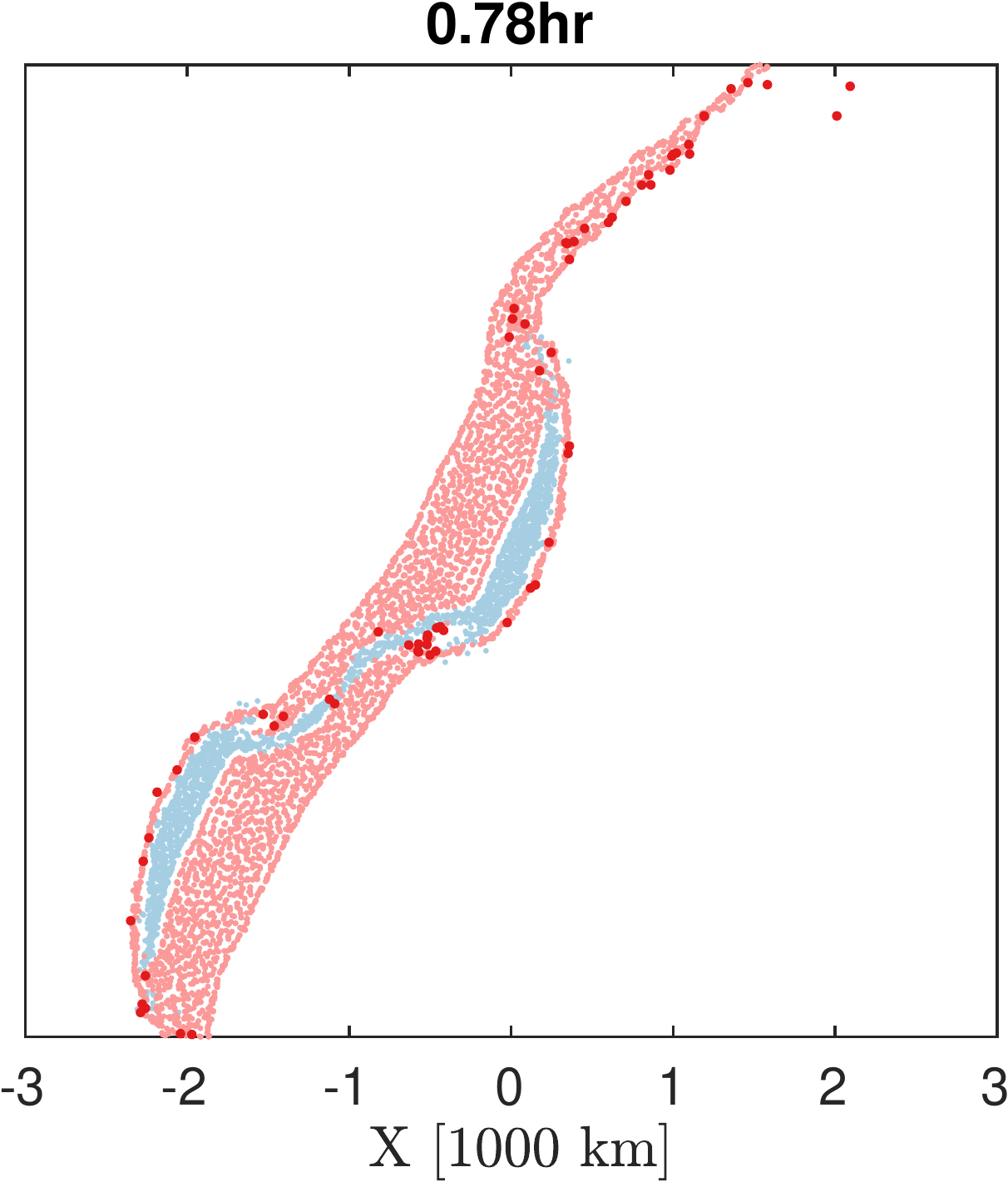} \includegraphics[height=0.2\linewidth]{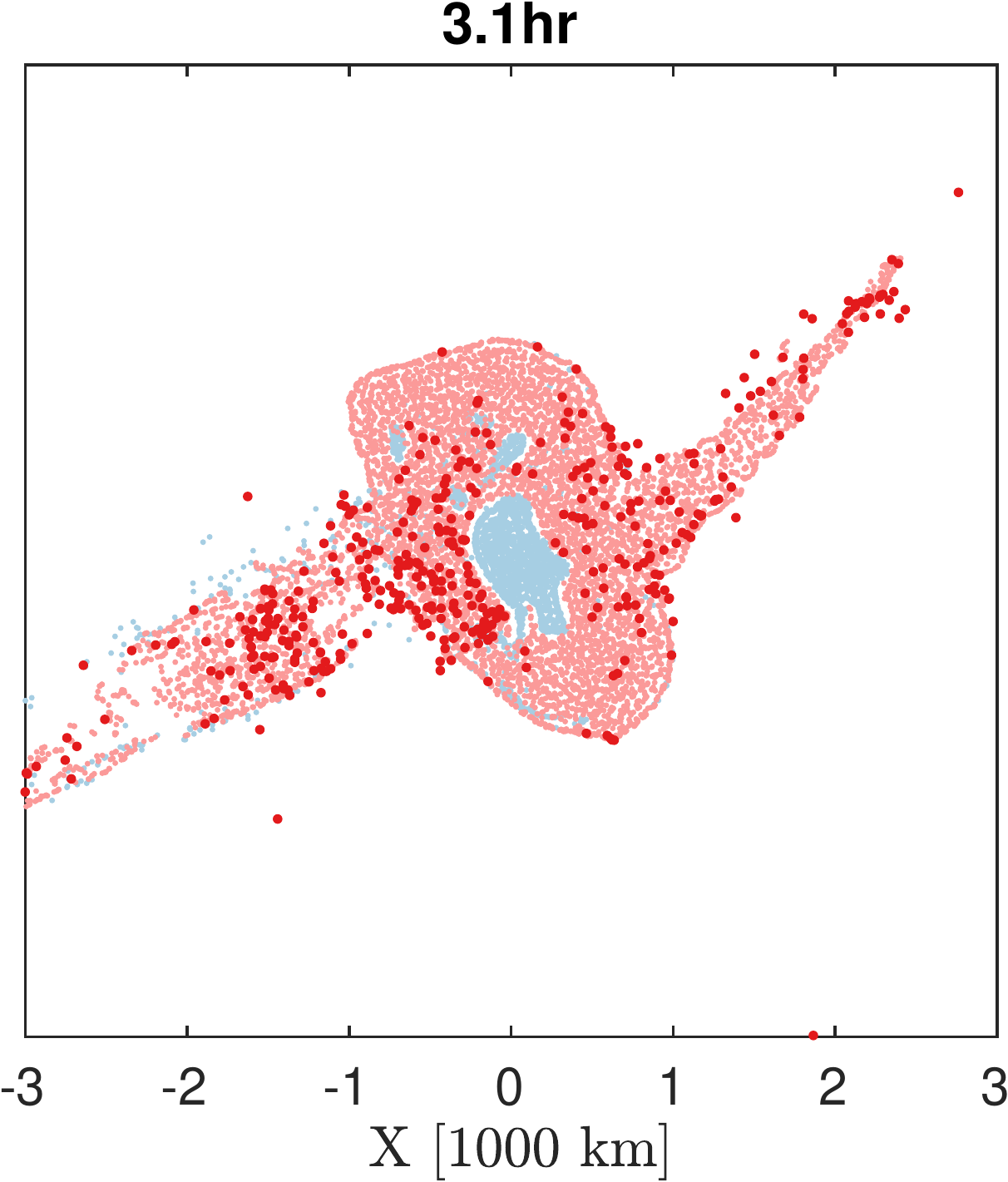} \includegraphics[height=0.2\linewidth]{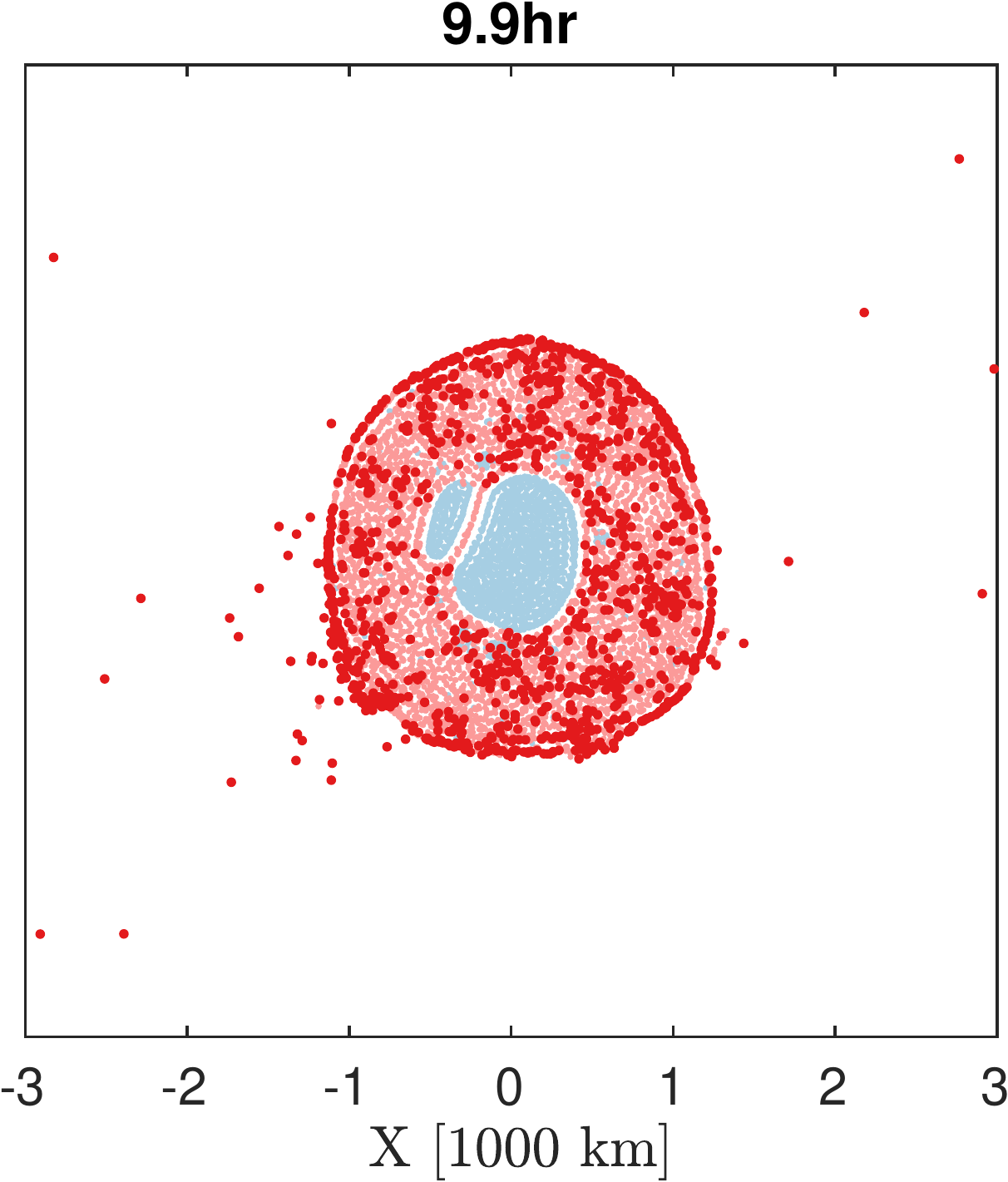}
\par\end{centering}
\begin{centering}
\includegraphics[height=0.2\linewidth]{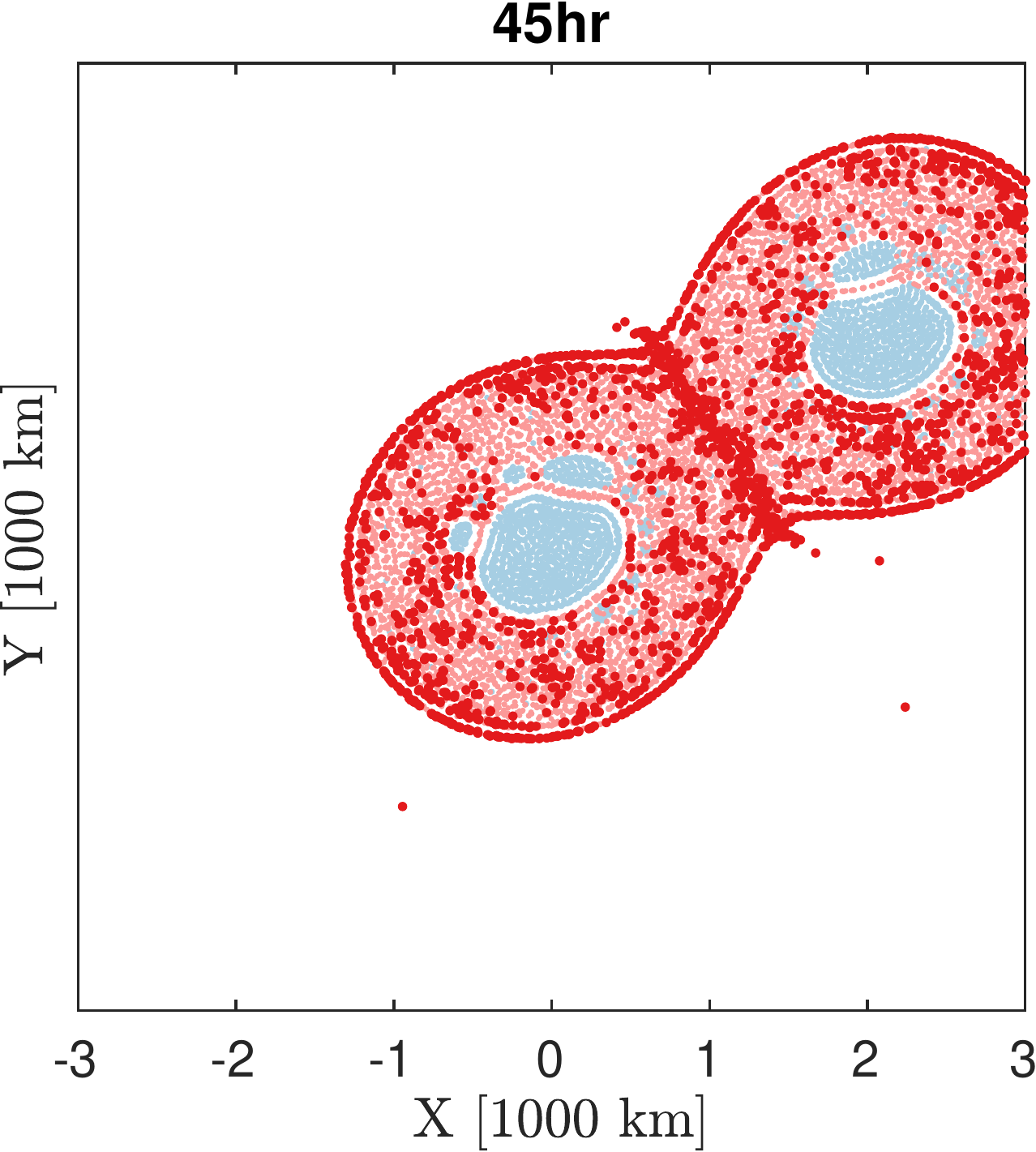} \includegraphics[height=0.2\linewidth]{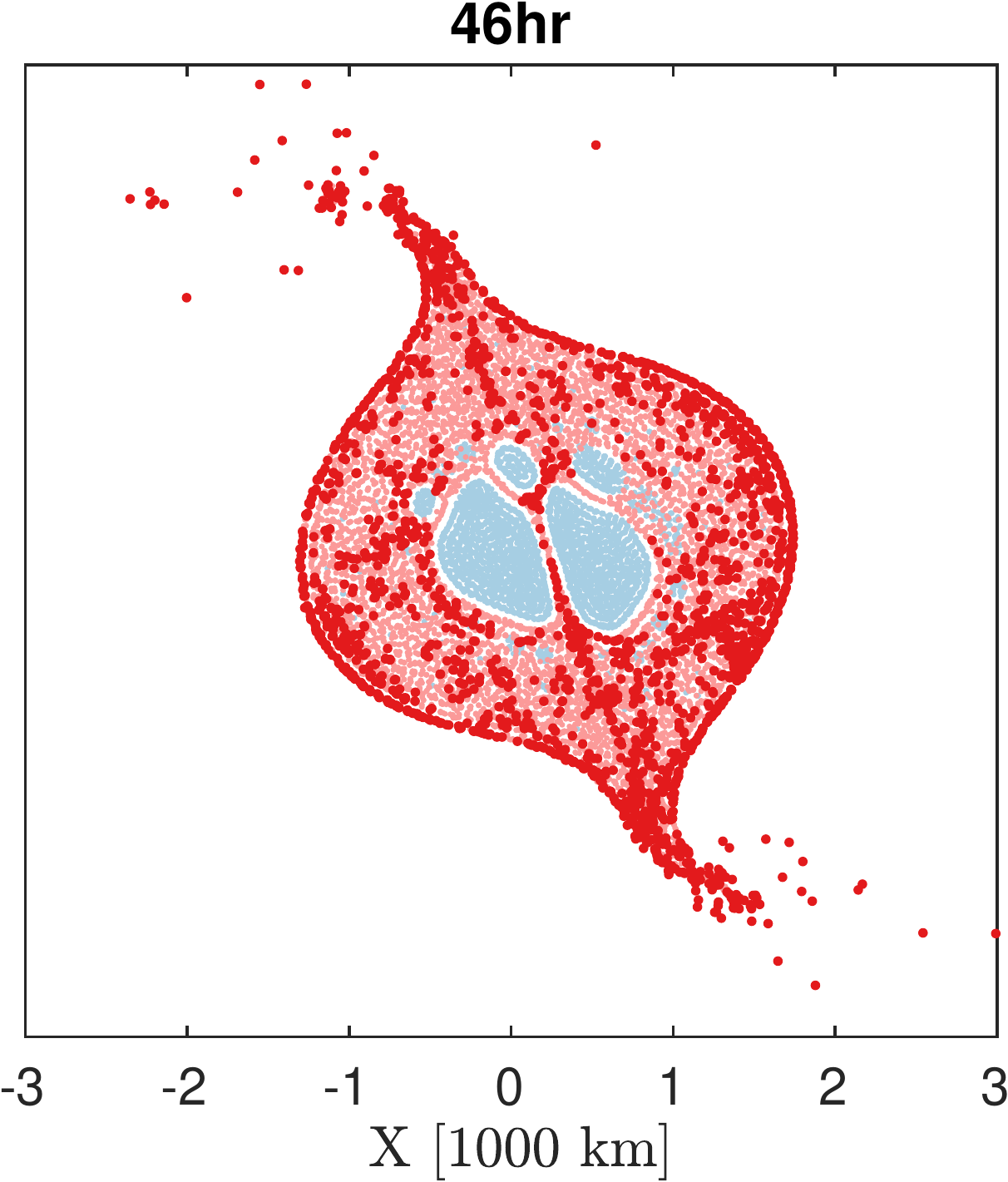} \includegraphics[height=0.2\linewidth]{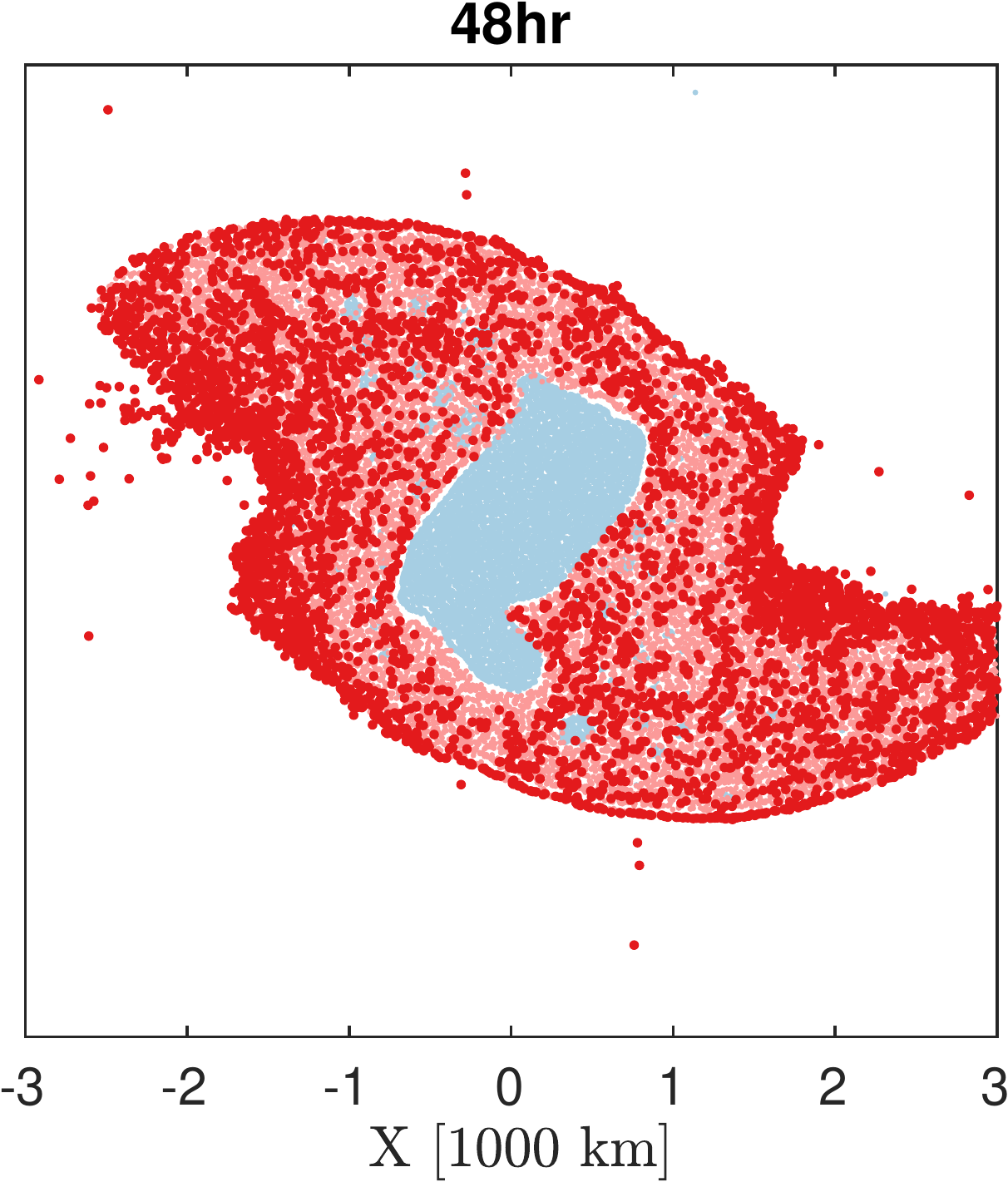} \includegraphics[height=0.2\linewidth]{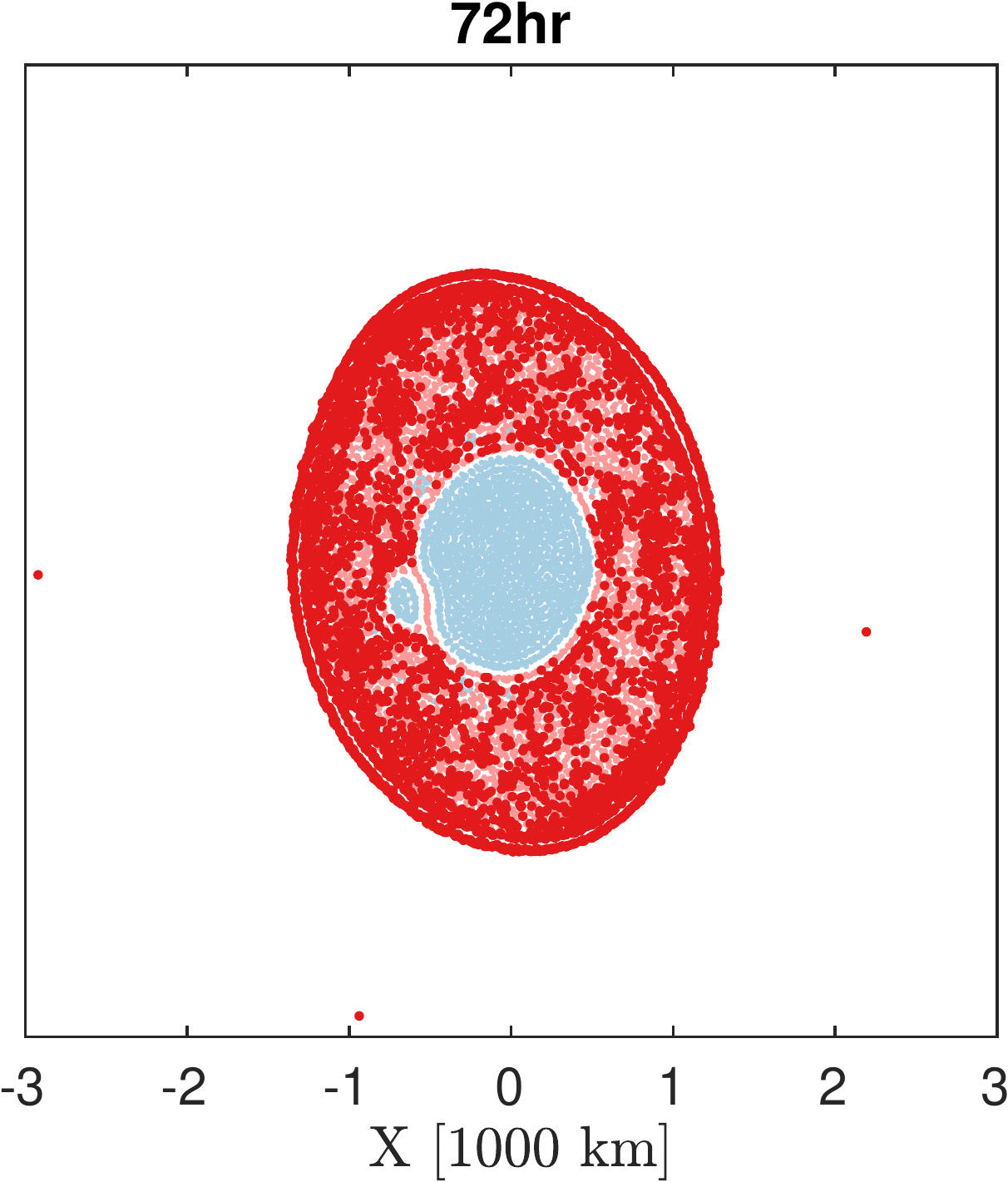}
\par\end{centering}
\caption[Figure 9]{\label{Figure 9} Melt evolution of a hit-and-run impact followed by accretionary impact. Sequence of snapshots from two colliding half moonlets at a distance of $5\ R_{{\rm Roche}}$ from the planet (first impact: $\beta=15^{o}$, $V_{{\rm imp}}=1.94\,V_{{\rm esc}}$; second impact: $\beta=10^{o}$, $V_{{\rm imp}}=0.87\,V_{{\rm esc}}$). The snapshots show a slice of width 100 km centered in the equatorial plane of the surviving moon. Note that for emphasizing purposes, the melted particles are larger. The blue particles represent iron material and light/dark red particles represent the unmelted/melted magma material. In the first  hit-and-run impact (first row), melt is mostly created in the secondary stage of the impact, where material falls back to the surface. In the second accretionary impact (second row), melt is mostly created during the deformation and redistribution of material.}
\end{figure}

Because the impact velocities are small ($\sim1\,{\rm km/s}$), melting induced by shock heating in early stages of the impact is limited. However, interactions of bodies of comparable size lead to greater mantle-redistribution. This in turn promotes melting by decompression, shear heating, and gravitational energy release from infalling material. High energy accretionary impacts can melt about $40\%$ of the final moon's mantle, corresponding to $\sim300\ {\rm km}$-deep magma ocean (assuming that all the melt is concentrated on the top layers of the moon; \nameref{Figure 7}). On the other hand, hit-and-run impacts will result in less than $<10\%$ melt, corresponding to $<100\ {\rm km}$-deep magma ocean. As discussed before, in the non-disruptive hit-and-run cases, we expect that components will impact again as they remain on intersecting orbits after the first impact. The reimpacting timescale is shorter (a few orbits $\sim{\rm hours-days}$) than the crust formation timescale ($1000\ {\rm yr}$; \citealt{Elkins2011}). Therefore, sequences of impacts (\textit{e.g. }two hit-and-run impacts, \nameref{Figure 8}, and a hit-and-run followed by an accretionary impact, \nameref{Figure 9}) add additional melt to the mantle of the surviving moonlet. Sequences of hit-and-run impacts do not transfer substantial mass between the impacts, but they can provide an additional heating source to the lunar magma ocean. 

For accretionary impacts with high mass ratios, a magma ocean of $>200\ {\rm km}$ is expected, which is consistent with the required magma ocean depth to form the lunar crust \citep{andrews2013ancient}. For impacts with small mass ratios, melt production is not efficient because these impacts are not energetic enough to disrupt and melt the body.

\subsection{Discussion and Conclusions}

We tested impacts between two orbiting moonlets. The classical phase space of two bodies in free space \citep{Leinhardt} is altered when the impacts occur under the influence of a  planetary potential. For comparable-sized impactors, hit-and-run impacts (bodies graze and transfer little amount of material) prevail over accretionary impacts \citep{asphaug2010similar}. This abundance is further increased in the tidal environment, as Hill spheres are smaller with decreasing distance to the planet and impacting bodies will require less energy to reach the point where they are not gravitationally bound. Similarly, erosion is increased in the tidal environment and the energy required for disruption (estimated by previous works, \textit{e.g. }\citealt{Movshovitz201685}) is smaller. Interactions between grazing moonlets can destabilize moonlets towards the Roche limit, where they will be disintegrated by tidal forces. The produced debris remain in orbit and can later accrete by the surviving moon, enhancing the surface mixing, or fall onto the planet. Additionally, generated debris disks can reform moonlets, similar to the scenario proposed by \cite{hesselbrock2017ongoing} for the formation of the Martian satellites. Overall, we find that moonlet merger efficiencies are lower than previously estimated by free-space simulations, therefore, dynamical studies of the evolution of multiple satellite should include some degree of fragmentation. 

\remove{In accretionary collisions, impact angular momentum induces differential rotation and enhances mixing, while in the hit-and-run regime, only small amounts of material are transferred between the bodies, therefore surfaces will be mostly sourced from a single body. However, as grazing moonlets remain on intersecting orbits, sequences of impacts are expected, each increasing the surface mixing.} Overall, if the last two components of the final Moon are comparable in size, mixing between the colliding bodies is efficient, and surface heterogeneities are not likely. \add{We note that these results represent the immediate post-impact surface variation, and emphasize that additional post-impact processes would decrease prior surface variations, further supporting our conclusions that surfaces resulting from the last-global-acrretionary events are well mixed.}\remove{If the last two components are not comparable in size, mixing is not as efficient and heterogeneities are more likely. Whereas, the mixing for the larger mass ratios occurs due to redistribution of material within the whole final moon, the mixing for small mass ratios occurs due to the disruption of the small component and reaccretion of its material on the top layers of the larger surviving moonlet.} The amount of mixing can be probed by future lunar samples (\textit{e.g, Chang'e 5} sample return mission) and compared with previously observed heterogeneities \citep{robinson2014heterogeneous, Robinson2016244}. Unfortunately, the lack of future observed heterogeneities could not preferentiate between scenarios with single or multiple moonlets impact, as the initial mixing efficiency is high. 

We estimated the typical amount of melting in moonlet impacts and their contribution to the lunar magma ocean. Overall, impacts of comparable-sized components can melt significant portions of the mantle, mainly due to redistribution of material. A smaller contribution occurs in the early stages of the impact, when shock heating is important. The observed anorthositic crust, which is created fast, after the solidification of $\sim80\%$ of the magma ocean \citep{Elkins2011} could reform after the merging impact. The impacts between moonlets extend the magma ocean phase of the final Moon. Additionally, our results show that the majority of small-sized impactors are not energetic enough to disrupt the moonlet and hence do not create enough melt to account for the observed anorthositic crust. Therefore, to avoid the disruption of the observed anorthositic lunar crust, impacts of smaller size components can occur only early in the evolution (less than $1000\ {\rm yr}$; \citealt{Elkins2011}) before crustal disruption is recorded. After crustal formation, resurfacing due to subsequent impacts is limited. We therefore conclude that in the context of the moonlet merger scenario, a relatively large last event is preferred in order to account for the global distribution of a floatation crust \citep{taylor2014lunar}.

\acknowledgments

We thank Dave Stevenson for helpful discussions. This project was supported by the Helen Kimmel Center for Planetary Science, the Minerva Center for Life Under Extreme Planetary Conditions, and by the I-CORE Program of the PBC and ISF (Center No. 1829/12). R.R. is grateful to the Israel Ministry of Science, Technology and Space for their Shulamit Aloni fellowship and NASA's SSERVI program for support. We thank \add{Simon Lock and an} anonymous reviewer for their thoughtful comments and suggestions that improved the final version of this manuscript. The modified version of GADGET-2 and EOS tables are available in the supporting information of \cite{cuk2012making}. A summary of the impact results is included in the supplementary material.



\begin{thebibliography}{55}
\providecommand{\natexlab}[1]{#1}
\expandafter\ifx\csname urlstyle\endcsname\relax
  \providecommand{\doi}[1]{doi:\discretionary{}{}{}#1}\else
  \providecommand{\doi}{doi:\discretionary{}{}{}\begingroup
  \urlstyle{rm}\Url}\fi

\bibitem[{\textit{Alvarellos}(2002)}]{Alvarellos_2002}
Alvarellos, J. (2002), Orbital evolution of impact ejecta from {G}anymede,
  \textit{Icarus}, \textit{160}(1), 108--123.

\bibitem[{\textit{Andrews-Hanna et~al.}(2013)\textit{Andrews-Hanna, Asmar,
  Head, Kiefer, Konopliv, Lemoine, Matsuyama, Mazarico, McGovern, Melosh
  et~al.}}]{andrews2013ancient}
Andrews-Hanna, J.~C., S.~W. Asmar, J.~W. Head, W.~S. Kiefer, A.~S. Konopliv,
  F.~G. Lemoine, I.~Matsuyama, E.~Mazarico, P.~J. McGovern, H.~J. Melosh,
  et~al. (2013), Ancient igneous intrusions and early expansion of the {M}oon
  revealed by {GRAIL} gravity gradiometry, \textit{Science},
  \textit{339}(6120), 675--678.

\bibitem[{\textit{Asphaug}(2010)}]{asphaug2010similar}
Asphaug, E. (2010), Similar-sized collisions and the diversity of planets,
  \textit{Chemie der Erde-Geochemistry}, \textit{70}(3), 199--219.

\bibitem[{\textit{Benz et~al.}(1989)\textit{Benz, Cameron, and
  Melosh}}]{BENZ1989113}
Benz, W., A.~Cameron, and H.~Melosh (1989), The origin of the {M}oon and the
  single-impact hypothesis {III}, \textit{Icarus}, \textit{81}(1), 113 -- 131.

\bibitem[{\textit{Bierhaus et~al.}(2012)\textit{Bierhaus, Dones, Alvarellos,
  and Zahnle}}]{Bierhaus_2012}
Bierhaus, E.~B., L.~Dones, J.~L. Alvarellos, and K.~Zahnle (2012), The role of
  ejecta in the small crater populations on the mid-sized {S}aturnian
  satellites, \textit{Icarus}, \textit{218}(1), 602--621.

\bibitem[{\textit{Brown et~al.}(2007)\textit{Brown, Barkume, Ragozzine, and
  Schaller}}]{Brown:2007aa}
Brown, M.~E., K.~M. Barkume, D.~Ragozzine, and E.~L. Schaller (2007), A
  collisional family of icy objects in the {K}uiper belt, \textit{Nature},
  \textit{446}(7133), 294--296.

\bibitem[{\textit{Cameron and Ward}(1976)}]{cameron1976origin}
Cameron, A.~G., and W.~R. Ward (1976), The origin of the {M}oon, in
  \textit{Lunar and Planetary Science Conference}, vol.~7.

\bibitem[{\textit{Canup}(2005)}]{Canup:2005aa}
Canup, R. (2005), A giant impact origin of {Pl}uto-{C}haron, \textit{Science},
  \textit{307}(5709), 546--550.

\bibitem[{\textit{Canup}(2004{\natexlab{a}})}]{Canup:2004aa}
Canup, R.~M. (2004{\natexlab{a}}), Dynamics of lunar formation, \textit{Annual
  Review of Astronomy and Astrophysics}, \textit{42}(1), 441--475.

\bibitem[{\textit{Canup}(2004{\natexlab{b}})}]{canup2004simulations}
Canup, R.~M. (2004{\natexlab{b}}), Simulations of a late lunar-forming impact,
  \textit{Icarus}, \textit{168}(2), 433--456.

\bibitem[{\textit{Canup}(2008)}]{canup2008lunar}
Canup, R.~M. (2008), Lunar-forming collisions with pre-impact rotation,
  \textit{Icarus}, \textit{196}(2), 518--538.

\bibitem[{\textit{Canup}(2012)}]{Canup23112012}
Canup, R.~M. (2012), Forming a {M}oon with an {E}arth-like composition via a
  giant impact, \textit{Science}, \textit{338}(6110), 1052--1055.

\bibitem[{\textit{Canup and Asphaug}(2001)}]{canup2001origin}
Canup, R.~M., and E.~Asphaug (2001), Origin of the {M}oon in a giant impact
  near the end of the {E}arth's formation, \textit{Nature}, \textit{412}(6848),
  708.

\bibitem[{\textit{Canup and Esposito}(1995)}]{canup1995accretion}
Canup, R.~M., and L.~W. Esposito (1995), Accretion in the {R}oche zone:
  Coexistence of rings and ringmoons, \textit{Icarus}, \textit{113}(2),
  331--352.

\bibitem[{\textit{Canup et~al.}(1999)\textit{Canup, Levison, and
  Stewart}}]{Canup1999}
Canup, R.~M., H.~F. Levison, and G.~R. Stewart (1999), Evolution of a
  terrestrial multiple-moon system, \textit{The Astronomical Journal},
  \textit{117}(1), 603.

\bibitem[{\textit{Charlier et~al.}(2018)\textit{Charlier, Grove, Namur, and
  Holtz}}]{charlier2018crystallization}
Charlier, B., T.~L. Grove, O.~Namur, and F.~Holtz (2018), Crystallization of
  the lunar magma ocean and the primordial mantle-crust differentiation of the
  {M}oon, \textit{Geochimica et Cosmochimica Acta}, \textit{234}, 50--69.

\bibitem[{\textit{Citron et~al.}(2018)\textit{Citron, Perets, and
  Aharonson}}]{citron2018role}
Citron, R.~I., H.~B. Perets, and O.~Aharonson (2018), The role of multiple
  giant impacts in the formation of the {E}arth-{M}oon system, \textit{The
  Astrophysical Journal}, \textit{862}(1), 5.

\bibitem[{\textit{{\'C}uk and Stewart}(2012)}]{cuk2012making}
{\'C}uk, M., and S.~T. Stewart (2012), Making the {M}oon from a fast-spinning
  {E}arth: A giant impact followed by resonant despinning, \textit{Science},
  \textit{338}(6110), 1047--1052.

\bibitem[{\textit{Elkins-Tanton et~al.}(2011)\textit{Elkins-Tanton, Burgess,
  and Yin}}]{Elkins2011}
Elkins-Tanton, L.~T., S.~Burgess, and Q.-Z. Yin (2011), The lunar magma ocean:
  Reconciling the solidification process with lunar petrology and
  geochronology, \textit{Earth and Planetary Science Letters}, \textit{304}(3),
  326--336.

\bibitem[{\textit{Emsenhuber et~al.}(2018)\textit{Emsenhuber, Jutzi, and
  Benz}}]{emsenhuber2018sph}
Emsenhuber, A., M.~Jutzi, and W.~Benz (2018), {SPH} calculations of
  {M}ars-scale collisions: The role of the equation of state, material
  rheologies, and numerical effects, \textit{Icarus}, \textit{301}, 247--257.

\bibitem[{\textit{Genda et~al.}(2012)\textit{Genda, Kokubo, and Ida}}]{Genda}
Genda, H., E.~Kokubo, and S.~Ida (2012), Merging criteria for giant impacts of
  protoplanets, \textit{The Astrophysical Journal}, \textit{744}(2), 137.

\bibitem[{\textit{Hartmann and Davis}(1975)}]{hartmann1975satellite}
Hartmann, W.~K., and D.~R. Davis (1975), Satellite-sized planetesimals and
  lunar origin, \textit{Icarus}, \textit{24}(4), 504--515.

\bibitem[{\textit{Hesselbrock and Minton}(2017)}]{hesselbrock2017ongoing}
Hesselbrock, A.~J., and D.~A. Minton (2017), An ongoing satellite--ring cycle
  of {M}ars and the origins of {P}hobos and {D}eimos, \textit{Nature
  Geoscience}, \textit{10}(4), 266.

\bibitem[{\textit{Housen and Holsapple}(1990)}]{housen1990fragmentation}
Housen, K.~R., and K.~A. Holsapple (1990), On the fragmentation of asteroids
  and planetary satellites, \textit{Icarus}, \textit{84}(1), 226--253.

\bibitem[{\textit{Ida et~al.}(1997)\textit{Ida, Canup, and
  Stewart}}]{ida1997lunar}
Ida, S., R.~M. Canup, and G.~R. Stewart (1997), Lunar accretion from an
  impact-generated disk, \textit{Nature}, \textit{389}(6649), 353--357.

\bibitem[{\textit{Jutzi}(2015)}]{jutzi2015sph}
Jutzi, M. (2015), {SPH} calculations of asteroid disruptions: The role of
  pressure dependent failure models, \textit{Planetary and space science},
  \textit{107}, 3--9.

\bibitem[{\textit{Jutzi and Asphaug}(2011)}]{Jutzi:2011aa}
Jutzi, M., and E.~Asphaug (2011), Forming the lunar farside highlands by
  accretion of a companion moon, \textit{Nature}, \textit{476}(7358), 69--72.

\bibitem[{\textit{Kokubo et~al.}(2006)\textit{Kokubo, Kominami, and
  Ida}}]{Kokubo2006}
Kokubo, E., J.~Kominami, and S.~Ida (2006), Formation of terrestrial planets
  from protoplanets. {I}. {S}tatistics of basic dynamical properties,
  \textit{The Astrophysical Journal}, \textit{642}(2), 1131.

\bibitem[{\textit{Laneuville et~al.}(2013)\textit{Laneuville, Wieczorek,
  Breuer, and Tosi}}]{Laneuville2013}
Laneuville, M., M.~A. Wieczorek, D.~Breuer, and N.~Tosi (2013), Asymmetric
  thermal evolution of the {M}oon, \textit{Journal of Geophysical Research:
  Planets}, \textit{118}(7), 1435--1452.

\bibitem[{\textit{Leinhardt and Stewart}(2012)}]{Leinhardt}
Leinhardt, Z.~M., and S.~T. Stewart (2012), Collisions between
  gravity-dominated bodies. {I}. {O}utcome regimes and scaling laws,
  \textit{The Astrophysical Journal}, \textit{745}(1), 79.

\bibitem[{\textit{Leleu et~al.}(2018)\textit{Leleu, Jutzi, and
  Rubin}}]{leleu2018peculiar}
Leleu, A., M.~Jutzi, and M.~Rubin (2018), The peculiar shapes of {S}aturn's
  small inner moons as evidence of mergers of similar-sized moonlets,
  \textit{Nature Astronomy}, \textit{2}(7), 555--561.

\bibitem[{\textit{Lock et~al.}(2018)\textit{Lock, Stewart, Petaev, Leinhardt,
  Mace, Jacobsen, and Cuk}}]{lock2018origin}
Lock, S.~J., S.~T. Stewart, M.~I. Petaev, Z.~Leinhardt, M.~T. Mace, S.~B.
  Jacobsen, and M.~Cuk (2018), The origin of the {M}oon within a terrestrial
  synestia, \textit{Journal of Geophysical Research: Planets}, \textit{123}(4),
  910--951.

\bibitem[{\textit{Marcus}(2011)}]{marcus2011role}
Marcus, R.~A. (2011), \textit{The role of giant impacts in planet formation and
  internal structure}, Harvard University.

\bibitem[{\textit{Marcus et~al.}(2009)\textit{Marcus, Stewart, Sasselov, and
  Hernquist}}]{Marcus2009}
Marcus, R.~A., S.~T. Stewart, D.~Sasselov, and L.~Hernquist (2009), Collisional
  stripping and disruption of super-earths, \textit{The Astrophysical Journal
  Letters}, \textit{700}(2), L118.

\bibitem[{\textit{Melosh}(2007)}]{melosh2007hydrocode}
Melosh, H. (2007), A hydrocode equation of state for {S}i{O}2,
  \textit{Meteoritics \& Planetary Science}, \textit{42}(12), 2079--2098.

\bibitem[{\textit{Melosh et~al.}(2017)\textit{Melosh, Kendall, Horgan, Johnson,
  Bowling, Lucey, and Taylor}}]{melosh2017south}
Melosh, H., J.~Kendall, B.~Horgan, B.~Johnson, T.~Bowling, P.~Lucey, and
  G.~Taylor (2017), {S}outh {P}ole--{A}itken basin ejecta reveal the {M}oon's
  upper mantle, \textit{Geology}, \textit{45}(12), 1063--1066.

\bibitem[{\textit{Meyer et~al.}(2010)\textit{Meyer, Elkins-Tanton, and
  Wisdom}}]{meyer2010coupled}
Meyer, J., L.~Elkins-Tanton, and J.~Wisdom (2010), Coupled thermal--orbital
  evolution of the early {M}oon, \textit{Icarus}, \textit{208}(1), 1--10.

\bibitem[{\textit{Movshovitz et~al.}(2016)\textit{Movshovitz, Nimmo,
  Korycansky, Asphaug, and Owen}}]{Movshovitz201685}
Movshovitz, N., F.~Nimmo, D.~Korycansky, E.~Asphaug, and J.~Owen (2016), Impact
  disruption of gravity-dominated bodies: New simulation data and scaling,
  \textit{Icarus}, \textit{275}, 85 -- 96.

\bibitem[{\textit{Ortiz et~al.}(2017)\textit{Ortiz, Santos-Sanz, Sicardy,
  Benedetti-Rossi, B{\'e}rard, Morales, Duffard, Braga-Ribas, Hopp, Ries
  et~al.}}]{Ortiz:2017aa}
Ortiz, J.~L., P.~Santos-Sanz, B.~Sicardy, G.~Benedetti-Rossi, D.~B{\'e}rard,
  N.~Morales, R.~Duffard, F.~Braga-Ribas, U.~Hopp, C.~Ries, et~al. (2017), The
  size, shape, density and ring of the dwarf planet {H}aumea from a stellar
  occultation, \textit{Nature}, \textit{550}(7675), 219.

\bibitem[{\textit{Pahlevan and Stevenson}(2007)}]{Pahlevan:2007aa}
Pahlevan, K., and D.~J. Stevenson (2007), Equilibration in the aftermath of the
  lunar-forming giant impact, \textit{Earth and Planetary Science Letters},
  \textit{262}, 438--449.

\bibitem[{\textit{Raymond et~al.}(2009)\textit{Raymond, O'Brien, Morbidelli,
  and Kaib}}]{Raymond2009644}
Raymond, S.~N., D.~P. O'Brien, A.~Morbidelli, and N.~A. Kaib (2009), Building
  the terrestrial planets: Constrained accretion in the inner solar system,
  \textit{Icarus}, \textit{203}(2), 644 -- 662.

\bibitem[{\textit{Robinson and Taylor}(2014)}]{robinson2014heterogeneous}
Robinson, K.~L., and G.~J. Taylor (2014), Heterogeneous distribution of water
  in the {M}oon, \textit{Nature Geoscience}, \textit{7}(6), 401.

\bibitem[{\textit{Robinson et~al.}(2016)\textit{Robinson, Barnes, Nagashima,
  Thomen, Franchi, Huss, Anand, and Taylor}}]{Robinson2016244}
Robinson, K.~L., J.~J. Barnes, K.~Nagashima, A.~Thomen, I.~A. Franchi, G.~R.
  Huss, M.~Anand, and G.~J. Taylor (2016), Water in evolved lunar rocks:
  Evidence for multiple reservoirs, \textit{Geochimica et Cosmochimica Acta},
  \textit{188}, 244 -- 260.

\bibitem[{\textit{Rufu and Canup}(2017)}]{RufuCanup}
Rufu, R., and R.~M. Canup (2017), Triton's evolution with a primordial
  {N}eptunian satellite system, \textit{The Astronomical Journal},
  \textit{154}(5), 208.

\bibitem[{\textit{Rufu et~al.}(2017)\textit{Rufu, Aharonson, and
  Perets}}]{Rufu:2017aa}
Rufu, R., O.~Aharonson, and H.~B. Perets (2017), A multiple-impact origin for
  the {M}oon, \textit{Nature Geoscience}, \textit{10}, 89 -- 94.

\bibitem[{\textit{Salmon and Canup}(2012)}]{salmon2012lunar}
Salmon, J., and R.~M. Canup (2012), Lunar accretion from a {R}oche-interior
  fluid disk, \textit{The Astrophysical Journal}, \textit{760}(1), 83.

\bibitem[{\textit{Salmon and Canup}(2014)}]{Salmon20130256}
Salmon, J., and R.~M. Canup (2014), Accretion of the {M}oon from non-canonical
  discs, \textit{Philosophical Transactions of the Royal Society of London A:
  Mathematical, Physical and Engineering Sciences}, \textit{372}(2024).

\bibitem[{\textit{Schlichting and Sari}(2009)}]{SchlichtingSari}
Schlichting, H.~E., and R.~Sari (2009), The creation of {H}aumea's collisional
  family, \textit{The Astrophysical Journal}, \textit{700}(2), 1242.

\bibitem[{\textit{Spath}(1985)}]{spath1985cluster}
Spath, H. (1985), \textit{The cluster dissection and analysis theory {f}ortran
  programs examples}, Prentice-Hall, Inc.

\bibitem[{\textit{Springel}(2005)}]{Springel2005}
Springel, V. (2005), The cosmological simulation code {GADGET}-2,
  \textit{Monthly Notices of the Royal Astronomical Society}, \textit{364}(4),
  1105--1134, \doi{10.1111/j.1365-2966.2005.09655.x}.

\bibitem[{\textit{Stewart and Leinhardt}(2009)}]{stewart2009velocity}
Stewart, S.~T., and Z.~M. Leinhardt (2009), Velocity-dependent catastrophic
  disruption criteria for planetesimals, \textit{The Astrophysical Journal
  Letters}, \textit{691}(2), L133.

\bibitem[{\textit{Taylor and Wieczorek}(2014)}]{taylor2014lunar}
Taylor, G.~J., and M.~A. Wieczorek (2014), Lunar bulk chemical composition: a
  post-gravity recovery and interior laboratory reassessment, \textit{Phil.
  Trans. R. Soc. A}, \textit{372}(2024), 20130,242.

\bibitem[{\textit{Thompson}(1990)}]{thompson1990aneos}
Thompson, S. (1990), Aneos---analytic equations of state for shock physics
  codes, sandia natl, \textit{Lab. Doc. SAND89-2951}.

\bibitem[{\textit{{Wood}}(1986)}]{Wood1986}
{Wood}, J. (1986), {Moon over Mauna Loa - A review of hypotheses of formation
  of Earth's Moon}, in \textit{Origin of the Moon}, edited by W.~{Hartmann},
  R.~{Phillips}, and G.~{Taylor}, pp. 17--55.

\bibitem[{\textit{Zuber et~al.}(2013)\textit{Zuber, Smith, Watkins, Asmar,
  Konopliv, Lemoine, Melosh, Neumann, Phillips, Solomon
  et~al.}}]{zuber2013gravity}
Zuber, M.~T., D.~E. Smith, M.~M. Watkins, S.~W. Asmar, A.~S. Konopliv, F.~G.
  Lemoine, H.~J. Melosh, G.~A. Neumann, R.~J. Phillips, S.~C. Solomon, et~al.
  (2013), Gravity field of the {M}oon from the gravity recovery and interior
  laboratory ({GRAIL}) mission, \textit{Science}, \textit{339}(6120), 668--671.

\end{thebibliography}
\end{document}


\supportinginfo{Impact Dynamics of Moons Inside a Planetary Potential}
\authors{R. Rufu\affil{1,2}and O. Aharonson\affil{1,3}}
\affiliation{1}{Weizmann Institute of Science, Department of Earth and Planetary Sciences, Rehovot, Israel}
\affiliation{2}{Southwest Research Institute, 1050 Walnut Street, Suite 300, Boulder, CO 80302, USA}
\affiliation{3}{Planetary Science Institute, Tucson, AZ, USA}

\correspondingauthor{R. Rufu}{raluca@boulder.swri.edu}

\section*{Contents}
\begin{enumerate}
\item Text S1 to S2
\item Figure S1 to S7
\item Caption Movie S1 to S2
\end{enumerate}

\section*{Text S1 - Variations due to Different Initial Thermal States}

\begin{figure*}

\begin{centering}
\includegraphics[height=5.2cm]{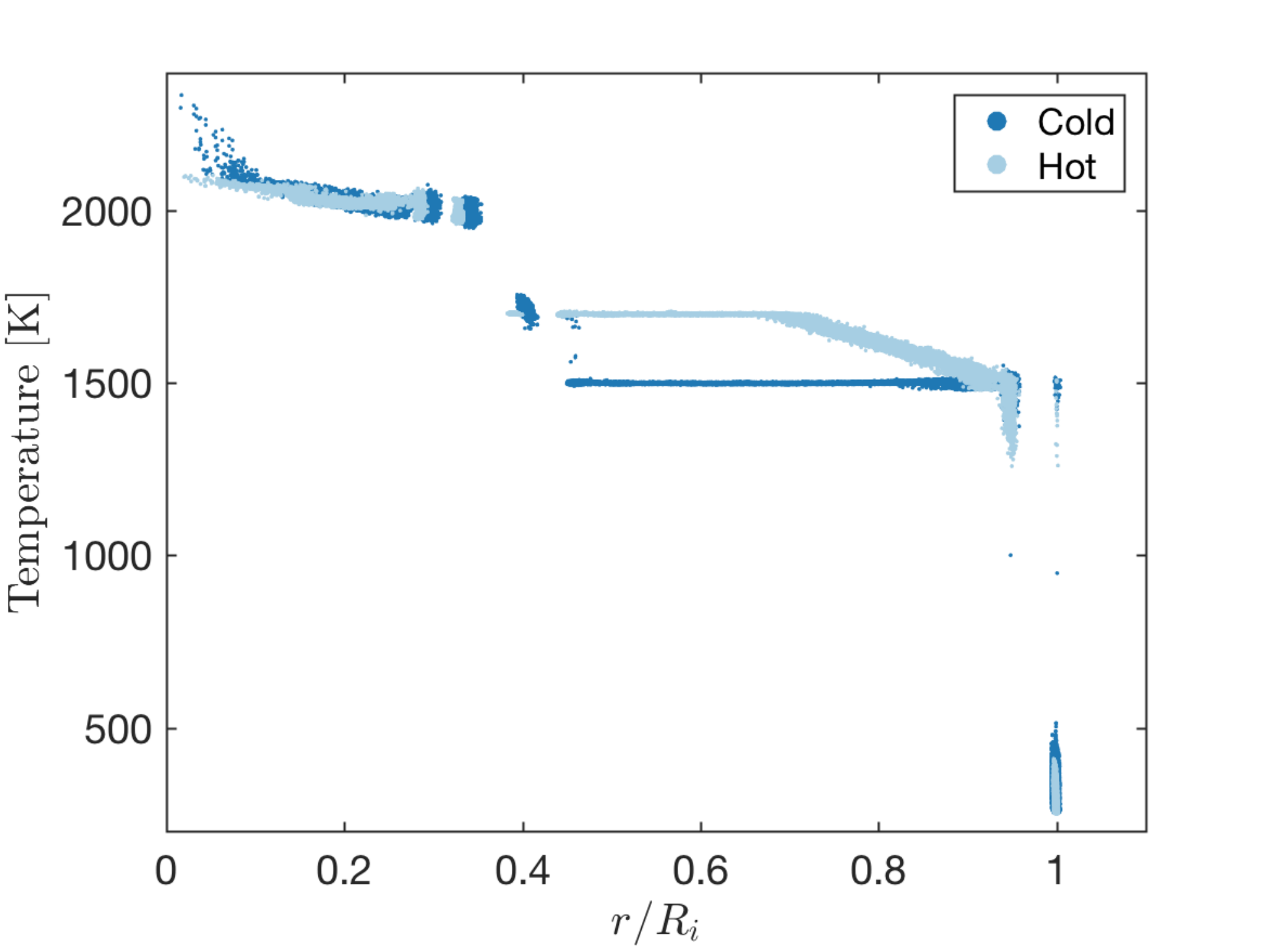}\par\end{centering} 
\caption{\label{fig:InitialTemp}{Initial profile temperatures as a function of the distance from the center, normalized by the radius of the moonlet, $R_i$.}}
\end{figure*}

Different lunar formation scenarios can predict different initial thermal states, depending on the exact masses and separation of the two moonlets, the impact timescale, and hence the cooling time before impact, may vary \citep{Canup1999}. For example, the canonical scenario \citep{Canup:2004aa} would predict a fully molten Moon \citep{barr2016origin}, that would require $10\,\rm{Myr}$ for total solidification \citep{Elkins2011}. Whereas, in the multiple impact hypothesis, the resulting moonlets are smaller (faster solidification) and the time between moonlet merger is longer, mostly depending on the moonlet formation timescale. To cover the range of possible Moon formation scenarios, we tested two extreme cases of initial thermal states, ``cold'' and ``hot'' states from previous studies (\citealp{Laneuville2013}; see Figure \ref{fig:InitialTemp}). For both initial states the surface temperature is set to $250\ \rm{[K]}$. Although, we chose two extreme cases, still the maximum temperature difference between the two moonlets is $200\ \rm K$, resulting in a maximum density (sound speed) difference of $0.02 \ \rm{g/cm^3}$ ($0.2\ \rm{km/s}$). We performed 48 pairs of impact simulations between two comparable-size bodies with different impact parameters, spanning the regimes found in subsection 3.1.
 
\add{In order to statistically quantify the surface variation in this sections, we choose to represent the overall surface variation as:}
\begin{equation}
   \Delta_{\rm surf}=\sigma-\sigma_{\rm rand}\label{Eq:SurfVar}
\end{equation}
\add{where, $\sigma$ is the standard variation of the surface particles values ($f_{\rm moon1}-f_{\rm moon2}$, as defined in section 3.2) and $\sigma_{\rm rand}$ is the standard variation of a random  distribution of the surface particles. Under this definition, small values of $\Delta_{\rm surf}$ are expected for surfaces that are are well mixed between two bodies (Figure S2-a) or surfaces that include material from one moonlet (Figure S2-b). Whereas, large values of $\Delta_{\rm surf}$ are expected for surfaces that include material from both bodies but their material is not mixed, creating regions where the material is unproportionally sourced (Figure S2-c, d). The statistic $\Delta_{\rm surf}$ should be regarded as an estimate of heterogeneity, designed to provide a convenient summary of the results, other definitions are certainly possible.}

The \add{the comparison between the two thermal states}\remove{results} shows that the surface variation between pairs of simulations is similar, therefore in this range of temperatures the surface variation is independent on the initial thermal state chosen (Figure \ref{fig:HotColdComp}). In all cases the two simulations resulted in close surface variation values ($|\Delta_{\rm surf,hot}-\Delta_{\rm surf,cold}|<0.1$). \par

The \add{majority of} mantle \add{particles }in the "hot" initial thermal state \change{is}{are} closer to the melting point, hence less energy is required for melting \add{(Figure }\ref{fig:HotColdMeltPar}). As expected, it has higher fractions of melt, compared to the "cold" initial state. However, for low velocity and hit-and-run impacts the difference between the two states is low (Figure \ref{fig:HotColdMelt}). For accretionary impacts ($V_{\rm imp}\sim V_{\rm esc}$ in Figure \ref{fig:HotColdMelt}), where deformation is abundant, the amount of melted material at the end of the simulation is significantly higher (highest difference value $\sim0.1$).
 As the differences between the two thermal states are smaller compared to the differences between the impacts parameters, we chose to focus on the "cold" initial state.


\begin{figure*}
\hspace{0.8cm} a) \hspace{2.5cm}  b) \hspace{2.5cm}  c) \hspace{2.5cm}  d)\hspace{2.5cm} \newline
\begin{centering} 
\includegraphics[height=3cm]{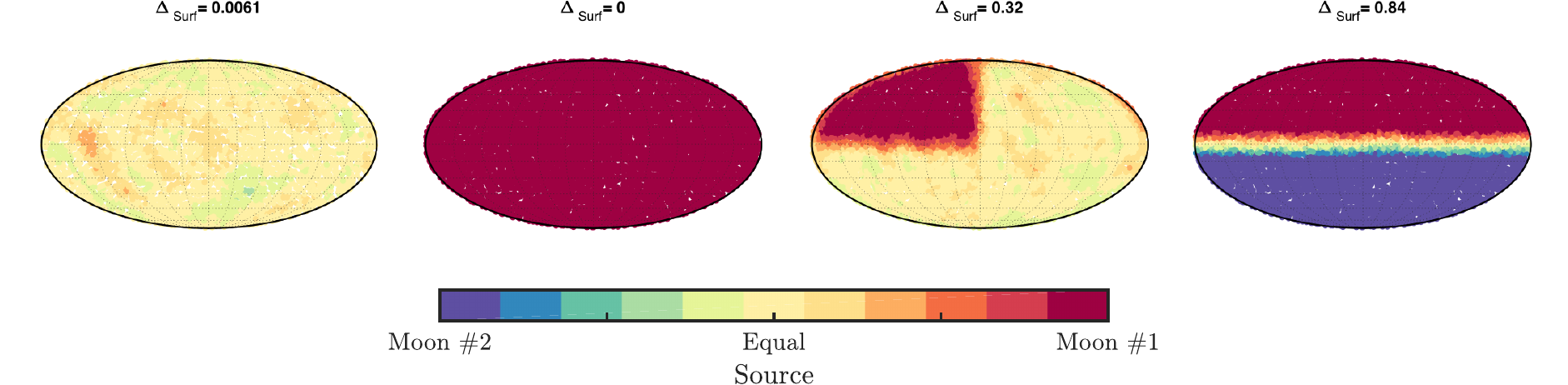}\par
\end{centering}
\caption{\label{fig:VariationVal}Surface variations across artificially generated surfaces. Small values of surface variation result from surfaces that are well mixed  between the two sources (a) or surfaces that are sourced from one body (b). Large values of surface variation result from surfaces that include material from both sources but their material is unmixed, creating regions where material is unproportionally sourced (c, d)}
\end{figure*} 

\begin{figure*}
\begin{tabular}{ll}
a) & b) \\
\begin{centering}
\includegraphics[height=5.2cm]{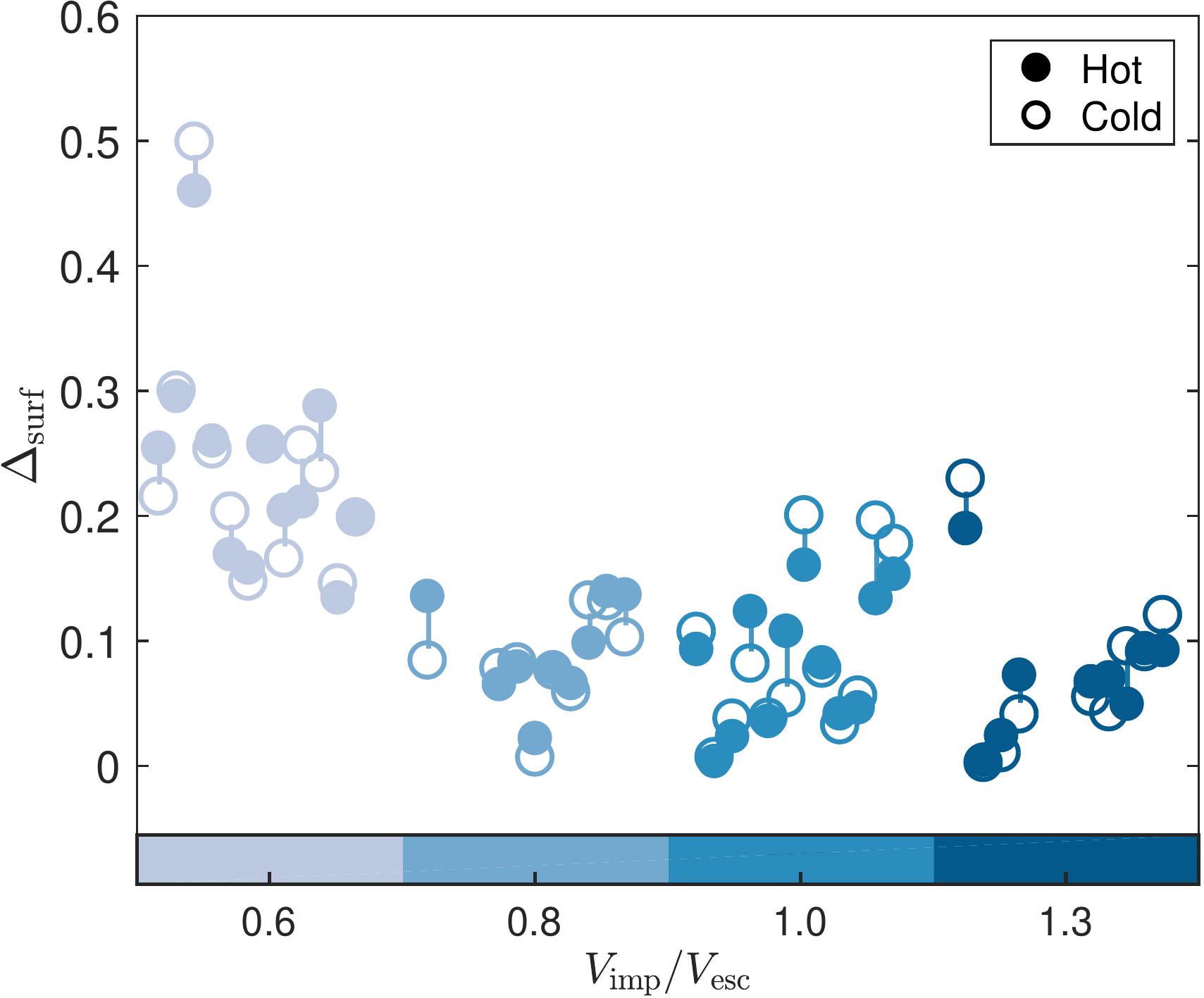}\end{centering} 
& 
\begin{centering}
\includegraphics[height=5.2cm]{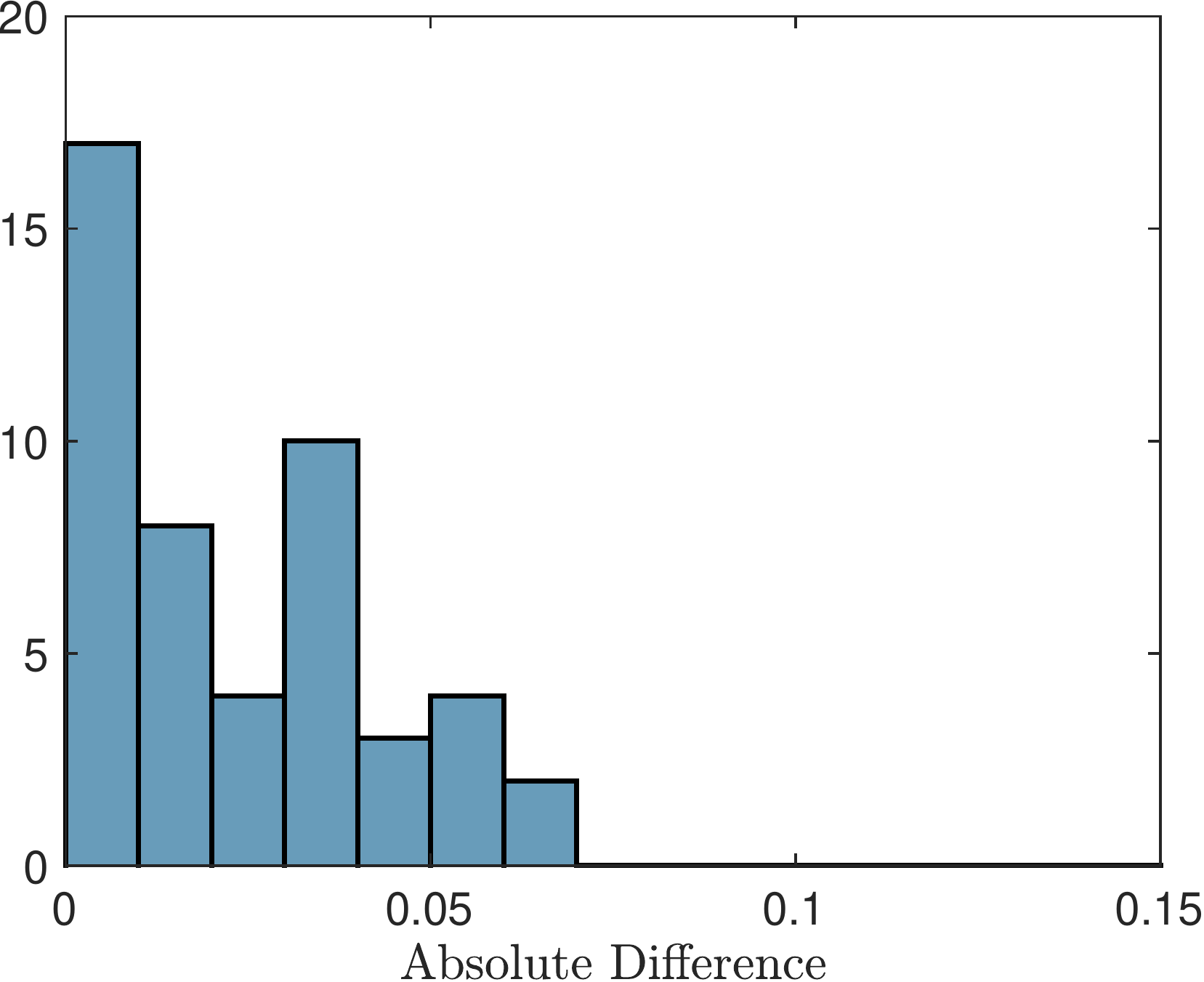}\end{centering}
\end{tabular}
\caption{\label{fig:HotColdComp}{Surface variation difference between two simulations with different thermal states. a) Variation values ($\Delta_{\rm surf}$) of same impact conditions but different initial thermal state. Filled/empty circles represent the hot/cold initial thermal  state. Colors represent the impact velocities. The vertical line represents the difference between the two simulations. b) Distribution of absolute surface heterogeneity difference between the two states ($|\Delta_{\rm surf,hot}-\Delta_{\rm surf,cold}|$).}}
\end{figure*} 

\begin{figure*}
\begin{tabular}{ll}
a) &  b) \\
\includegraphics[height=5.2cm]{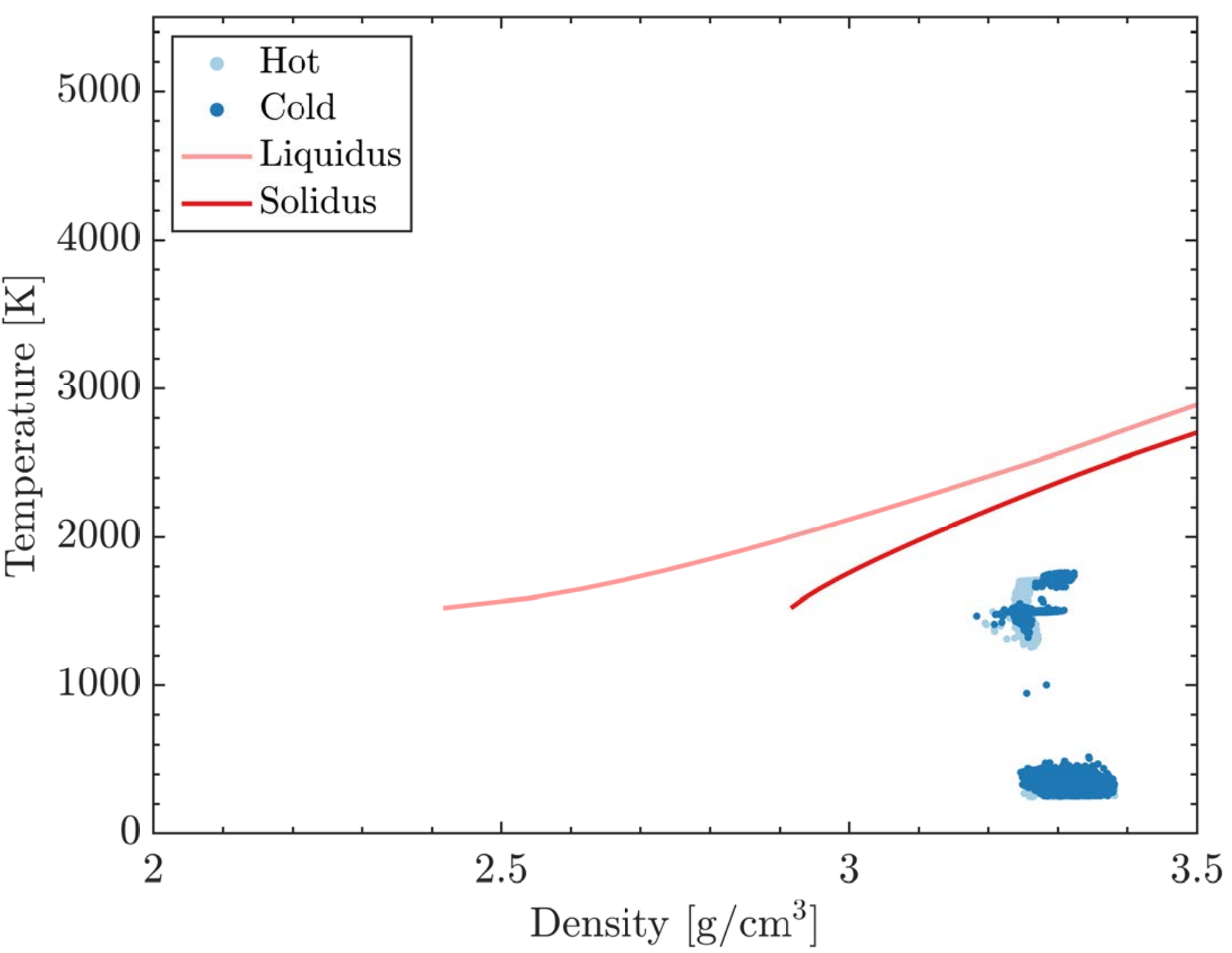}
&
\includegraphics[height=5.2cm]{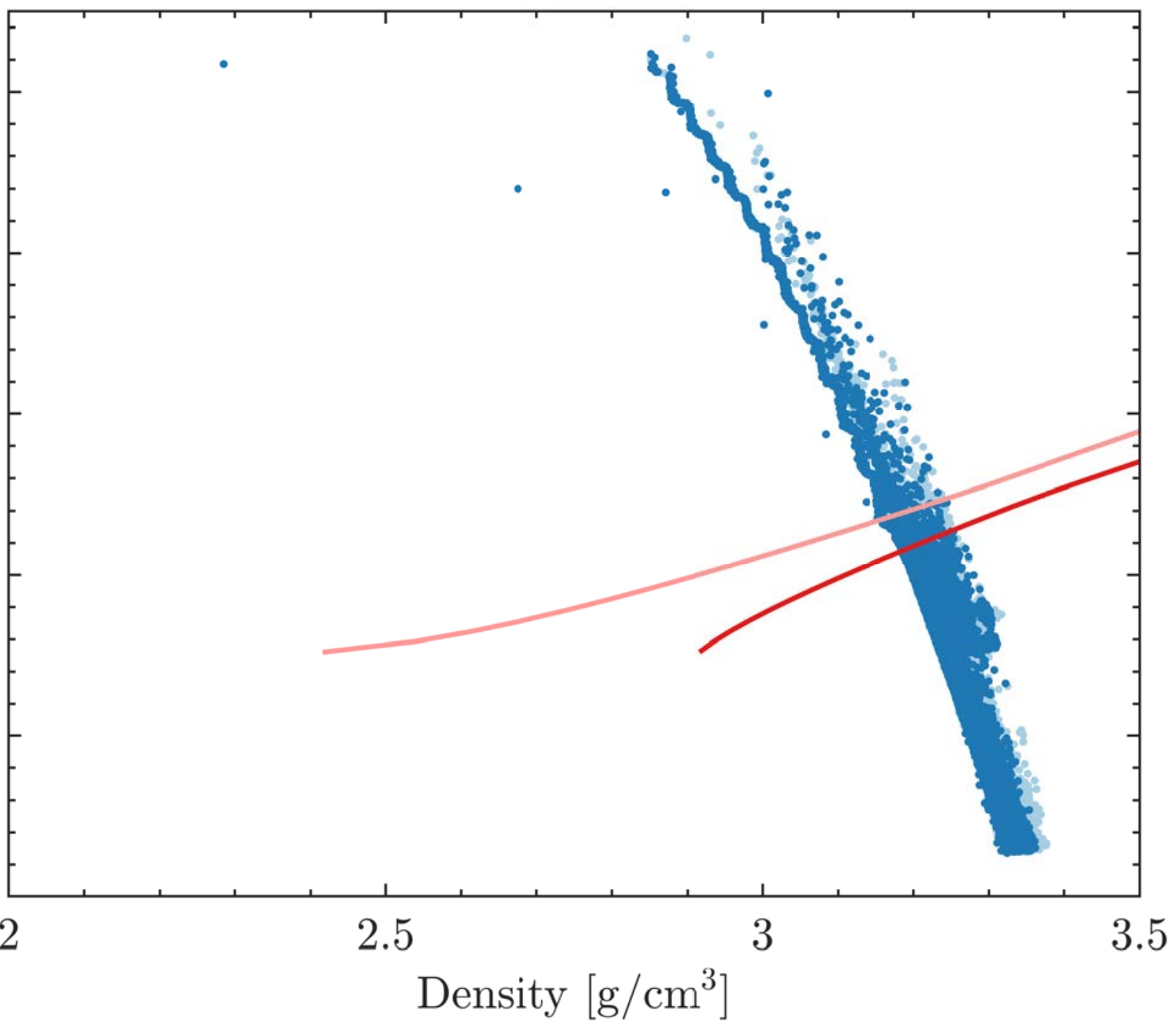}

\end{tabular}
\caption{\label{fig:HotColdMeltPar}{Thermal state of SPH particles. Temperature as a function of density at the a) beginning; b) end of simulation for the largest surviving remnant in a hit-and-run simulation. The dark/light red curves represent the solidus/liquidus curve, calculated by the M-ANEOS code \citep{melosh2007hydrocode} (input parameters provided by J. Melosh) .}}
\end{figure*} 

\begin{figure*}

\begin{centering}
\includegraphics[height=5.2cm]{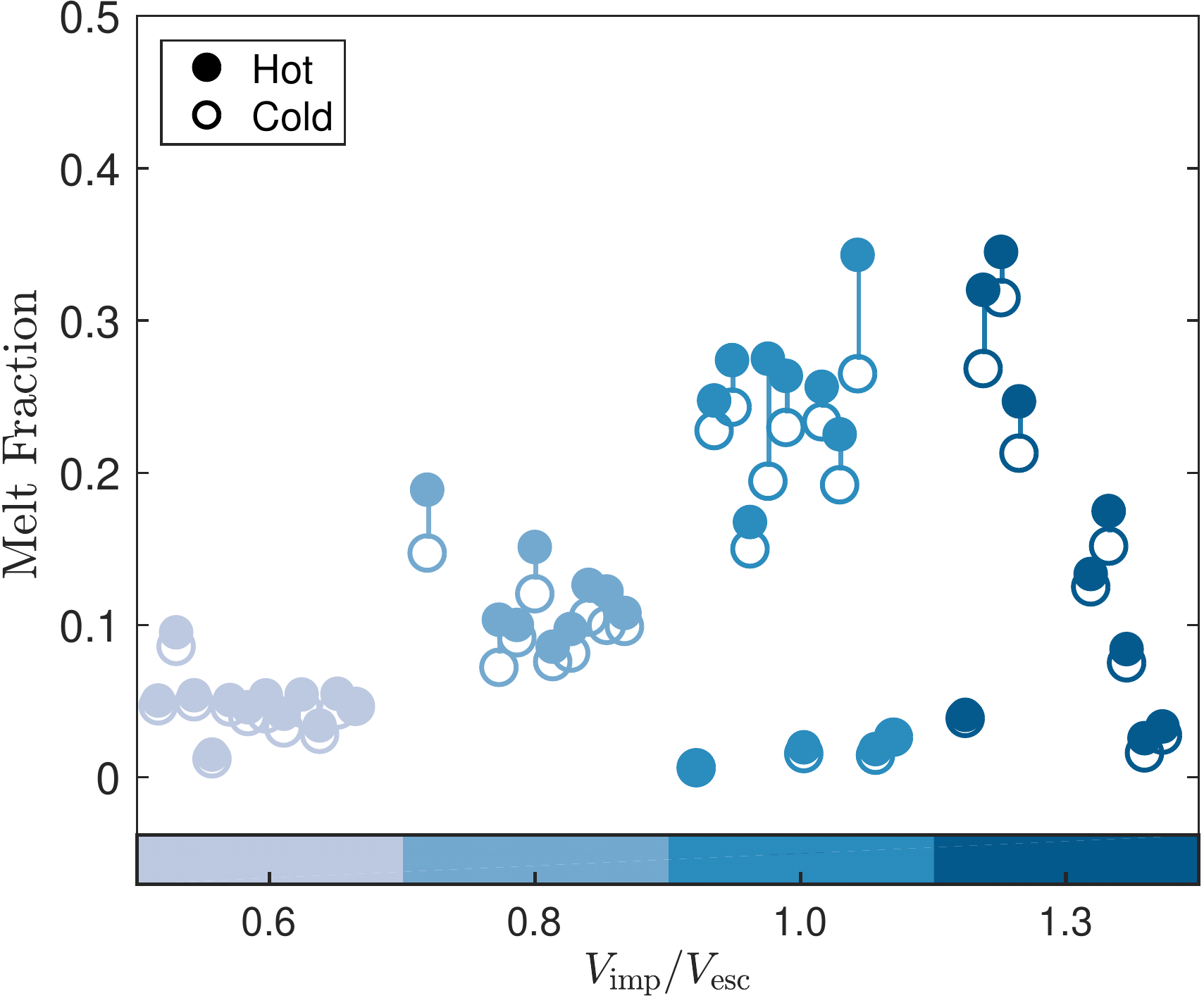}\par\end{centering} 
\caption[Melting fractions for different initial state difference]{\label{fig:HotColdMelt}Melting fraction difference between two simulation with two different initial states but same impact conditions. Filled/empty circles represent the hot/cold initial thermal  state. Colors represent the impact velocities. The vertical line represents the difference between the two simulations.}
\end{figure*} 
\section*{Text S2 - Surface Variation Sensitivity Tests}

To find the suitable neighborhood length that describes the large scale surface variations (differences between quadrants), we  calculated the surface variation for the case depicted in Figure 6-d, assuming a range of values, $L_{\rm neighbour}=50-1000\,\rm km$. Using small $L_{\rm neighbour}$,  order of magnitude of smoothing length, most particles have small number of neighbours, therefore large scale surface variations are not resolved (\textit{e.g}, Figure S6 - b $L_{\rm neighbour}= 50\,\ \rm km$). By increasing $L_{\rm neighbour}$ each particle samples more neighbours (\textit{e.g.}, for $L_{\rm neighbour} = 200\,\rm km$ case, typically each particle has 20-30 neighbours) and larger scale heterogeneities are resolved (\textit{e.g}, Figure S6-b for $L_{\rm neighbour}= 200\,\rm km$, difference in source material between equatorial and pole region). Surface variation values between  $250-700\,\rm km$ have a small dependence on the $L_{\rm neighbour}$ chosen, whereas for larger values of the neighborhood length the  heterogeneity values decrease (Figure S6-a). This is not surprising as when the neighbourhood length approaches similar scales to the radius of the moonlet, particles sample too many neighbours, and dichotomies between quadrants are barely resolved. As heterogeneity values do not depend strongly on the neighborhood length chosen for $L_{\rm neighbour} = 250-700\,\rm km$, we chose to use the $L_{\rm neighbour} = 400\,\rm km$ value for our surface variation analysis.

\add{In order to ensure that the the nominal resolution (number of SPH particles, $N_{\rm par}\sim2\cdot10^5$) is appropriate for estimating the surface variations, we performed additional impact simulations with varying number of SPH particles ($N_{\rm par}\sim2\cdot10^4$, $1\cdot10^5$ and $4\cdot10^5$). For ease of comparison, we chose to reproduce one of the impacts that resulted in some degree of heterogeneity (impact between two-half-moon sized bodies with a low impact velocity, $V_{\rm imp}=0.9V_{\rm esc}$, and angle, $\beta=21^o$).}
\add{These tests show that the pattern of the surface mixing variations remains constant with number of particles $\ge10^5$ (Figure S7).  The amplitude of the mixing variations is somewhat reduced in the highest resolution case ($N\sim4\cdot10^5$), but here too, the pattern is maintained (Figure~}\ref{fig:Resulution}\add{-left panel - first row; surface material at high latitudes is unmixed). We conclude that simulations that show a high degree of mixing between the moonlets will remain so at higher resolution; increasing the number of particles is likely only to strengthen this result.}

\add{We note that, in the main text we have defined the surface particles as the upper $5\%$ particles in each longitude-latitude rectangle of $15^o\times15^o$. According to this definition, the surface is represented by $\mathcal{O}(10^3)$ particles, and sample particles as deep as $80\ \rm km$ for the nominal resolution. It allows a fast calculation of neighbouring particles and hence the surface mixing. However, it does not systematically sample the same depth for each longitude-latitude rectangle. Therefore, we made an additional test and calculated the mixing of all particles in the upper $50\ \rm{km}$  layer. According to this definition, the surface is represented by $\mathcal{O}(10^4)$ particles in the nominal resolution ($2\cdot10^5$ particles), and ensures that the sampling depth is consistent when the number of particles varies. However this technique, drastically increases the computational time of the neighboring particle search. We verified that the two surface definitions are consistent by comparing the $\Delta_{\rm surf}$ values on a subset of our nominal resolution ($\gamma=0.5$, 90 simulations), and find that the surface variations values are similar within $|\Delta_{\rm surf}<0.1|$. Finally, the variations among the different resolutions (Figure }\ref{fig:Resulution}\add{; second row) show a similar trend as before.}

\begin{figure*}
a)\newline
\begin{centering}
\includegraphics[height=5.2cm]{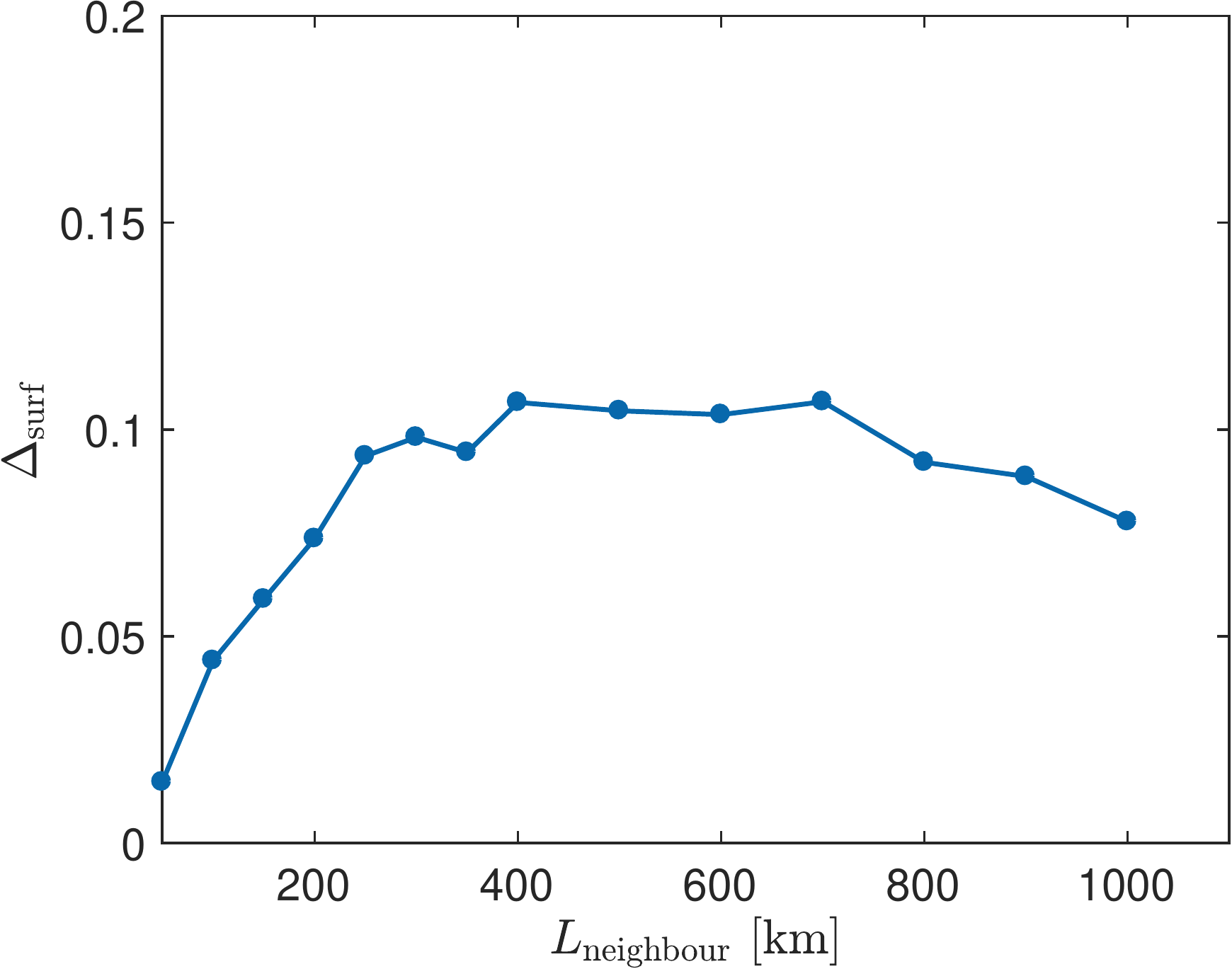}\par\end{centering} 
b)\newline
\begin{centering}
\includegraphics[height=5.2cm]{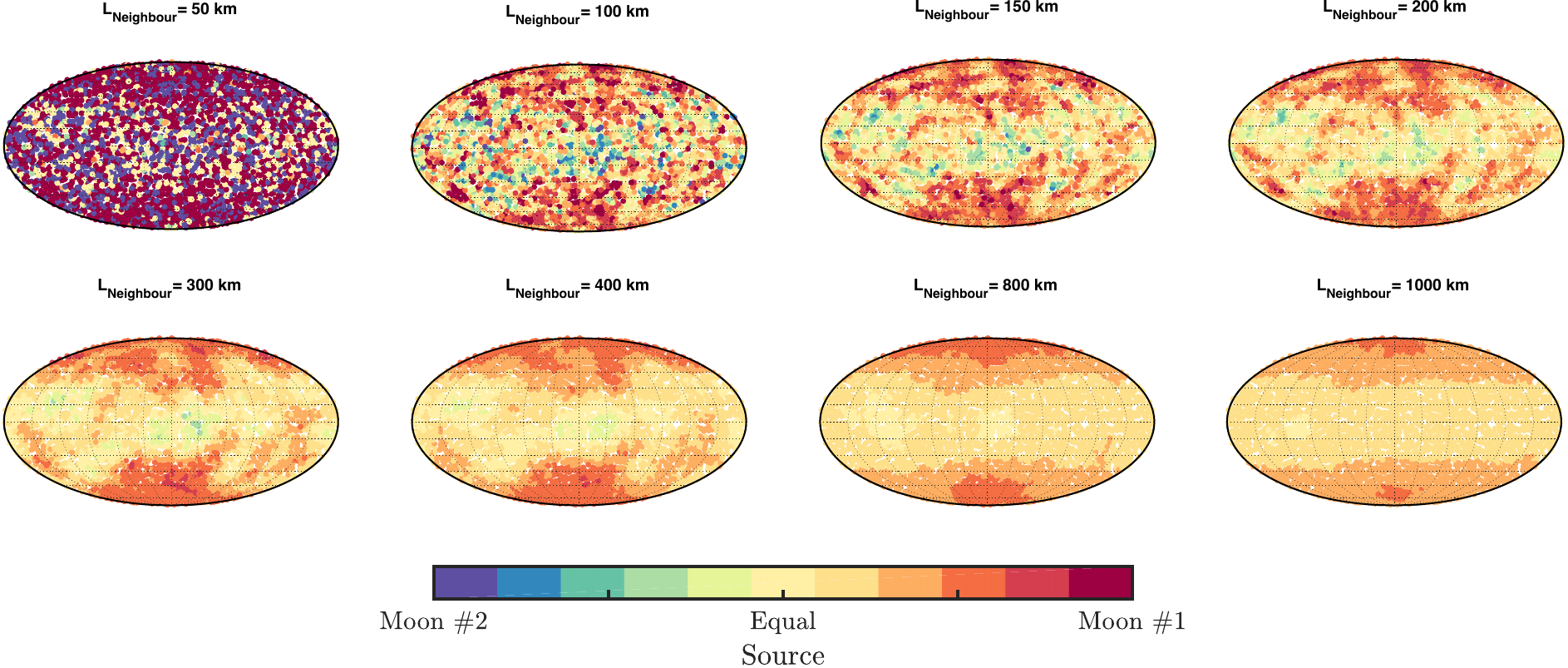}\par\end{centering} 
\caption{\label{fig:MixSens}Sensitivity analysis on cases depicted in Figure 6-c4. a) Surface variation values defined by Eq. 1 vs. the neighbourhood length, $L_{\rm neighbour}$; b) Source material from each moonlet for every surface particle. For the main manuscript we use the $L_{\rm neighbour}=400\ \rm km$}
\end{figure*} 

\begin{figure*}
\begin{centering}
\includegraphics[height=5cm]{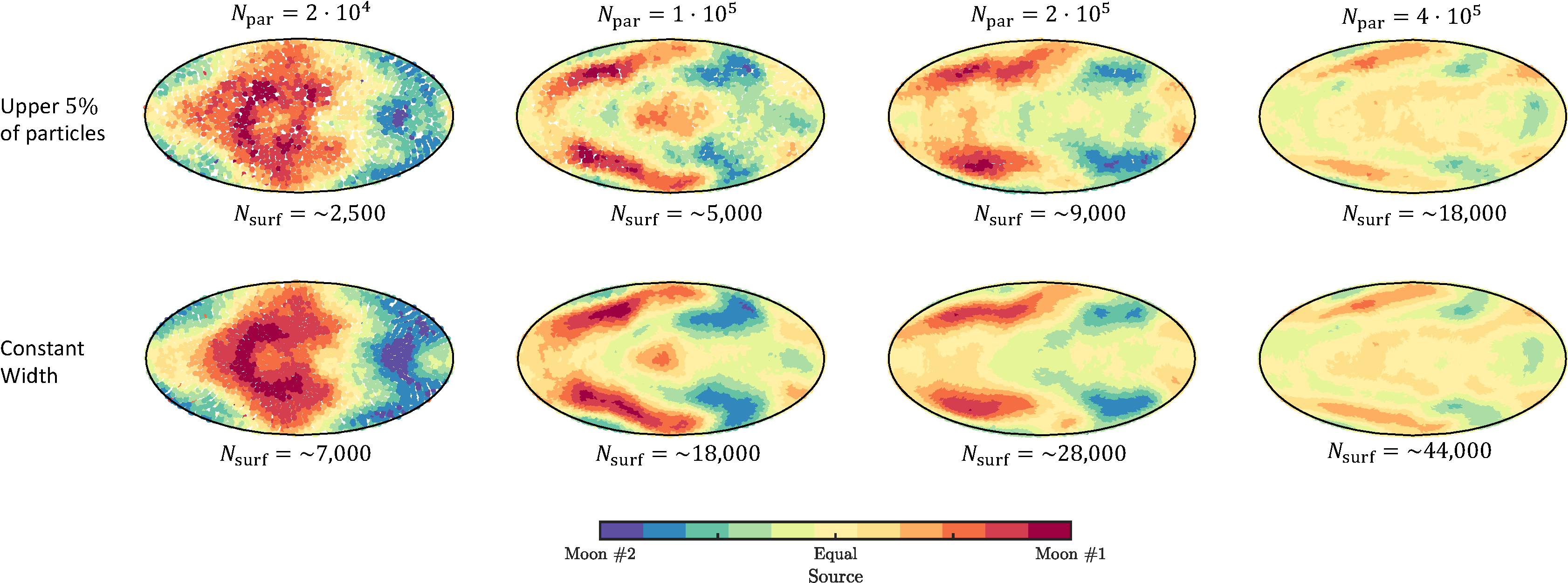}
\end{centering} 
\caption{\label{fig:Resulution} \add{Resolution analysis. Source material on final moon, resulting from an impact between two-half-moon sized bodies with a low impact velocity and angle ($V_{\rm imp}=0.9\,V_{\rm esc}$; $\beta=21^o$) with varying number of SPH particles, $N_{\rm par}$ (detailed on top of each column). The surface is defined as the upper $5\%$ particles in each longitude-latitude rectangle of $15^o\times15^o$ (first row) or the upper $50\ \rm{km}$ layer (second row). The surface variation value, $\Delta_{\rm surf}$, and the number of surface particles, $N_{\rm surf}$, is detailed below each subpanel.}}
\end{figure*} 
\newpage
\section*{Caption Movie S1}
Evolution of a sequence of two hit-and-run impacts. The  two-half moonlets impact at a distance of $1.5\,R_{{\rm Roche}}$ from the planet (first impact: $\beta=40^{o}$; $V_{{\rm imp}}=1.66\,V_{{\rm esc}}$; second impact: $\beta=44^{o}$; $V_{{\rm imp}}=1.56\,V_{{\rm esc}}$). The planet (Roche limit) is represented by the green circle (dashed line), whereas the different color bars represent the entropy of the material originating from different moonlets. After the first impact, the two moonlets remain on similar, interacting orbits and the impact again at $\sim 57\ \rm{hr}$. After the second impact, moon\#2 is scattered towards the Roche limit, where it disrupts. All projections are on the equatorial plane with one hemisphere removed. The upper left panel focuses on the surviving moonlet (Moon \#1).

\section*{Caption Movie S2}
Melt evolution of a sequence of two hit-and-run impacts. Snapshots of two colliding half moonlets at a distance of $1.5\,R_{{\rm Roche}}$ from the planet (first impact: $\beta=40^{o}$; $V_{{\rm imp}}=1.66\,V_{{\rm esc}}$; second impact: $\beta=44^{o}$; $V_{{\rm imp}}=1.56\,V_{{\rm esc}}$). The movie shows a slice of width 100 km centered in the equatorial plane of the largest surviving moon. The blue particles represent iron material and light/dark red particles represent the unmelted/melted magma material. Note that for emphasizing purposes, the melted particles are larger. In the first and second impact, melt is created mainly in the secondary stage of the impact, where material falls back to the surface.